\documentclass{aa}

\usepackage{verbatim}
\usepackage{graphicx}
\usepackage{epstopdf}
\usepackage{subfigure}
\usepackage{mathtools}
\usepackage{color,soul}

\newcommand{\mdot}{M$_\odot$ yr$^{-1}$}

\newcommand{\kms}{km~s$^{-1}$}

\title{Stellar Winds on the Main-Sequence I: Wind Model}

\titlerunning{Stellar Winds on the Main-Sequence I}

\author{C. P. Johnstone\inst{\ref{vienna}} \and M. G\"{u}del\inst{\ref{vienna}} \and T. L\"{u}ftinger\inst{\ref{vienna}} \and G. Toth\inst{\ref{michigan}} \and I. Brott\inst{\ref{vienna}}}

\institute{
University of Vienna, Department of Astrophysics, T\"{u}rkenschanzstrasse 17, 1180 Vienna, Austria \label{vienna} \and
Center for Space Environment Modeling, University of Michigan, Ann Arbor, MI 48109, USA \label{michigan}
}

\abstract{}{
We develop a method for estimating the properties of stellar winds for low-mass main-sequence stars between masses of 0.4~M$_\odot$ and 1.1~M$_\odot$ at a range of distances from the star.
}{
We use 1D thermal pressure driven hydrodynamic wind models run using the \emph{Versatile Advection Code}.
Using \emph{in situ} measurements of the solar wind, we produce models for the slow and fast components of the solar wind. 
We consider two radically different methods for scaling the base temperature of the wind to other stars: in Model~A, we assume that wind temperatures are fundamentally linked to coronal temperatures, and in Model~B, we assume that the sound speed at the base of the wind is a fixed fraction of the escape velocity. 
In Paper~II of this series, we use observationally constrained rotational evolution models to derive wind mass loss rates.
}{
Our model for the solar wind provides an excellent description of the real solar wind far from the solar surface, but is unrealistic within the solar corona. 
We run a grid of 1200 wind models to derive relations for the wind properties as a function of stellar mass, radius, and wind temperature . 
Using these results, we explore how wind properties depend on stellar mass and rotation.
}{
Based on our two assumptions about the scaling of the wind temperature, we argue that there is still significant uncertainty in how these properties should be determined.
Resolution of this uncertainty will probably require both the application of solar wind physics to other stars and detailed observational constraints on the properties of stellar winds.
In the final section of this paper, we give step by step instructions for how to apply our results to calculate the stellar wind conditions far from the stellar surface. 
}

\begin{document}

\maketitle


\section{Introduction}

Based on analogies with the Sun, and observations of stellar rotational evolution, it is known that all low-mass main-sequence stars lose mass through stellar winds.
It is thought that the winds of other low-mass stars are analogous to the solar wind, which is fully ionised and flows away from the Sun in all directions at supersonic speeds.
These winds can significantly influence the environments surrounding their host stars.
Currently, the properties of low-mass stellar winds are poorly understood, both theoretically and observationally. 
In this paper, we construct a model for the solar wind and develop methods for scaling it to other stars.
This paper is accompanied by a second paper in which we couple our wind model to a rotational evolution model and explore how stellar wind properties evolve on the main-sequence for stars with a range of masses.  
Our motivation for these papers is to develop an understanding of the stellar environments of potentially habitable planets, and how stellar activity influences the evolution of planetary atmospheres.

It is important that we understand the properties of stellar wind so that we can predict how they influence planetary atmospheres.
Winds can influence planets both directly and indirectly. 
By removing angular momentum, the winds cause their host stars to spin down with time (\citealt{1967ApJ...148..217W}; \citealt{1967ApJ...150..551K}; \citealt{1972ApJ...171..565S}). 
Since rotation is the most important parameter that determines the strength of a star's magnetic dynamo, this spin down leads to a decrease in magnetic activity of low-mass stars as they age (\citealt{1972ApJ...171..565S}; \citealt{1997ApJ...483..947G}; \citealt{2014MNRAS.441.2361V}), and a corresponding decrease in the emission of UV and X-ray radiation (\citealt{1997ApJ...483..947G}; \citealt{2005ApJ...622..680R}).
Planetary atmospheres are highly sensitive to the level of stellar high-energy radiation (\citealt{2010AsBio..10...45L}).
In particular, absorption of EUV radiation causes planetary atmospheres to expand, and can lead to significant hydrodynamic escape  (\citealt{2003ApJ...598L.121L}; \citealt{2005ApJ...621.1049T}; \citealt{2008JGRE..113.5008T}; \citealt{2010Icar..210....1L}; \citealt{2014MNRAS.439.3225L}).
The winds can also directly influence planetary atmospheres. 
Planets are exposed to a continuous flow of supersonic electrons and protons from the wind which compress their magnetospheres and lead to significant non-thermal escape and erosion of the upper atmospheres (\citealt{2008Natur.451..970H}; \citealt{2012ApJ...744...70K}; \citealt{2014A&A...562A.116K}).
In addition, winds can strongly influence the number of galactic cosmic rays that reach the inner regions of stellar systems, changing the flux of cosmic rays incident on a planet's atmosphere (\citealt{2012ApJ...760...85C}). 
Charge exchange between heavy ions in the wind and planetary neutrals produces X-ray radiation which can be very significant for hot Jupiters (\citealt{2015ApJ...799L..15K}).
Therefore, it is clear that an understanding of the evolution of stellar rotation and winds is necessary for a proper understanding of the formation of habitable planetary environments.

This is the first in a series of papers studying the properties and evolution of stellar winds on the main-sequence between masses of 0.4~M$_\odot$ and 1.1~M$_\odot$. 
In Section~\ref{sect:summary}, we summarise the current observational and theoretical knowledge of stellar winds. 
In Section~\ref{sect:solarwindmodel}, we develop a model for the slow and fast components of the solar wind using the \emph{Versatile Advection Code}. 
In Section~\ref{sect:scaling}, we discuss methods for scaling the solar wind model to other stars and develop two separate models based on different assumptions about the scaling of the wind temperature.
In Section~\ref{sect:results}, we present a grid of wind models with different stellar masses, stellar radii, and wind temperatures and discuss their properties. 
In Section~\ref{sect:summaryfinal}, we summarise our calculations, explain how our model can be applied by others in a simple way, and discuss possible solutions to the open questions in our model.

\section{Stellar Winds in Theory and Practice} \label{sect:summary}


The majority of what we know about stellar winds comes from our knowledge of the solar wind. 
The solar wind is known to be generated by the solar magnetic field, though the mechanisms responsible are poorly understood (e.g. \citealt{2009LRSP....6....3C}). 
Unfortunately, given this uncertainty, how the example of the solar wind should be applied to other stars is highly uncertain. 

Most of the detected magnetic field on the Sun is in the photosphere and is contained within discrete \mbox{kG-strength} field structures such as sunspots and pores that cover a small fraction of the solar surface.
The rest of the photosphere is covered by a much more complex small-scale field (\citealt{1973SoPh...32...41S}; \citealt{1993SSRv...63....1S}; \citealt{2011A&A...529A..42S}).
The photospheric magnetic field extends outwards, through the chromosphere and transition region, and into the corona where it is responsible for heating the plasma to MK temperatures.
Low in the corona, the \mbox{plasma-$\beta$} (i.e. the ratio of the thermal pressure to the magnetic pressure) is very low, meaning that the magnetic field dominates the dynamics of the plasma.
This leads to most of the lower corona being covered by regions of closed magnetic field where the magnetic field prevents the coronal plasma from expanding away from the Sun.
The rest of the corona is covered in regions of open magnetic field, where the geometry of the field lines means that the coronal plasma is able to expand into interplanetary space.
Typically, regions of open field cover between 5\% and 20\% of the solar surface (\citealt{1990ApJ...355..726W}).
The geometry of the Sun's global magnetic field changes periodically with the solar cycle, and has a simple axisymmetric structure at cycle minimum and a more complex non-axisymmetric structure at cycle maximum.
This leads to the Sun having large coronal holes at the poles and regions of closed field around the equator during cycle minimum, and complex distributions of open and closed field over the surface during cycle maximum (\citealt{1978SoPh...56..161B}; \citealt{1990ApJ...365..372W}).

\begin{figure}
\includegraphics[trim = 13mm 8mm 10mm 4mm, clip=true,width=0.49\textwidth]{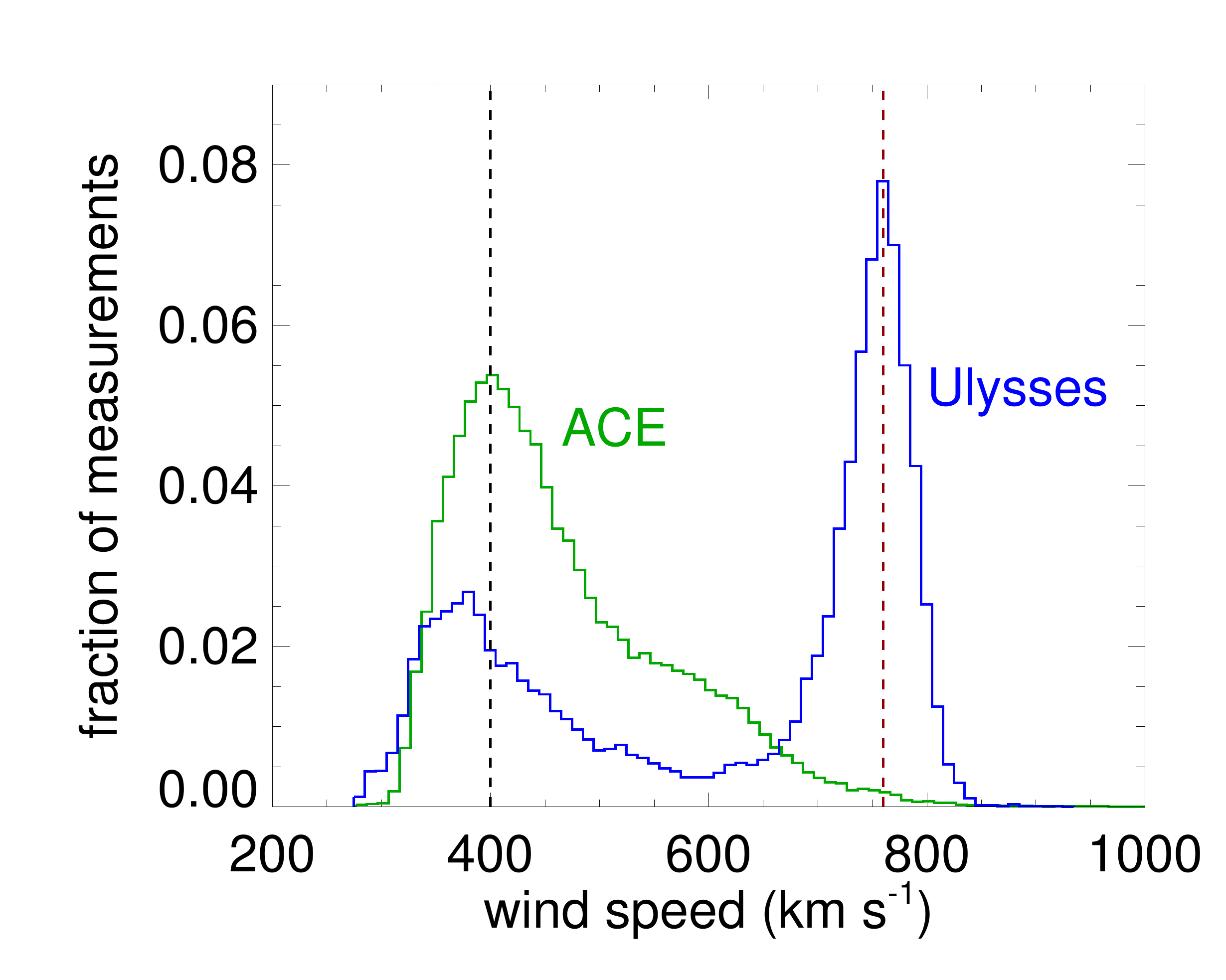}
\caption{
Histogram showing hourly average solar wind speeds measured by the spacecraft \emph{ACE} (\emph{green}) and \emph{Ulysses} (\emph{blue}).
The vertical dashed black and red lines show typical slow and fast wind velocities respectively. 
We have only considered \emph{Ulysses} measurements from when the spacecraft was less than 2.5~AU from the Sun. 
The distributions from the two spacecraft are different because they probed the solar wind at different latitudes.
The orbit of \emph{ACE} is in the equatorial plane, and therefore \emph{ACE} measured mostly the slow component of the solar wind.
The orbit of \emph{Ulysses} is almost perpendicular to the equatorial plane, and therefore \emph{Ulysses} measured both the slow and fast components of the solar wind.
}
 \label{fig:ACEandUlyssesSpeeds}
\end{figure}

 
Closely connected to the solar magnetic field is the solar wind. 
The solar wind expands away from the Sun in all directions at all times with a mass loss rate of \mbox{$\sim2 \times 10^{-14}$~\mdot} and can be roughly broken down into two distinct components based on the wind speed, as can be seen in Fig.~\ref{fig:ACEandUlyssesSpeeds}. 
The slow wind travels at typical speeds of $\sim$400~\kms \hspace{0.0mm} and the fast wind travels at typical speeds of $\sim$760~\kms.
The wind speed correlates well with the geometry of the Sun's coronal magnetic field, with fast wind originating from coronal holes and slow wind originating from regions above closed field lines or the edges of coronal holes (\citealt{1973SoPh...29..505K}; \citealt{1976SoPh...46..303N}).
 \citet{1990ApJ...355..726W} and \citet{2000JGR...10510465A} showed that the wind speed can be predicted based on the radial expansion of the magnetic field in the corona, such that slow wind emanates from regions where the magnetic field expands quickly and fast wind emanates from regions where the magnetic field expands slowly. 
A consequence of this connection is that the structure of the solar wind changes over the solar cycle (e.g. \citealt{2009JGRA..114.1109E}).
At cycle minimum, the wind has a simple structure with fast wind emanating from the poles and slow wind emanating from the equator.
At cycle maximum, when the magnetic field is complex, the wind has a much more complex structure, with slow and fast winds emanating from all latitudes. 
By comparing \mbox{\emph{in situ}} measurements of the solar wind with 3D extrapolations of photospheric magnetograms, \citet{2010ApJ...715L.121W} showed that over the solar surface, the mass flux in the wind is approximately proportional to the magnetic field strength.
However, due to the non-uniform expansion of the magnetic field with radial distance from the photosphere, the mass flux far from the Sun is approximately uniform in all directions, and is approximately constant in time (\citealt{2011MNRAS.417.2592C}). 


\citet{1958ApJ...128..664P} showed that a supersonic wind of a few hundred \kms \hspace{0.0mm} arises naturally from thermal pressure gradients if the coronal plasma is heated to MK temperatures.
The original model of \citet{1958ApJ...128..664P} assumed that the gas is isothermal, which implicitly leads to heating of the wind as it expands.
Heating of the wind within and above the corona is necessary for the expansion to take place; without it, the wind will not have enough energy to be lifted out of the star's gravitational potential well without having unrealistically high temperatures at the base of the corona (\citealt{1999isw..book.....L}; \citealt{2000JASTP..62.1515G}).
It is now recognised that other processes, such as Alfv\'{e}n wave pressure, likely also contribute to the wind acceleration (\citealt{2009LRSP....6....3C}).
In a purely hydrodynamic pressure driven wind, the amount of heat that is given to the gas and the spatial distribution of this heating are the most important parameters for determining the wind properties. 
Heating that takes place in the lower subsonic part of the wind contributes mostly to the mass flux.
Heating that takes place in the supersonic part of the wind is unable to change the mass flux, and instead contributes to the wind speed.

There are good observational reasons to believe that close to the Sun, the heating of the wind is sufficient to keep it approximately isothermal (\citealt{1977ApJ...217..296S}; \citealt{2003ApJ...595L..57R}). 
Far from the Sun, \emph{in situ} measurements by multiple spacecraft have shown that the wind is hotter than would be expected from adiabatic expansion, and therefore must continue to be heated out to several AU (\citealt{1979SSRv...23..217S}; \citealt{1995JGR...100...13T}; \citealt{2009JGRA..114.1109E}). 
Unfortunately, how the plasma is heated to such temperatures is currently poorly understood.

The isothermal wind model of \citet{1958ApJ...128..664P} leads to unrealistic acceleration of the wind at large distances from the Sun. 
A more general model assumes that the wind can be described by a polytropic equation of state (e.g. \citealt{1965SSRv....4..666P}; \citealt{1999A&A...343..251K}; \citealt{2011AdSpR..48.1958J}). 
This assumption, discussed in more detail in Section~\ref{sect:solarwindmodel}, gives a good description of the solar wind properties.
Alternatively, solar wind models have been developed that use heating functions that ignore the physical mechanisms responsible for heating the wind, but instead are empirically constrained (\citealt{2000JGR...10525053G}; \citealt{2004JGRA..109.1102M}). 
Recently, solar wind models have been developed that do not rely on \emph{ad hoc} assumptions about the heating of the wind, but instead produce the wind in a more self-consistent way based on models of the possible physical mechanisms that drive the solar wind (e.g. \citealt{2005ApJ...632L..49S}; \citealt{2007ApJS..171..520C}; \citealt{2010ApJ...708L.116V}).


Unlike the strong radiatively driven winds of hot luminous stars, the winds of low-mass stars are very difficult to detect due to their small mass loss rates (the mass loss rates of high-mass stars can be ten orders of magnitude higher than that of the current solar wind). 
The presence of winds on stars other than the Sun was predicted by \citet{1960ApJ...132..821P}, and from the observed rotational evolution of stars on the main-sequence, it is known that all low-mass main-sequence stars have magnetised winds.
However, other than the solar wind, there has been no direct detection of a wind from a low-mass star. 
There have been several attempts to directly observe free-free thermal radiation at radio wavelengths, all of which have resulted in non-detections, putting important upper limits on the mass loss rates (\citealt{1990ApJ...361..220B}; \citealt{1993ApJ...406..247D}; \citealt{1997A&A...319..578V}; \citealt{2000GeoRL..27..501G}).
The most sensitive limits on the winds of solar mass stars were derived for three young solar analogues by \citet{2000GeoRL..27..501G}, who found upper limits on the mass loss rates of \mbox{$\sim 5 \times 10^{-11}$~\mdot}.
In addition to non-detections of wind emission, upper limits on the strengths of winds can be derived from the detection of coronal radio flares, since such detections imply that the winds are optically thin all the way down to the stellar surface (\citealt{2002ARA&A..40..217G}).
For example, \citet{1996ApJ...462L..91L} used this criterion to derive an upper limit of \mbox{$10^{-12}$~\mdot\hspace{0mm}} on the mass loss rate for the M-dwarf YZ~CMi.
A more indirect method for determining wind mass loss rates for M-dwarfs with white dwarf binary companions was applied to six systems by \citet{2006ApJ...652..636D}.
They attempted to measure accretion rates onto the surfaces of the white dwarfs, and use that to constrain the mass loss rates of their M-dwarf companions.
For the three tight binaries in their sample, they found mass loss rates below $10^{-14}$~\mdot, and for the three systems with larger separations, they found values above $10^{-10}$~\mdot (which they considered unreasonably high).  
An alternative method for detecting stellar winds is to look for X-ray emission produced by charge exchange interactions between neutral interstellar hydrogen and the ionised wind.
\citet{2002ApJ...578..503W} were unable to detect such emission around the slowly rotating M-dwarf Proxima~Centauri, putting an upper limit on the mass loss rate of \mbox{$\sim 3 \times 10^{-13}$~\mdot}.

Another consequence of these charge exchange interactions is the build-up of a wall of hot neutral hydrogen at the edge of a stellar system's astrosphere (analogous to the Sun's heliosphere).
This requires that the stellar system is embedded in a region of the interstellar medium (ISM) that contains a large enough quantity of neutral hydrogen (\citealt{2001ApJ...547L..49W}).
The Ly$\alpha$ emission line is the most important feature of the UV spectrum of low-mass stars (\citealt{2012ApJ...750L..32F}; \citealt{2014ApJ...780...61L}), but due to strong absorption by interstellar neutral hydrogen, only a small fraction of the emitted Ly$\alpha$ flux is observed.
In addition to the strong interstellar absorption, the astrospheric hydrogen walls cause additional absorption of the Ly$\alpha$ line. 
Measuring this extra absorption requires reconstructing the intrinsic stellar Ly$\alpha$ emission line and then modelling ISM absorption.
Once measured, the extra absorption can be compared with the results of hydrodynamic models of \mbox{wind-ISM} interactions to estimate wind mass fluxes.
\citet{2001ApJ...547L..49W} applied this technique and predicted mass loss rates for the binary system $\alpha$~Centauri of 2~$\dot{\text{M}}_\odot$, and an upper limit for Proxima~Centauri of 0.2~$\dot{\text{M}}_\odot$, where $\dot{\text{M}}_\odot$ is the mass loss rate of the current solar wind.
Measurements using this technique are now available for several low-mass stars (\citealt{2002ApJ...574..412W}; \citealt{2005ApJ...628L.143W}; \citealt{2014ApJ...781L..33W}).
The general trend for the majority of the main-sequence stars is that at a given stellar radius, mass loss rate scales with age as $\dot{M}_\star \propto t^{-2.33}$ (\citealt{2005ApJ...628L.143W}).
Combining this with the well known result of \citet{1972ApJ...171..565S} that stars spin down with age according to $\Omega_\star \propto t^{-1/2}$ implies that $\dot{M}_\star \propto \Omega_\star^{4.66}$. 
However, this relation appears to break down for the most active stars in the sample, which all show mass loss rates that are much lower than would be predicted.
For example, \citet{2014ApJ...781L..33W} estimated a mass loss rate for the $\sim$300~Myr old solar analogue $\pi^1$~UMa of 0.5~$\dot{\text{M}}_\odot$.


Modelling the winds of low-mass stars is a tricky business.
The lack of clear observational constraints means that it remains unclear which, if any, of the existing models are reliable (including the model presented in this paper).
Most wind models can be divided roughly into two categories: those that try to apply existing knowledge of the physics of the solar wind to other stars (\citealt{2011ApJ...741...54C}; \citealt{2013PASJ...65...98S}), and those that scale the solar wind to other stars by assuming scaling relations between wind properties and stellar parameters, such as age, rotation rate, or coronal X-ray properties (e.g. \citealt{1992SvA....36...70B}; \citealt{2004A&A...425..753G}; \citealt{2007A&A...463...11H}; \citealt{2014arXiv1409.1237S}).
An example of the latter type of model was used by \citet{2004A&A...425..753G} and \citet{2007P&SS...55..618G}. 
Using a scaling relation between wind speed and stellar age derived by \citet{1980asfr.symp..293N} and the relation between wind ram pressure and rotation rate from \citet{2002ApJ...574..412W}, they predicted wind densities and velocities as a function of age for the Sun.
\citet{2007P&SS...55..618G} used these constraints as input into an isothermal Parker wind model to predict the wind properties as a function of distance from the star.  
\citet{2007A&A...463...11H} produced a model for the winds of low-mass stars based on an analytic polytropic solar wind model. 
They made predictions for the wind properties of other stars by assuming power-law dependences of wind base temperature and density on stellar angular velocity.

Due to recent advances in the understanding of the solar wind, stellar wind models have been developed that derive wind properties by considering the physics of the solar wind. 
\citet{2011ApJ...741...54C} derived an Alfv\'{e}n wave and MHD turbulence driven model that determines the mass flux from the stellar surface based on energy balance considerations in the transition region.
Their model combines this with observational knowledge of stellar magnetic fields to predict wind mass loss rates from low-mass stars based on a few basic stellar parameters.
For solar mass stars, this model predicts that at low rotation rates, the mass loss rate depends on stellar angular velocity as \mbox{$\dot{M}_\star \propto \Omega_\star^{1.6}$} (\citealt{2013A&A...556A..36G}), which leads to an approximate time dependence of the mass loss rate of \mbox{$\dot{M}_\star \propto t^{-0.9}$}. 
At high rotation rates, the mass loss rate saturates due to saturation in the magnetic field. 
Similarly, a model was presented by \citet{2013PASJ...65...98S} who used 1D magnetohydrodynamic simulations that self-consistently propagate energy from the stellar photosphere into the corona by Alfv\'{e}n waves and dissipate this energy, leading to an expansion of the coronal plasma and the formation of a wind. 
They predicted that the mass loss rates from solar mass stars vary with age as \mbox{$\dot{M}_\star \propto t^{-1.23}$} during the least active stages of their lives, but are saturated at the most active stages.
Unlike in the model of \citet{2011ApJ...741...54C}, this saturation happens due to enhanced radiative losses leading to a smaller fraction of the energy input into the magnetic field being converted into kinetic energy.


Although the application of the physics of the solar wind driving to other stars represents a significant advance in the study of stellar winds, there are still many uncertainties in both our understanding of the physics of the solar wind and in our understanding of how to apply solar wind models to other stars.
An example of the latter type of uncertainty can be seen in the \citet{2011ApJ...741...54C} model, which is heavily dependent on estimates of the magnetic flux filling factor, $f$ (commonly seen in the expression for the surface averaged field strength $fB$). 
It is unclear to what extent this parameter can be derived from observations of low-mass main-sequence stars (\citealt{2012LRSP....9....1R}; \citealt{2014IAUS..302..156R}) and how well it can be predicted from stellar parameters such as mass and rotation rate.


The above models are, at most, one dimensional and do not take into account the complex dynamical interplay between the star's coronal magnetic field and the wind. 
With the recent increase in available computational resources, it has become possible to study winds using numerical magnetohydrodynamic simulations (e.g. \citealt{2008ApJ...678.1109M}; \citealt{2009ApJ...699..441V}; \citealt{2014ApJ...783...55C}; \citealt{2015ApJ...798..116R}).
Such simulations can now take into account real stellar magnetic field structures reconstructed using the \mbox{Zeeman-Doppler Imaging} technique (\citealt{2011MNRAS.412..351V}; \citealt{2013MNRAS.431..528J}; \citealt{2014MNRAS.438.1162V}; \citealt{2014ApJ...790...57C}).
These models are a major advance in the study of stellar winds, though they mostly contain similar free parameters to the simpler scaling models discussed above which need to be set before they can predict the wind speeds and mass loss rates.
However, they do have important advantages; probably the most significant of these is their ability to self-consistently predict the wind torque on the star once the other wind properties have been set, which cannot be done with simpler models. 
For example, \citet{2012ApJ...754L..26M} produced a grid of 2D MHD models assuming dipole field geometries and derived a formula for the wind torque as a function of stellar parameters, magnetic field strength, and wind mass loss rate. 
Models run in 3D using realistic magnetic field structures are also being used to study wind torques in excellent detail (\citealt{2014MNRAS.438.1162V}; \citealt{2014ApJ...783...55C}).


It has been suggested that a separate type of wind might be operating on more active stars. 
Based on the observed correlations between solar flares and coronal mass ejections (CMEs), and the correlations between stellar magnetic activity and flare rates, highly active stars could have winds that are dominated by CMEs (\citealt{2007AsBio...7..167K}; \citealt{2012ApJ...760....9A}; \citealt{2013ApJ...764..170D}).
CME dominated winds, if they exist, could have significant impacts on the magnetospheres and atmospheres of planets (\citealt{2007AsBio...7..167K}; \citealt{2007AsBio...7..185L}; \citealt{2013AsBio..13.1030K}). 
However, the existence of such strong CME activity remains controversial (\citealt{2014MNRAS.443..898L}) and is not discussed further in this paper.

\section{Solar Wind Model} \label{sect:solarwindmodel}

\subsection{Numerical model} \label{sect:numericalmodel}


The isothermal stellar wind model of \citet{1958ApJ...128..664P} has been shown to provide an acceptable description of the solar wind within 5~AU (\citealt{1999A&A...348..614M}), but it leads to unrealistic heating and too much acceleration far from the solar surface.
Polytropic wind models provide a more realistic description of the solar wind.
In this section, we develop a 1D polytropic model for the solar wind with a spatially varying polytropic index that we use as the basis for our stellar wind model. 
Since such models cannot be solved analytically, we use hydrodynamic simulations run with the \emph{Versatile Advection Code} (VAC).  
VAC was developed by \citet{1996ApL&C..34..245T} and \citet{1997JCoPh.138..981T} as a general tool for performing hydrodynamic and magnetohydrodynamic simulations in 1D, 2D, and 3D. 
The code has been used extensively for modelling the solar wind (e.g. \citealt{1999A&A...343..251K}; \citealt{2007ApJ...671L..77V}; \citealt{2008JGRA..113.8107Z}; \citealt{2011AdSpR..48.1958J}).
Our numerical hydrodynamic simulations are run using a 1D grid in spherical coordinates.
Our grid consists of 1000 cells with sizes in the radial direction that increase by a factor of 100 from the solar surface to the outer edge of the computational domain at 1~AU.
This allows us to have a good resolution close to the Sun while only using a relatively small number of cells. 
This setup is used in all simulations presented in this paper.


The fundamental driving mechanism for our wind is thermal pressure gradients. 
This is a simplification of the real solar wind, which is likely driven by a combination of thermal pressure gradients and other forces, such as wave pressure. 
\citet{2004ESASP.575..154C} used observations of the wind properties in coronal holes to estimate what fraction of the wind acceleration comes from thermal pressure for the fast wind. 
Very close to the Sun (\mbox{$r\lesssim4$~R$_\odot$}), thermal pressure acceleration dominates, but the influence of other forces becomes stronger further from the surface, and by $r\sim10$~R$_\odot$, thermal pressure contributes about half of the acceleration. 
Since in our fast wind models, most of the acceleration of the wind happens in the region close to the star where thermal pressure forces are dominant, our assumption that the winds are entirely pressure driven is an acceptable approximation.
However, since other forces likely play an appreciable role in the real solar wind, this represents a genuine limitation of our model, and of any other model that drives the wind with thermal pressure gradients only. 

Another approximation in our model is that the wind can be described as a single temperature collisional fluid.
This is true low in the solar corona, but due to the rapid decrease in plasma density as the wind expands, the wind quickly becomes collisionless. 
In a collisionless fluid, thermodynamic equilibrium is unlikely to be maintained, and the temperatures of different species evolve differently as the wind expands into the heliosphere.
One consequence of this is the decoupling of electron and proton temperatures.
Spacecraft measurements of the solar wind have shown that the temperatures of the two species can be a factor of a few different far from the Sun (e.g. \citealt{1968JGR....73.4999M}; \citealt{1998JGR...103.9553N}; \citealt{2009ApJ...702.1604C}).
The temperatures for individual species are also anisotropic, with $T_\perp \ne T_\parallel$, where $T_\perp$ and $T_\parallel$ are the temperatures perpendicular and parallel to the ambient magnetic field (\citealt{1975JGR....80.4181F}; \citealt{2003ApJ...585.1147S}).
Properly taking into account all of these properties of the solar wind would require multi-fluid kinetic models.
However, such complex models would be inappropriate in this paper. 
For simplicity, we assume that $T = T_p = T_e$, and we constrain the wind temperatures using measured proton temperatures. 
This assumption is likely to be reasonable given that the wind acceleration in our model happens mostly close to the Sun where the proton and electron temperatures are similar.

An important part of any solar wind model is that the wind is heated as it expands. 
We do this implicitly by assuming a polytropic equation of state, which relates the gas pressure to the density by

\begin{equation} \label{eqn:polytropic}
p = K \rho^\alpha,
\end{equation}

\noindent where $\alpha$ is the polytropic index and we refer to $K$ as the `polytropic constant'.
The value of $K$ is set by the wind density, $\rho$, and temperature, $T$, by 

\begin{equation} \label{eqn:polyK}
K = \frac{k_{\text{B}} \rho^{1-\alpha}}{\mu m_{\text{p}}} T,
\end{equation}

\noindent where $k_{\text{B}}$ is the Boltzmann constant, $m_{\text{p}}$ is the proton mass, and $\mu$ is the average mass per particle in units of $m_{\text{p}}$.
Based on the solar wind, we set \mbox{$\mu=0.6$} in all our models.
In general, the polytropic equation of state assumes that both $K$ and $\alpha$ are held uniform in space and constant in time. 
This arises naturally if the wind densities and temperatures have power-law dependences on the distance from the central star (if \mbox{$\rho \propto r^{\beta}$} and \mbox{$T \propto r^{\delta}$}, then Eqn.~\ref{eqn:polytropic} can be derived with \mbox{$\alpha = 1 + \delta/\beta$}).
Assuming a polytropic equation of state is equivalent to assuming that the wind temperature depends on mass density by $T\propto \rho^{\alpha-1}$.
By making this assumption, it is unnecessary to include the energy equation in our hydrodynamic simulations. 
The parameter $\alpha$ determines the temperature profiles for the expanding wind.
For a monatomic wind, a value of $\alpha$ of 5/3 is equivalent to a wind that is expanding adiabatically. 
A value of $\alpha < 5/3$ means that there is implicit heating of the expanding wind, with $\alpha = 1$ corresponding to an isothermal wind. 
A value of $\alpha < 1$ corresponds to an increasing temperature with radial distance from the Sun. 
The parameter $\alpha$ is not the adiabatic index $\gamma = C_p/C_V$, and Eqn.~\ref{eqn:polytropic} should not be confused with the correlation between pressure and density for an adiabatic process, where $p \propto \rho^\gamma$. 
When we assume a polytropic equation of state, we are assuming that the wind is not adiabatic, but is being heated in such a way that Eqn.~\ref{eqn:polytropic} is reproduced on large scales.
We are not making an assumption about the microscopic physics of the wind plasma.
It is possible to mimic the effects of the polytropic equation of state in simulations that do include the energy equation by varying $\gamma$, so long as $\gamma > 1$.

Our solar wind model uses a value of $\alpha$ that varies with distance from the Sun.
In the real solar wind, within radii of about 1.15~R$_\odot$, the temperature of the expanding wind increases with height (\citealt{1998A&A...336L..90D}) and therefore, a value of $\alpha$ less than unity should be used.
However, as we show in the following section, our wind model is not physically realistic close to the solar surface, and we therefore choose a simpler model.
To crudely mimic this increase and subsequent decrease in temperature with radial distance from the solar surface, we assume that the wind is isothermal ($\alpha=1.0$) within 1.3~R$_\odot$ (i.e. within 0.3~R$_\odot$ of the solar surface). 
Beyond 1.3~R$_\odot$, we assume the wind close to the Sun is described by $\alpha_{\text{in}}$ and the wind far from the Sun is described by $\alpha_{\text{out}}$. 
To ensure a smooth transition in the temperature structure between the inner and outer regions, we vary $\alpha$ linearly from $\alpha_{\text{in}}$ to $\alpha_{\text{out}}$ between 15~R$_\odot$ and 25~R$_\odot$. 
Although there is little observational constraint on exactly where $\alpha$ should change from $\alpha_{\text{in}}$ to $\alpha_{\text{out}}$, these values represent reasonable guesses.
For the wind to have a reasonable temperature structure in the region where $\alpha$ varies, it is necessary to also vary the polytropic constant, $K$ (from Eqn.~\ref{eqn:polytropic}), though we do this in a way that is determined by the variations in $\alpha$ and does not add extra free parameters into the model.
The algorithm that we use to vary $K$ with radius is described in Appendix~\ref{appendix:polyK}.

\begin{figure*}
\includegraphics[trim = 13mm 5mm 10mm 0mm, clip=true,width=0.49\textwidth]{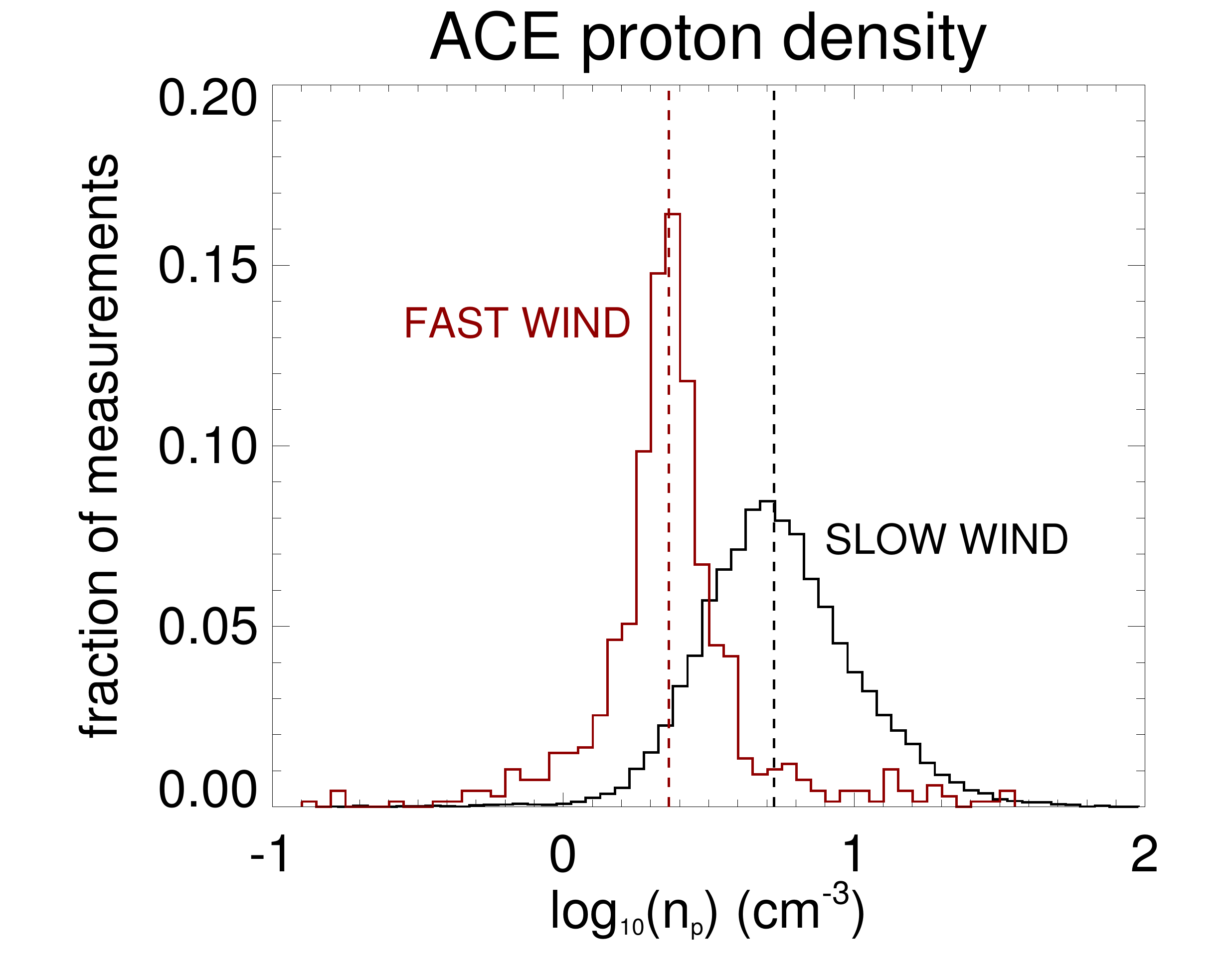}
\includegraphics[trim = 13mm 5mm 10mm 0mm, clip=true,width=0.49\textwidth]{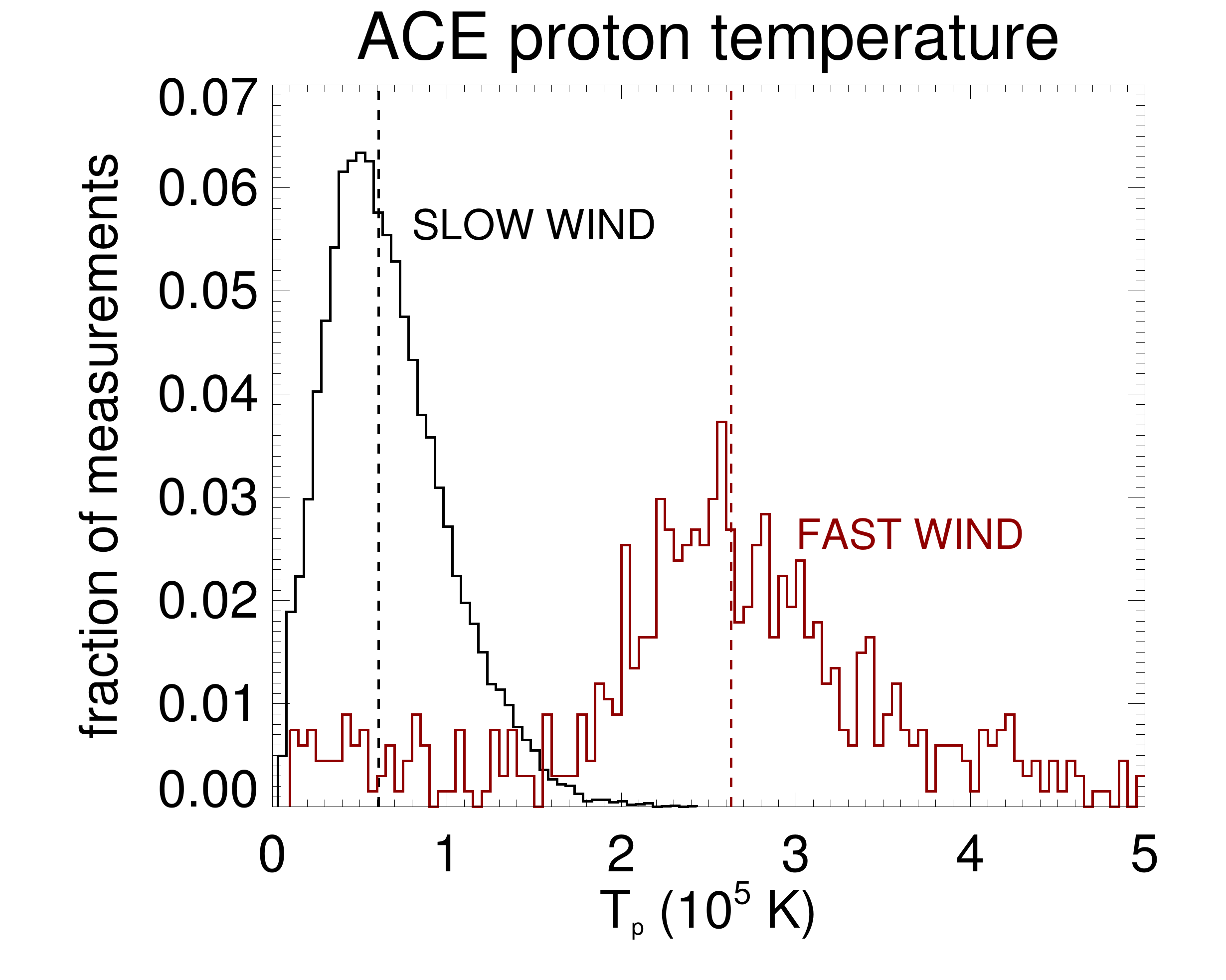}
\includegraphics[trim = 13mm 5mm 10mm 0mm, clip=true,width=0.49\textwidth]{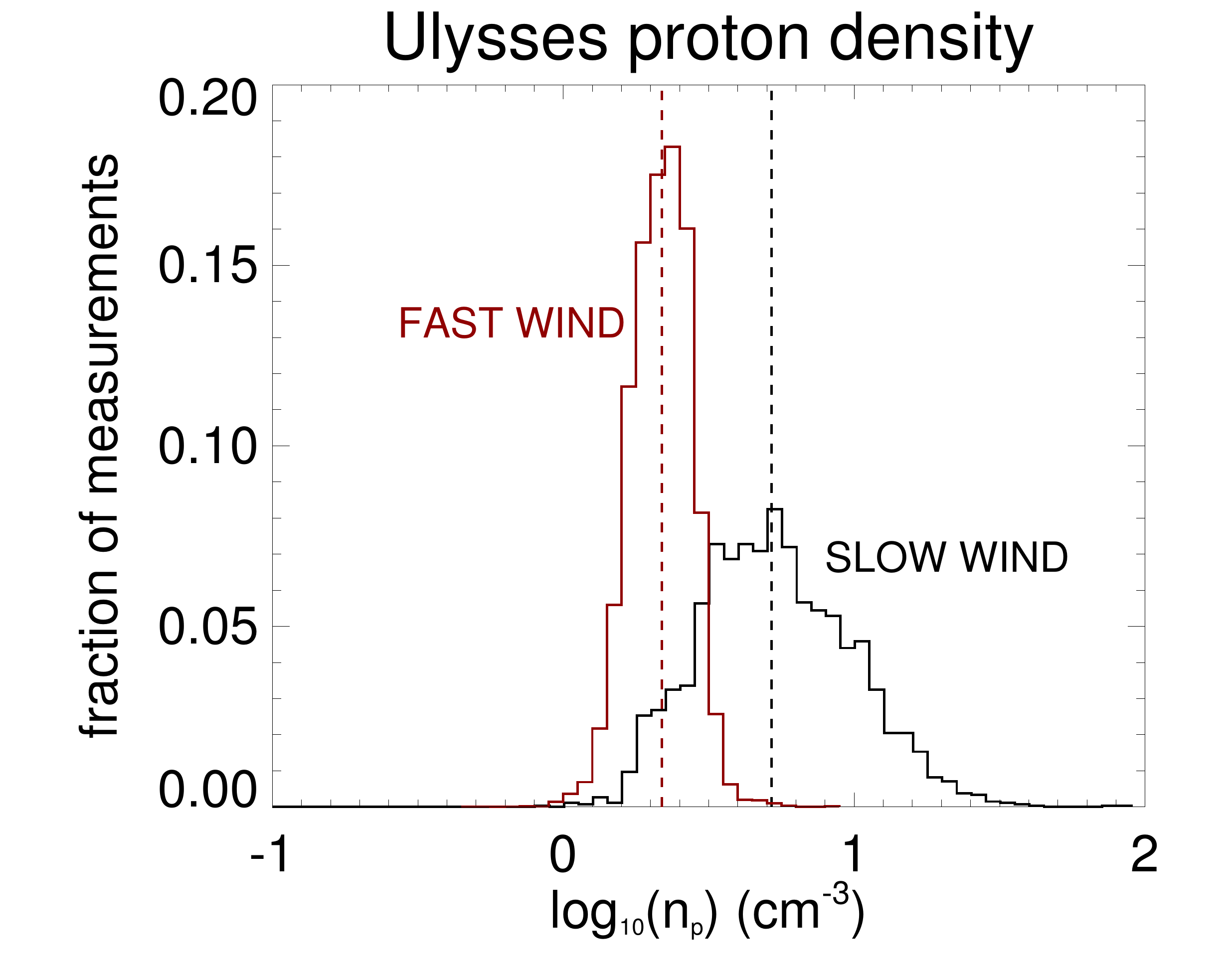}
\hspace{2.5mm}
\includegraphics[trim = 13mm 5mm 10mm 0mm, clip=true,width=0.49\textwidth]{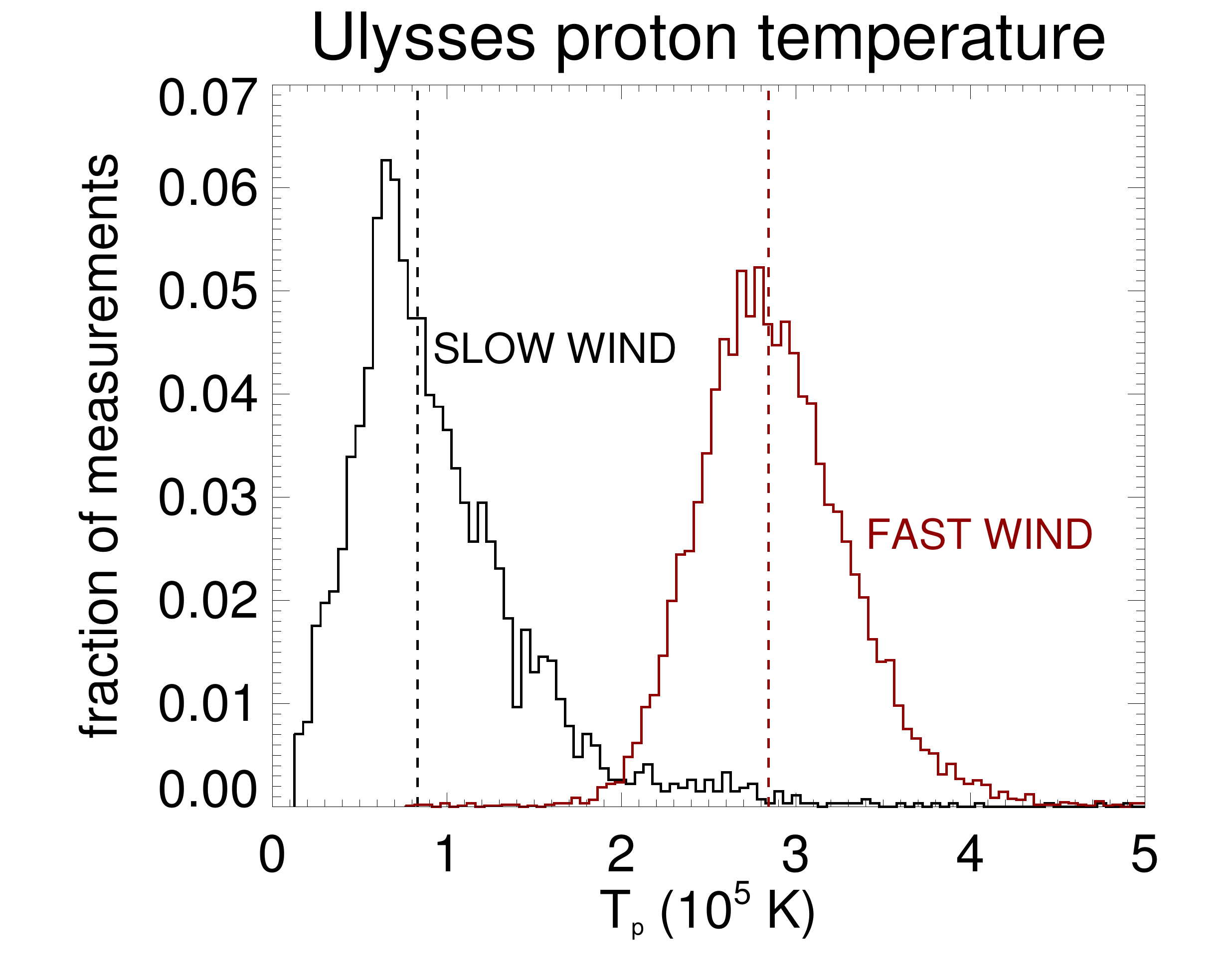}
\caption{
Histograms showing the proton densities (\emph{left column}) and proton temperatures (\emph{right column}) of the solar wind at 1~AU. 
The values are measured \emph{in situ} by the \emph{ACE} spacecraft (\emph{upper row}) and the \emph{Ulysses} spacecraft (\emph{lower row}). 
The black lines show the measurements of the slow wind, defined as all wind with speeds at 1~AU between 380~\kms\hspace{0mm} and 420~\kms.
The red lines show the measurements of the fast wind, defined as all wind with speeds at 1~AU between 740~\kms\hspace{0mm} and 780~\kms.
The vertical dashed lines show the median values of each quantity, as given in Table~\ref{tbl:ACEandUlysses}. 
\emph{Ulysses} was on an eccentric orbit between 1~AU and 5~AU, and therefore to construct the histograms for the density and temperature at 1~AU, we only consider data taken from when the spacecraft was less than 2.5~AU from the Sun, which we extrapolate back to 1~AU, as described in Section~\ref{ref:solarwindconstraints}.
}
 \label{fig:ACEandUlysses}
\end{figure*}

\begin{table*}
\begin{tabular}{c|ccccccc}
\hline
 		& mean $n_\text{p}$ (cm$^{-3}$) 	& median $n_\text{p}$ (cm$^{-3}$) 	& $\sigma_{n_\text{p}}$ (cm$^{-3}$) 		& mean $T_\text{p}$ (K) 		& median $T_\text{p}$ (K) 				& $\sigma_{T_\text{p}}$ (K) \\
\hline
Slow wind: \\
ACE 		& 6.79 					& $\mathbf{5.31}$			& 5.39 						& $ 6.61 \times 10^{4}$ 	& $\mathbf{6.10 \times 10^{4}}$ 	& $3.42 \times 10^{4}$ \\
Ulysses 	& 6.54 					& 5.20 					& 4.95 						& $9.60 \times 10^{4}$ 	& $8.33 \times 10^{4}$ 			& $5.57 \times 10^{4}$ \\
Fast wind:\\
ACE 		& 3.04 					& 2.31 					& 3.55 						& $2.81 \times 10^{5}$ 	& $2.63 \times 10^{5}$ 			& $1.39 \times 10^{5}$ \\
Ulysses 	& 2.22 					& $\mathbf{2.18}$ 			& 0.53 						& $2.88 \times 10^{5}$ 	& $\mathbf{2.84 \times 10^{5}}$ 	& $4.52 \times 10^{4}$ \\
\hline
\end{tabular}
\caption{
Average densities and temperatures of the slow and fast components of the solar wind at 1~AU as measured by the \emph{ACE} and \emph{Ulysses} spacecraft. 
The slow wind is defined as wind with speeds between 380~\kms\hspace{0mm} and 420~\kms, and the fast wind is defined as wind with speeds between 740~\kms\hspace{0mm} and 780~\kms.
Details for how these quantities were derived can be found in Section~\ref{sect:solarwindmodel} and the caption of Fig.~\ref{fig:ACEandUlysses}.
We mark with bold font the quantities we use to constrain the free parameters in the slow and fast wind models.
}
\label{tbl:ACEandUlysses}
\end{table*}

\begin{figure}
\centering
\includegraphics[trim = 8mm 8mm 7mm 0mm, clip=true,width=0.49\textwidth]{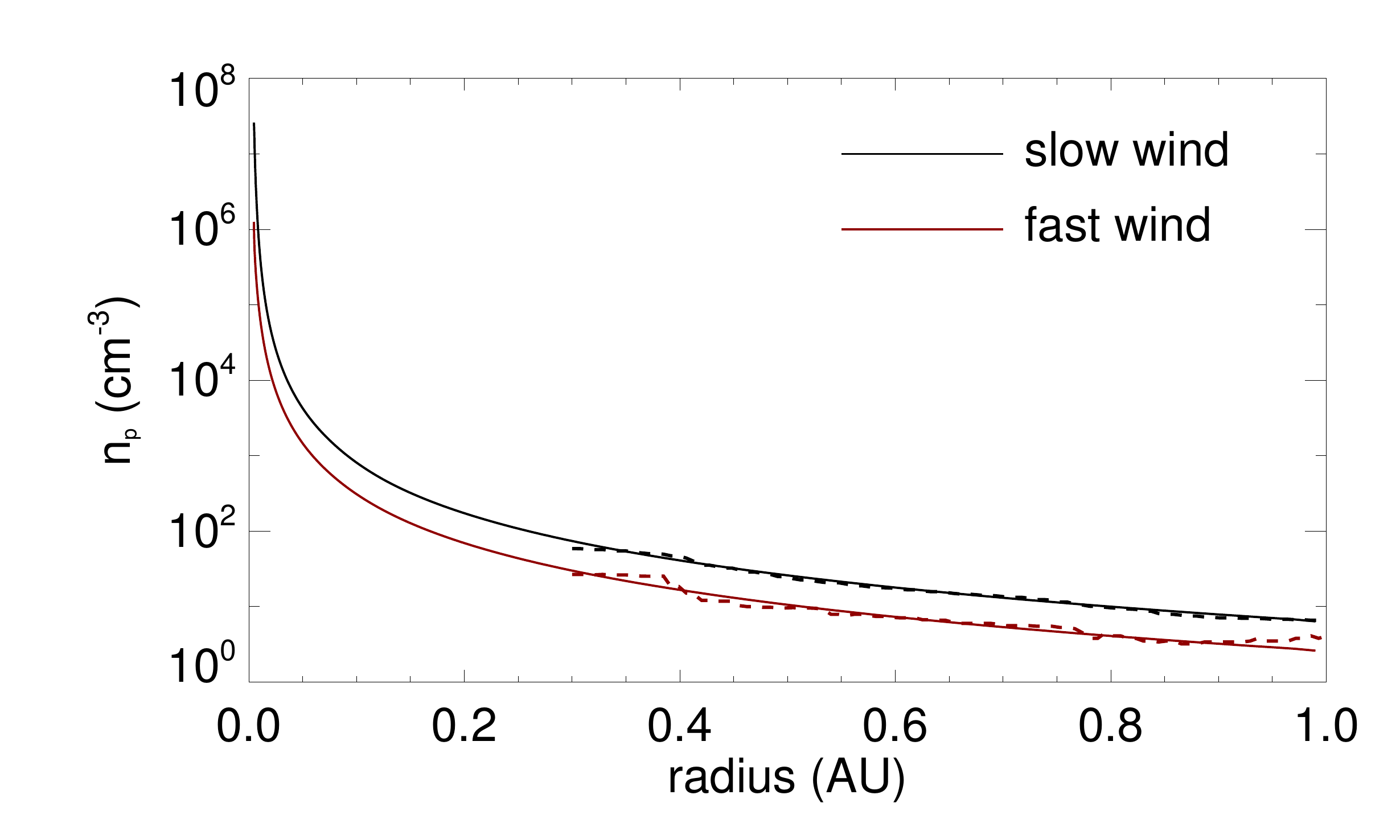}
\includegraphics[trim = 8mm 8mm 7mm 0mm, clip=true,width=0.49\textwidth]{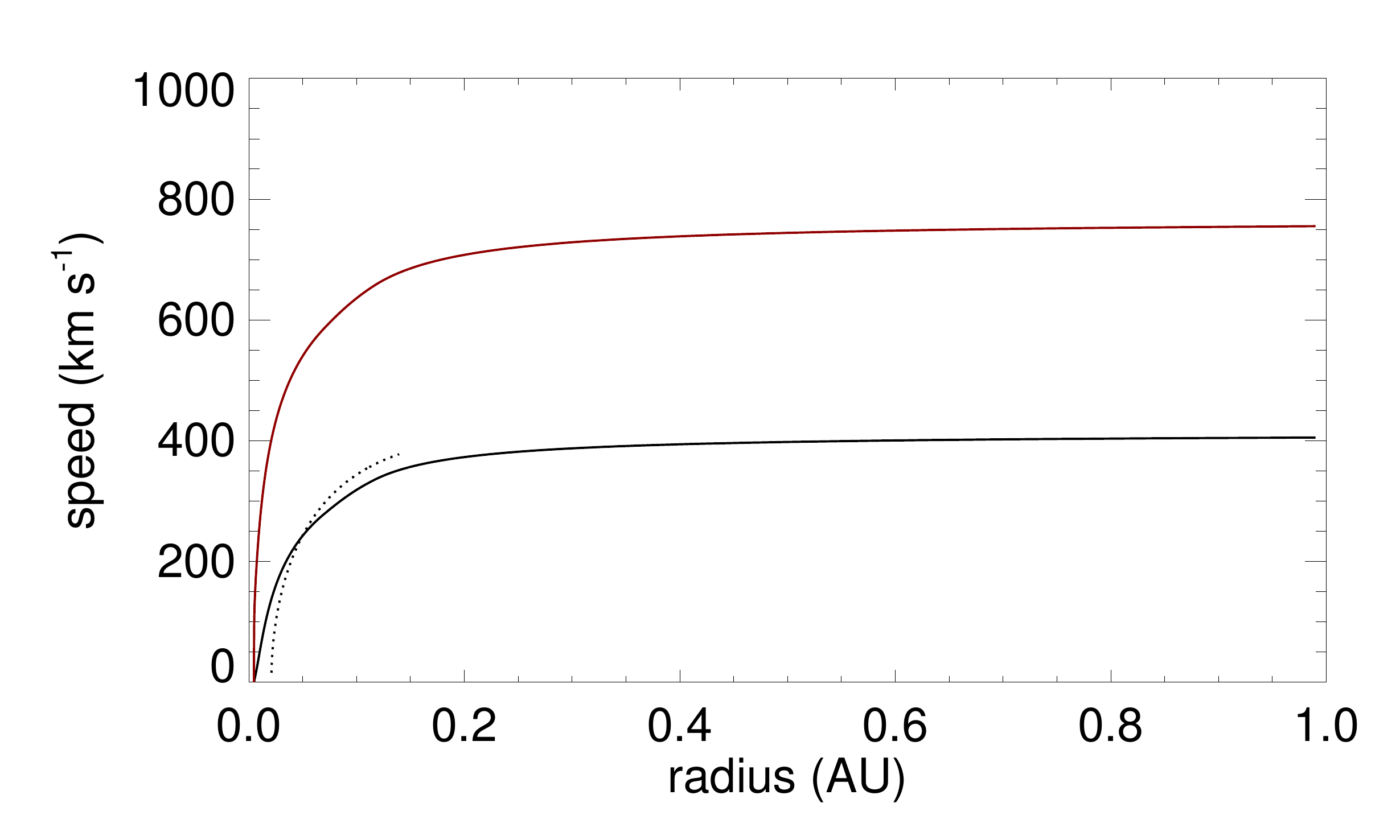}
\includegraphics[trim = 8mm 8mm 7mm 0mm, clip=true,width=0.49\textwidth]{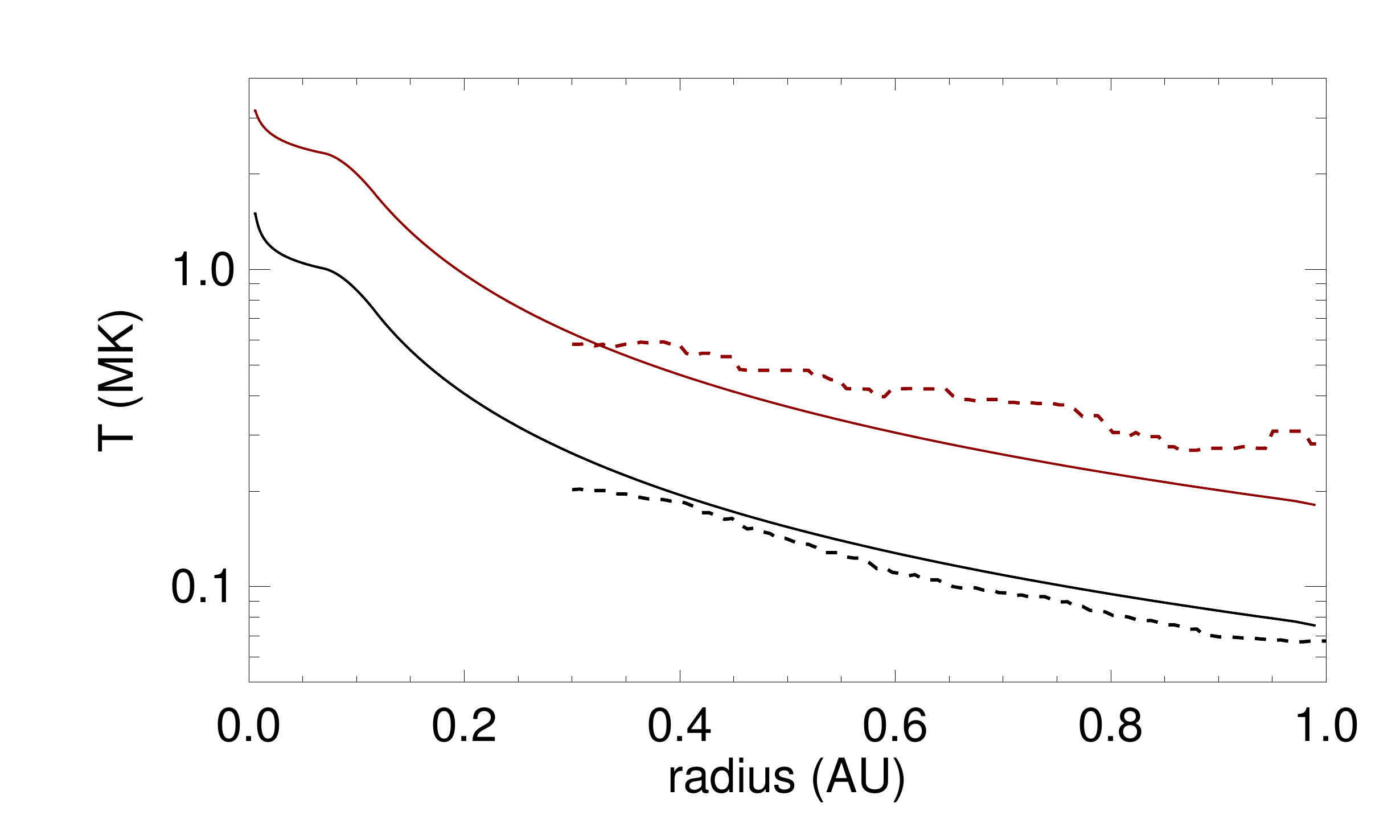}
\caption{
Plots showing the proton number density (\emph{upper panel}), wind speed (\emph{middle panel}), and temperature (\emph{lower panel}) of our slow and fast wind models.
As in Fig.~\ref{fig:ACEandUlysses}, the black and red lines corresponds to the slow and fast winds respectively.
The dashed lines in the upper and lower panels show the Helios~A and Helios~B measurements of the slow and fast wind properties, as described in Section~\ref{ref:solarwindconstraints}. 
The dotted line in the middle panel shows the radial speed profile of the slow wind estimated by \citet{1997ApJ...484..472S}.
The line corresponds to $v = \sqrt{v_\text{a}^2 \left( 1 - e^{-(r-r_1)/r_\text{a}} \right)}$, where $v_\text{a}=418.7$~\kms, $r_\text{a} = 15.2$~R$_\odot$, and $r_1=4.5$~R$_\odot$. 
}
 \label{fig:solarfastslowwinds}
\end{figure}


The value of $\alpha_{\text{out}}$ is relatively simple to determine since the radial structure of the solar wind has been measured \emph{in situ} by several spacecraft. 
Using data from the Helios~A spacecraft, which orbited the Sun on an eccentric orbit between 0.3~AU and 1.0~AU, \citet{1995JGR...100...13T} measured radial variations in wind density and temperature to determine a value of $\alpha$ of 1.46 with no clear dependence of $\alpha$ on wind speed.  
Other determinations of $\alpha$ in the literature are broadly consistent with this, though it is unclear if different values are needed for the slow and fast winds\footnotemark. 
\footnotetext{
For example, \citet{2009JGRA..114.1109E} measured radial gradients of various wind parameters using Ulysses data, which cover distances of approximately 1.0~AU to 5.0~AU from the Sun.
If density and temperature vary with radial distance as \mbox{$n \propto r^{\beta}$} and \mbox{$T \propto r^{\delta}$}, then the polytropic index is \mbox{$\alpha = 1 + \delta/\beta$}. 
For the fast wind, \citet{2009JGRA..114.1109E} obtained $\beta=-1.86$ and $\delta=-0.97$, giving \mbox{$\alpha = 1.52$}. 
For the slow wind, they obtained $\beta=-1.93$ and $\delta=-0.68$, giving \mbox{$\alpha = 1.35$}.
These values are broadly consistent with the \mbox{$\alpha = 1.46$} from \citet{1995JGR...100...13T}, with the differences possibly being a result of Helios~A and Ulysses probing different distances from the Sun and different heliolatitudes.
}
It is also likely that different values of $\alpha$ are needed for electrons and protons, given the collisionless nature of the wind far from the Sun.
We assume for our model that $\alpha_{\text{out}} = 1.51$, which is justified in the next section.
Determining $\alpha_{\text{in}}$ is more difficult because \emph{in~situ} measurements of the solar wind do not extend closer to the Sun than 0.3~AU, but all indications suggest that the solar corona and the inner regions of the wind are almost isothermal. 
For example, \citet{1977ApJ...217..296S} analysed observations of a polar coronal hole between 2~R$_\odot$ and 5~R$_\odot$ and found a value of $\alpha$ of 1.05 gives a good fit.  
This value has been used extensively in solar and stellar wind simulations (e.g. \citealt{1993MNRAS.262..936W}; \citealt{2008ApJ...678.1109M}).
Larger values, closer to 1.1 are also common in the literature (e.g. \citealt{2003ApJ...595L..57R}; \citealt{2012MNRAS.423.3285V}), and there is also evidence that $\alpha$ varies with position in the corona (\citealt{2007ApJ...654L.163C}).
For our models, we assume $\alpha_{\text{in}} = 1.05$ for the entire inner region of the wind.


Our solar wind model now has three free parameters: these are the base temperature, $T_0$, the base density, $n_0$, and $\alpha_{\text{out}}$. 
The base temperature determines the wind speed far from the star as well as having a strong influence on the mass flux and the radial temperature structure. 
Since the kinetic energy that wind particles gain as they accelerate comes originally from thermal energy in our model, higher base temperatures lead to higher wind speeds and larger mass loss rates.
The value of $\alpha_{\text{out}}$ has some influence on the wind speed far from the star, since there is still some acceleration of the wind beyond 20~R$_\odot$, but it primarily determines how quickly the wind temperature decreases as the wind expands. 
The base density has no influence on the wind speed in our model; instead, it influences strongly the mass loss rate of the wind.
When all other parameters are set, the density at every point in the wind, as well as the mass loss rate, is proportional to the base density.
We produce separate models for the slow solar wind and for the fast solar wind by choosing different values of $T_0$ and $n_0$.
For $\alpha_{\text{out}}$, we choose the value that gives the best fit to both the slow and fast winds.

\begin{table*}
\centering
\begin{tabular}{c|cccccccc}
\hline
 			& $n_0$ (cm$^{-3}$) 	& $T_0$ (MK) 		&$c_\text{s} / v_{\text{esc}}$	 	& $r_s$ (R$_\odot$) 	 	& $n_{\text{p},1\text{AU}}$ (cm$^{-3}$) 	& $v_{1\text{AU}}$ (km s$^{-1}$)	& $T_{1\text{AU}}$ (K) \\
\hline
slow wind		& $2.60 \times 10^7$	& 1.8				&0.329					&6.2					& 5.3					& 405					& $7.5 \times 10^4$\\ 
fast wind		& $1.05 \times 10^6$	& 3.8				&0.478					&2.4 					& 2.2					& 755					& $1.8 \times 10^5$\\ 
\hline
\end{tabular}
\caption{
Properties of our slow and fast solar wind models.
From left to right, the columns show the base proton density, the base temperature, the fraction of the base sound speed to the surface escape velocity, the radius of the sonic point, the proton density at 1~AU, the wind speed at 1~AU, and the temperature at 1~AU.
The first two columns represent free parameters in the model and the rest show results of the model.  
}
\label{tbl:solarwindmodelparams}
\end{table*}

\subsection{Constraints on the free parameters} \label{ref:solarwindconstraints}

In order to constrain the free parameters in our model, we use \emph{in situ} measurements of the solar wind from four spacecraft. 
These are the \emph{Advanced Compositional Explorer} (\emph{ACE}), \emph{Ulysses}, and the two Helios spacecraft\footnotemark. 
\emph{ACE} orbits the Sun at 1~AU close to the L1 Lagrange point and has been measuring the solar wind properties since 1998.
Given its position in the ecliptic plane, \emph{ACE} primarily measures the slow solar wind.
\emph{Ulysses} on the other hand orbited the Sun on an eccentric orbit that took it between 1.3~AU and 5.4~AU and up to \mbox{$80^\circ$} above and below the ecliptic plane.
\emph{Ulysses} measured the solar wind for almost 19 years, allowing it to sample the fast wind very well. 
In this paper, we only consider Ulysses data from when the spacecraft was less than 2.5~AU from the Sun and assume that the wind speed does not change significantly between 1~AU and 2.5~AU.
That this is a reasonable assumption can be seen from the fact that the average wind properties that we derive from \emph{Ulysses} are very similar to the average wind properties derived from \emph{ACE}.
There is also no evidence in the \emph{Ulysses} data for significant acceleration or deceleration of the wind between 1~AU and 5~AU.
For each \emph{Ulysses} density measurement, $n_{\text{p}}(r)$, we calculate the corresponding density at 1~AU, $n_{\text{p},1\text{AU}}$, by assuming the wind is expanding spherically with a constant speed, and therefore $n_\text{p}(r) = n_{\text{p},1\text{AU}} r^{-2}$, where $r$ is the radial distance from the Sun in AU. 
For the temperature, we assume additionally that the wind is described by a polytropic equation of state, and therefore $T_{1\text{AU}} = T(r) r^{2(\alpha-1)}$, where we take $\alpha = 1.46$.
Since the Ulysses proton temperature data comes in the form of a minimum temperature, $T_{\text{min}}$, and a maximum temperature, $T_{\text{max}}$, we use the method of \citet{2009ApJ...702.1604C} and calculate the proton temperature as $T_\text{p} = \sqrt{T_{\text{min}} T_{\text{max}}}$.
In Fig.~\ref{fig:ACEandUlyssesSpeeds}, we show histograms of hourly average wind speeds as measured by the \emph{ACE} and \emph{Ulysses} spacecraft. 
The bimodal distribution in wind speed is clearly visible, with the slow and fast winds displaying typical velocities of 400~\kms\hspace{0mm} and 760~\kms \hspace{0mm} respectively.

Histograms showing hourly-average wind densities and temperatures at 1~AU for the slow and fast winds are shown in Fig.~\ref{fig:ACEandUlysses}.
The results from the two spacecraft agree excellently, with the slow wind in general having a higher density and a lower temperature than the fast wind.
The only notable disagreement is that the proton temperatures of the fast wind as measured by \emph{ACE} show a much broader distribution than what is seen in the \emph{Ulysses} data. 
This is likely due to \emph{ACE} having many fewer measurements of the fast wind given its position in the ecliptic plane.
Since \emph{ACE} samples better the slow wind and \emph{Ulysses} samples better the fast wind, we use the \emph{ACE} results to constrain our slow wind model and the \emph{Ulysses} results to constrain our fast wind model.
For both components of the wind, we calculate typical densities and temperatures using median values from all measurements within 20~\kms\hspace{0mm} of the typical wind speeds.
We use median values in order to reduce the influence of outlying measurements from transient features in the wind such as CMEs. 
These values, shown as vertical lines in Fig.~\ref{fig:ACEandUlysses}, are summarised in Table~\ref{tbl:ACEandUlysses}.
We therefore constrain the base densities in our wind models by assuming that the 1~AU proton densities in the slow and fast winds are 5.31~cm$^{-3}$ and 2.18~cm$^{-3}$ respectively.
  
\footnotetext{
The data from \emph{ACE} was obtained from \mbox{\url{www.srl.caltech.edu/ACE/ASC/}}.
The data from \emph{Ulysses}, Helios A, and Helios B were obtained from \mbox{\url{cohoweb.gsfc.nasa.gov}}.
For all four spacecraft, we use hourly average values for each wind parameter. 
 }

Since neither \emph{ACE} nor \emph{Ulysses} data probe the properties of the solar wind closer to the Sun than 1~AU, we additionally use data from \mbox{Helios~A} and Helios~B. 
Both spacecraft orbited the Sun in the ecliptic plane on eccentric orbits, taking them in as close as 0.3~AU from the Sun and out as far as 1.0~AU.
Since they were in the ecliptic plane, both spacecraft sampled the slow wind better than the fast wind. 
In order to compare the results of our wind simulations to data from the Helios spacecraft, we break the data down into radial bins of width 0.2~AU and calculate median densities and temperatures for the slow and fast winds separately.

In order to constrain the base temperature, $T_0$, for each model, we choose a value that leads to the desired 1~AU wind speeds of 400~\kms\hspace{0mm} and 760~\kms\hspace{0mm} for the slow and fast winds respectively, initially assuming the value of $\alpha_{\text{out}}$ of 1.46 from \citet{1995JGR...100...13T}. 
When the base temperature has been set, we scale the value of the base proton density, $n_0$, to give the measured proton density at 1~AU. 
Finally, we calculate the value of $\alpha_{\text{out}}$ that gives the best fit to the temperature structures measured by \mbox{Helios A} and Helios~B, varying $T_0$ and $n_0$ by small amounts to recover the required temperatures and densities at 1~AU when necessary. 
We find for the fast wind model, $\alpha_{\text{out}}$ of 1.46 provides a good fit to the measured temperature structure, but for the slow wind model, a larger value of 1.56 gives a better fit.
Since we want the two models to differ only by the base temperature and density, we assume that for both winds, $\alpha_{\text{out}} = 1.51$.
This ensures that the wind models both provide good fits to the measurements, while greatly simplifying the analysis in the rest of the paper. 
For the slow wind, we obtain a base temperature of 1.8~MK and a base proton density of  $2.60 \times 10^{7}$~cm$^{-3}$. 
For the fast wind, we obtain a  base temperature of 3.8~MK and a base proton density of  $1.05 \times 10^{6}$~cm$^{-3}$. 
Although at 1~AU, the fast wind density is only a factor of two lower than the slow wind density, we require a base density for the fast wind that is more than an order of magnitude lower than the value for the slow wind because of the former's higher base temperature.
The properties of the two models are summarised in Table~\ref{tbl:solarwindmodelparams}.

\begin{figure}
\includegraphics[trim = 13mm 4mm 5mm 4mm, clip=true,width=0.49\textwidth]{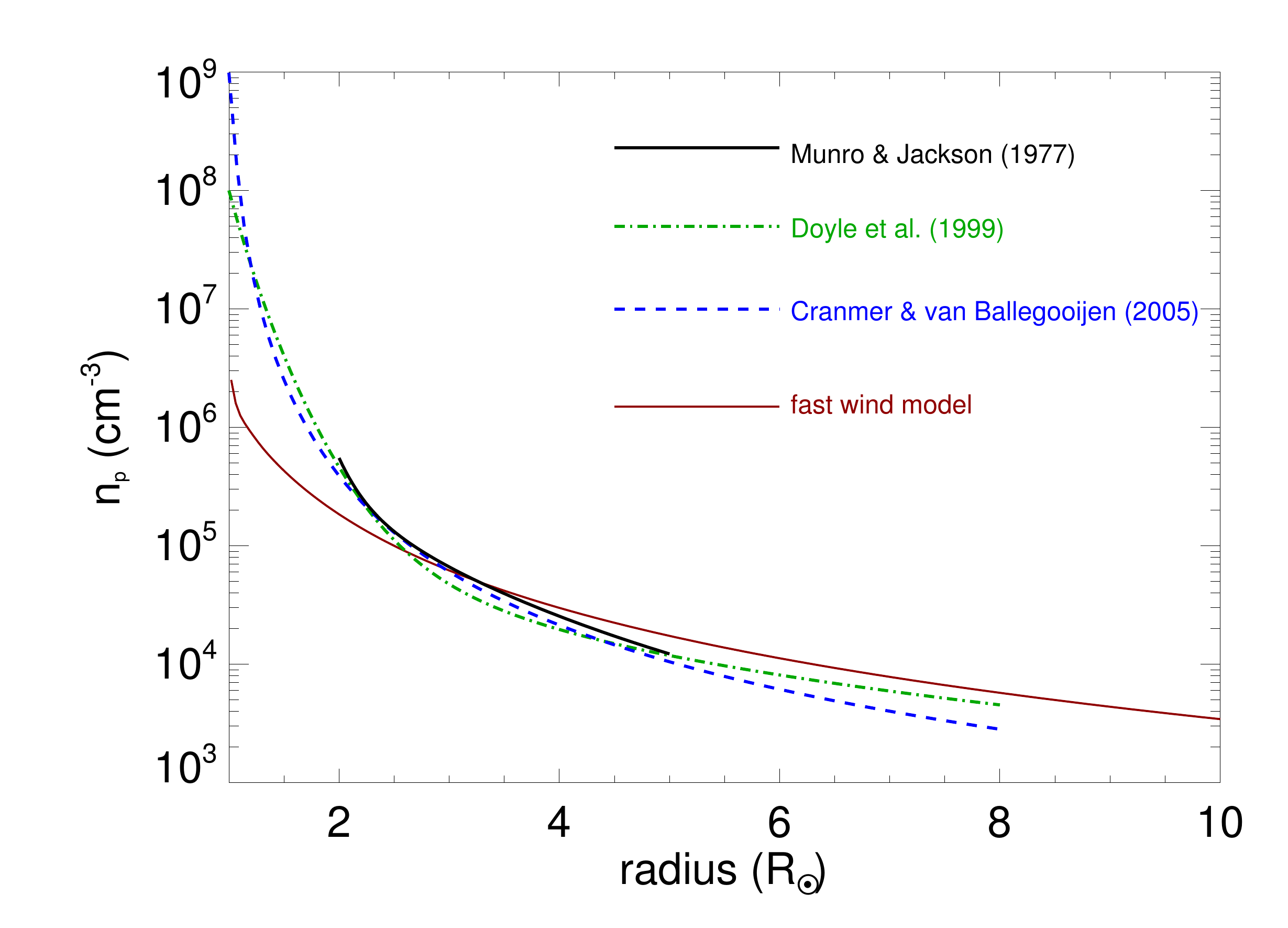}
\includegraphics[trim = 13mm 4mm 5mm 4mm, clip=true,width=0.49\textwidth]{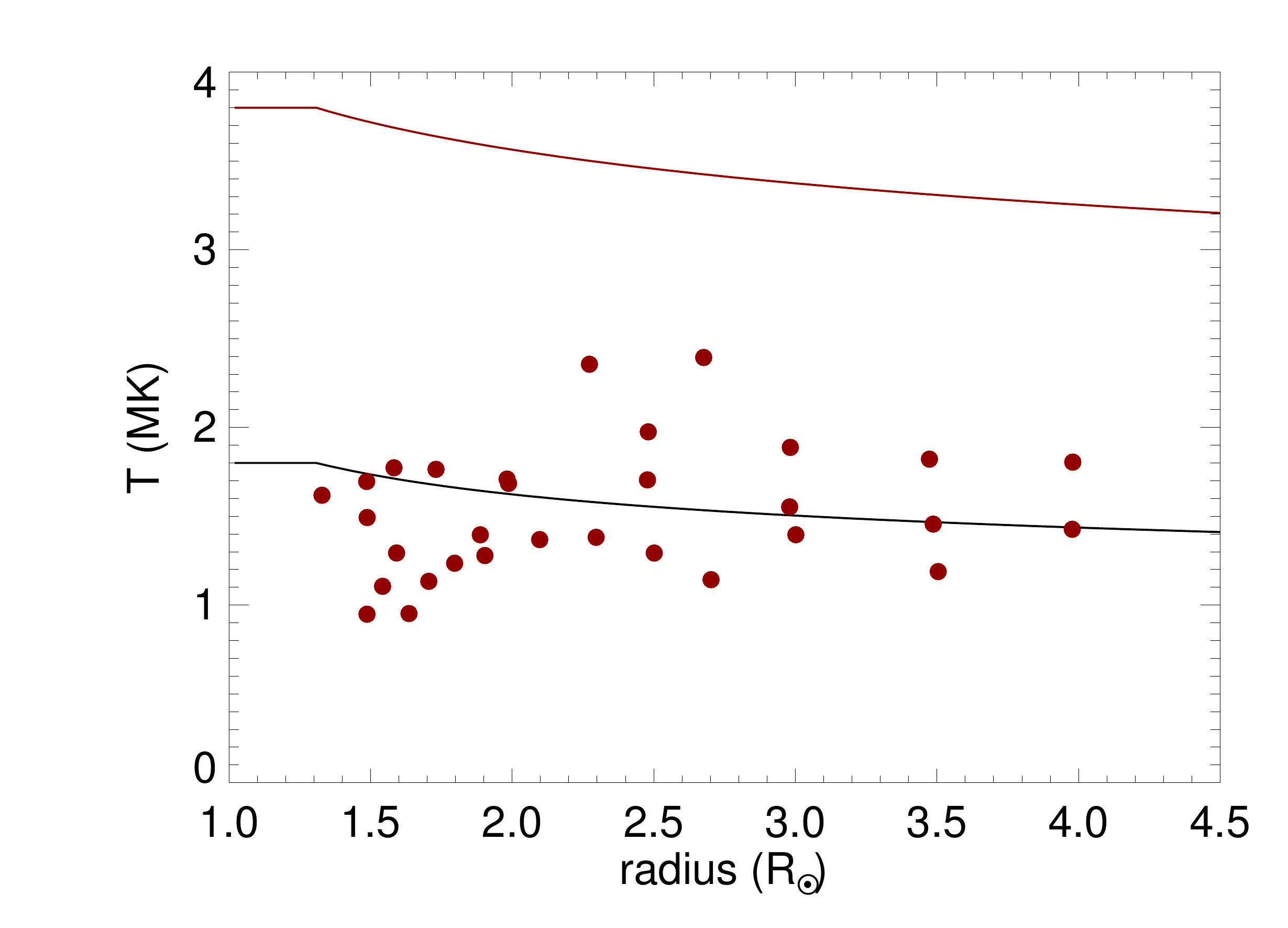}
\caption{
Figures comparing our fast solar wind model to observed properties of the real fast solar wind close to the solar surface. 
The upper panel shows proton densities within 10~R$_\odot$ of the solar surface in our fast wind model compared to observational electron density profiles inside and above coronal holes from \citet{1977ApJ...213..874M}, \citet{1999A&A...349..956D}, and \citet{2005ApJS..156..265C}.
The lower panel shows wind temperature within 4.5~R$_\odot$ compared to measurements of wind temperatures in coronal holes (\emph{red circles}) compiled by \citet{2004ESASP.575..154C}.
As in Fig.~\ref{fig:solarfastslowwinds}, the red and black lines in the lower panel show the fast and slow wind simulations respectively. 
}
 \label{fig:fastdens}
\end{figure}

The difference in the base temperatures that we find for our slow and fast wind models is similar to the model of \citet{2000JASTP..62.1515G}.
They performed 3D MHD simulations of the solar wind based on a dipole magnetic field.
In order to reproduce the slow and fast winds, they assumed base temperatures of 4.99~MK inside coronal holes (leading to fast wind) and 2.85~MK outside coronal holes (leading to slow wind). 
The differences in the base temperatures between our winds and their winds is probably mostly a result of different assumptions about the heating of the wind, and also because their simulations include the influence of the solar magnetic field on the wind dynamics.

\subsection{Results: consistency with solar wind observations} \label{sect:solarwindresults}

In Fig.~\ref{fig:solarfastslowwinds}, we show the structures of our slow and fast wind models. 
For the slow wind, we get a temperature at 1~AU of \mbox{$7.5 \times 10^{4}$~K}, which is similar to the measured median proton temperature of \mbox{$6.1 \times 10^{4}$~K}. 
For the fast wind, we get a temperature at 1~AU of \mbox{$1.8 \times 10^{5}$~K}, which is similar to the measured median proton temperature of \mbox{$2.8 \times 10^{5}$~K}.
The dashed lines in the upper and lower panels of Fig.~\ref{fig:solarfastslowwinds} show the densities and temperatures measured by \mbox{Helios A} and Helios~B.
Our models give excellent fits to the real density structure and a good fit to the temperature structure far from the Sun. 
However, we stress that a direct comparison between the temperatures in our model and the real solar wind temperatures should not be taken too seriously given our assumption that the wind can be described with one temperature that is isotropic and the same for all species. 
We have also experimented with extending our models out to 5~AU, and we find similarly good correspondence between our models and the real wind properties as measured by \emph{Ulysses}.

\begin{figure}
\includegraphics[trim = 13mm 4mm 10mm 4mm, clip=true,width=0.49\textwidth]{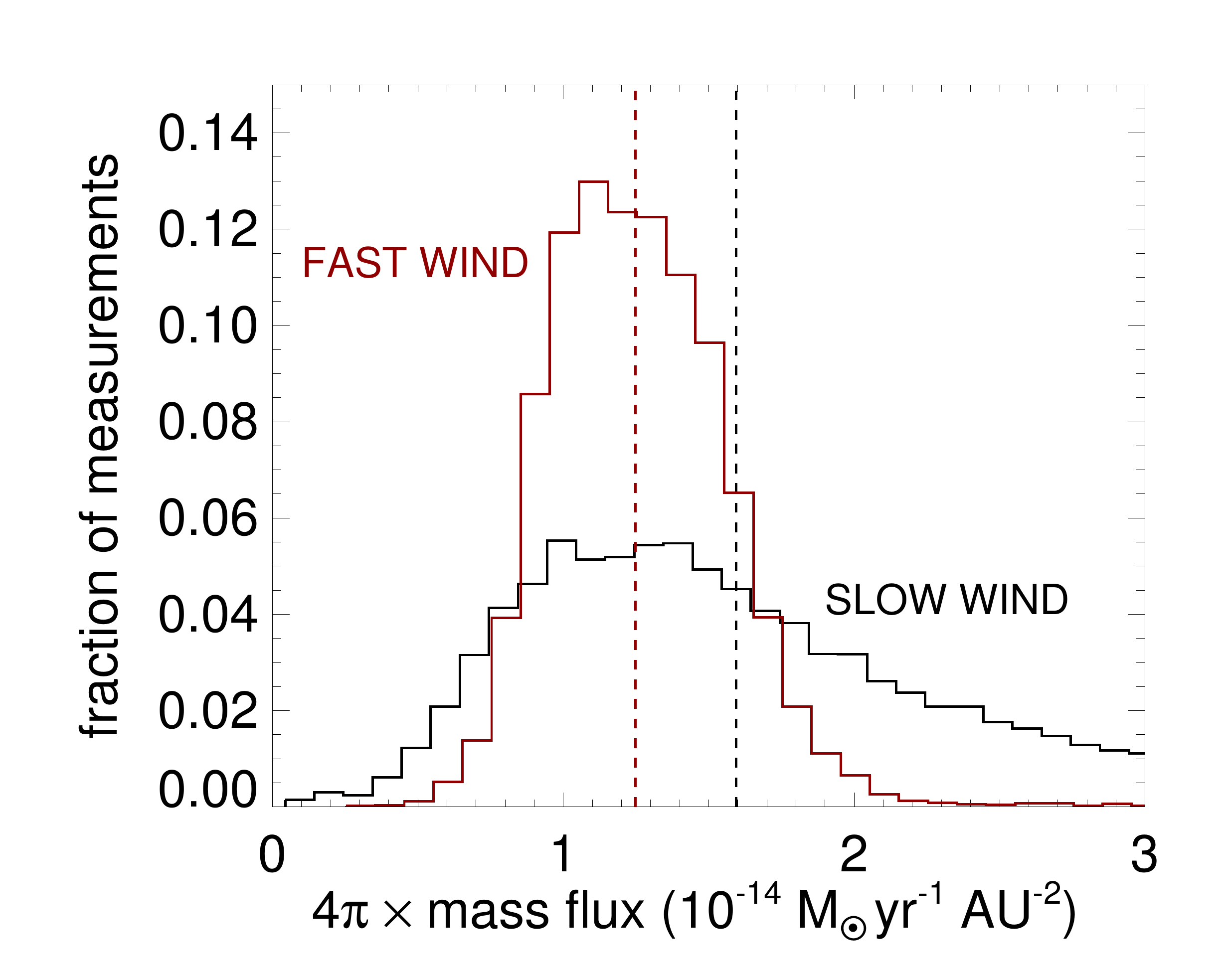}
\caption{
Histogram showing the mass flux at 1~AU in the solar wind based on hourly average measurements from \emph{ACE} for the slow wind (\emph{black}) and \emph{Ulysses} for the fast wind (\emph{red}). 
The values are represented in units of \mbox{M$_\odot$ yr$^{-1}$ AU$^{-2}$} and have been multiplied by \mbox{$4\pi$} so that they can be compared with the total mass loss rate of the solar wind.
The vertical dashed lines show mass loss rates derived from the wind models of \mbox{$1.6 \times 10^{-14}$ M$_\odot$ yr$^{-1}$} and \mbox{$1.2 \times 10^{-14}$ M$_\odot$ yr$^{-1}$} for the slow and fast winds respectively.
These are very similar to the medians for the two distributions shown in this figure.
}
 \label{fig:solarwindmassflux}
\end{figure}

Clearly our model gives a realistic description of the solar wind far from the solar surface, but we would not expect that a 1D hydrodynamic model of this sort would be realistic within a few solar radii of the surface where the real solar wind is in a low plasma-$\beta$ environment, and therefore the dynamics of the plasma is strongly influenced by the solar magnetic field. 
It is therefore important to understand where our model becomes realistic. 
Several studies have derived the electron density structure in and above coronal holes (for a summary, see Fig.~1 of \citealt{1999ApJ...521L.145E}).  
In Fig.~\ref{fig:fastdens}, we compare our fast wind model within 10~R$_\odot$ to observationally constrained proton densities.
Our model provides a good fit to the observed density of the fast wind beyond 3~R$_\odot$, especially to the density structure measured by \citet{1999A&A...349..956D}, but underestimates the density very close to the surface by more than an order of magnitude.
This is at least partly due to the fact that we assume the wind is expanding radially with $r^2$, whereas in reality, the fast wind is expanding much quicker close to the surface due to the superradial expansion of the magnetic field within coronal holes. 
Another notable contradiction between our models and observed solar wind properties is the temperature of the fast wind close to the solar surface.
This is shown in the lower panel of Fig.~\ref{fig:fastdens}.
Several studies have measured fast wind temperatures within coronal holes and found that neither the protons nor the electrons reach temperatures significantly above 2~MK (see Fig.~6 of \mbox{\citealt{2009LRSP....6....3C}} for a summary).
On the other hand, our model predicts temperatures in coronal holes of almost 4~MK.
This discrepancy is possibly due to our assumption that the wind is driven entirely by thermal pressure gradients.
A model that includes additional driving mechanisms, such as wave pressures, would require a lower temperature to accelerate the fast wind to the desired speeds.

Another useful comparison between our models and the real solar wind is shown as the dotted line in the middle panel of Fig.~\ref{fig:solarfastslowwinds}.
The line corresponds to the radial profile for the slow wind speed derived by \citet{1997ApJ...484..472S} by tracking inhomogeneities in white light images of the corona between a few R$_\odot$ and 30~R$_\odot$. 
The inhomogeneities were seen to originate from above helmet streamers and travel radially away from the Sun with speeds that are dependent on the distance from the top of the helmet streamers.
Our simulation of the slow solar wind corresponds well to the measured velocity profile, though there is some indication that in our model, the wind accelerates too slowly relative to the real solar wind.

\begin{figure}
\includegraphics[trim = 5mm 4mm 5mm 4mm, clip=true,width=0.49\textwidth]{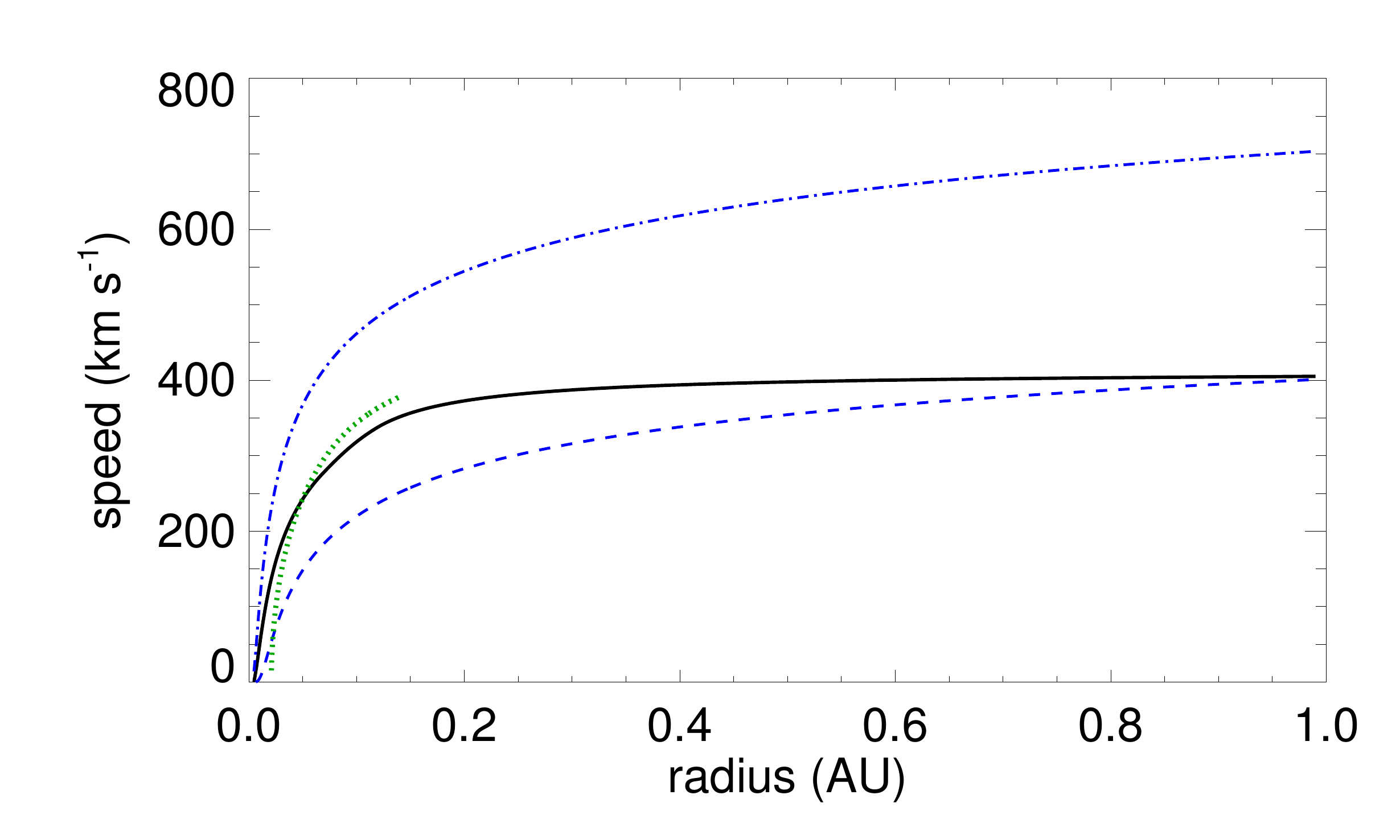}
\caption{
Figure comparing the velocity structure of our slow solar wind model with the isothermal Parker wind model. 
The solid black line represents our slow wind model and the blue lines show isothermal winds. 
The blue dot-dashed line shows an isothermal wind with the same temperature as the base temperatures of our slow wind and the blue dashed line shows an isothermal wind with the temperature reduced to a value that leads to the same wind speed at 1~AU as in our wind model. 
The dotted green line shows the slow wind velocity profile estimated by \mbox{\citet{1997ApJ...484..472S}} described in the caption of Fig.~\ref{fig:solarfastslowwinds}.
}
 \label{fig:comparewithparker}
\end{figure}

Both winds become supersonic very close to the Sun, with sonic points at radii of 6.2~R$_\odot$ and 2.4~R$_\odot$ for the slow and fast winds respectively. 
Assuming a 3~G photospheric dipolar magnetic field (see Paper~II) that falls off with $r^{-3}$, we calculate that the slow wind becomes superalfv\'{e}nic at $\sim$2.6~R$_\odot$.
This corresponds to the largest radius where the magnetic field is able to hold onto the coronal plasma and beyond which the magnetic field is torn open by the wind (\citealt{2009ApJ...699..441V}). 
For the Sun, this radius is typically $\sim2.5$~R$_\odot$ (\citealt{1984PhDT.........5H}), which gives further confirmation that the velocity structure of the wind in our simulations close to the Sun is approximately correct. 
By 0.2~AU, both wind models have reached speeds of beyond 90\% of their 1.0~AU values.
Beyond this point, the wind speeds increase approximately linearly, with average accelerations of 41~km~s$^{-1}$~AU$^{-1}$ and 61~km~s$^{-1}$~AU$^{-1}$ for the slow and fast winds respectively.

In Fig.~\ref{fig:solarwindmassflux}, we show histograms of the mass flux at 1~AU in the slow and fast winds as measured by \emph{ACE} and \emph{Ulysses} and compare these histograms to the values calculated from our wind models.
Clearly the wind models accurately reproduce the mass fluxes in the wind, with the well known result that the slow wind has a slightly higher mass flux than the fast wind. 
The mean mass flux in the fast wind is $65\%$ of the mean mass flux in the slow wind, as determined from both the measurements and our model.
A wind consisting entirely of the slow wind would have a mass loss rate of \mbox{$1.6\times10^{-14}$~M$_\odot$~yr$^{-1}$} and a wind consisting entirely of the fast wind would have a mass loss rate of \mbox{$1.2\times10^{-14}$~M$_\odot$~yr$^{-1}$}.
In order to estimate the mass loss rate from the entire solar wind, we need to know what fractions of the wind are made up of slow and fast wind. 
This is likely to change over the course of the solar cycle with the structure of the magnetic field, but we assume crudely that the two components contribute equally to the mass loss rate at all times.
This assumption is reasonable (\citealt{2010JGRA..115.4102T}), and is likely to have little effect on our calculations. 
Therefore, in our model, the solar wind mass loss rate is \mbox{$1.4\times10^{-14}$~M$_\odot$~yr$^{-1}$}.

Since the analytic isothermal wind model derived by \citet{1958ApJ...128..664P} is commonly used in the literature (\citealt{2007P&SS...55..618G}; \citealt{2014arXiv1409.1237S}), it is interesting to compare it to our model.
In Fig.~\ref{fig:comparewithparker}, we compare the velocity structure of our slow wind model to the predictions of the isothermal model. 
The main difference in the shapes of the profiles is that instead of quickly accelerating to approximately its terminal velocity, the acceleration in an isothermal wind is much more distributed over all radii.
For the isothermal wind model with the same base temperature as our slow solar wind (1.8~MK), shown as the dot-dashed blue line in Fig.~\ref{fig:comparewithparker}, this leads to an unrealistically fast wind at 1~AU.
In order to reproduce the desired slow wind speed at 1~AU, we need to reduce the wind temperature to 0.75~MK.
This is shown as the blue dashed line in Fig.~\ref{fig:comparewithparker}.
In this model, the 1~AU wind speed is correct, but the wind speeds close to the solar surface are underestimated.
For a comparison, the green dotted line in Fig.~\ref{fig:comparewithparker} shows the velocity profile for the real slow wind estimated by \mbox{\citet{1997ApJ...484..472S}} described in the caption of Fig.~\ref{fig:solarfastslowwinds}.
Clearly our model provides a much more realistic description of the real solar wind, though the isothermal model is still a good first approximation.

\section{Stellar Wind Model} \label{sect:scaling}

We now have a model that provides an excellent description of the solar wind far from the solar surface.
How this model should be applied to other stars is unclear.
In this section, we develop methods for scaling our solar wind model to other stars.
Our final model can be used to derive the wind properties within a few AU of a star based on the stellar mass, radius, and rotation rate only. 
The free parameters in the wind model are the base temperature, $T_0$, the base density, $n_0$, and the radial structure of the polytropic index, $\alpha (r)$.
Since we do not have \emph{in situ} measurements of the winds of other stars, we have no way to constrain $\alpha (r)$, and so we assume that the values derived for the solar wind apply to other stars. 
This is unlikely to be a large source of uncertainty since the wind properties are much more sensitive to $T_0$ and $n_0$ than to the detailed structure of $\alpha(r)$ (as can be seen by the fact that the isothermal model gives results that are similar to our solar wind model, as discussed above). 
In Section~\ref{sect:scalingT0} and Section~\ref{sect:scalingn0}, we discuss how we scale the solar value of $T_0$ and $n_0$ to other stars.

\subsection{Scaling base temperature} \label{sect:scalingT0}

In this section, we develop two methods for scaling the base temperatures of the solar wind to other stars. 
In any model that does not try to calculate the stellar wind properties self-consistently from the physical mechanisms that drive the solar wind, the base temperature is the most important free parameter.
How to determine this parameter is currently an unsolved problem and various methods have been used in the literature.

Observationally, it is known that the temperature of a star's corona scales with its level of X-ray emission, such that more active stars have hotter coronae (\citealt{1990ApJ...365..704S}; \citealt{1997ApJ...483..947G}; \citealt{2005ApJ...622..653T}).
\citet{2007A&A...463...11H} and \citet{2014MNRAS.438.1162V} assumed that coronal temperatures can be used as a proxy for the wind temperature. 
This would imply that the solar wind was in the past hotter, and therefore faster, than the present solar wind. 
This type of model is attractive since stellar coronae are easily observable in X-rays.
The link between coronal and wind temperatures has some plausibility given the similarity between the temperatures of the closed and open regions of the solar corona, and the fact that the energy that heats both the wind and the corona comes from the same source (i.e. energy in convective motions in the photosphere being transferred to higher altitudes by the magnetic field).

However, it is unclear to what extent the mechanisms that transfer the energy from the source and dissipate the energy in the corona are the same for open and closed field regions. 
For example, it could be that the solar wind is heated by the dissipation of Alfv\'{e}n waves whereas the solar corona is heated by magnetic reconnection events (\citealt{2010ApJ...720..824C}).
Furthermore, \citet{2011MNRAS.417.2592C} pointed out that while the solar X-ray luminosity, $L_\text{X}$, varies by over an order of magnitude during the solar cycle, the mass loss rate does not change in any significant way, and suggested therefore that the mass loss rates of other stars are independent of their X-ray properties. 

The lack of a correlation between solar $L_\text{X}$ and wind mass loss rate is suggestive, but we should be cautious when interpreting such observations.
The reason for the constancy of the solar mass loss rate is that, for reasons that are currently unclear, the two components that dominate the solar wind have approximately equal mass fluxes (see Fig.~\ref{fig:solarwindmassflux}).
Over the course of the solar cycle, the spatial distributions of slow and fast winds change in response to changes in the coronal magnetic field structure, but the mass flux remains approximately constant in all directions. 
It is possible that when going to more rapidly rotating active stars, the mass fluxes of each component scale up with \mbox{X-ray} activity. 
The Sun's $L_\text{X}$ varies over the solar cycle due mostly to the addition and subtraction of active regions (\citealt{2001ApJ...560..499O}).
However, covering the entire solar surface with normal solar active regions leads to an $L_\text{X}$ more than an order or magnitude below the $L_\text{X}$ values of the most active stars (\citealt{1984A&A...133..117V}; \citealt{2004A&ARv..12...71G}).
Furthermore, \citet{2005ApJ...622..653T} showed that the plasma temperatures \emph{averaged} over the X-ray emitting coronae of young solar analogues can exceed 10~MK, well above the temperatures of normal solar active regions. 
Also, the long-term (evolutionary) decline of magnetic activity is due to the decay of the rotation-induced magnetic dynamo as a consequence of stellar spin down whereas the short-term cyclic variations are {\it not} related to changes in rotation. 
For the behaviour of winds, evolutionary (or rotation-related) trends are poorly understood in stars and inaccessible on the Sun.
\emph{The rotation-dependent differences between the coronae of stars with different levels of activity are not necessarily analogous to the rotation-independent differences in the solar corona at different times in the solar cycle.} 
It is therefore unclear how much we can learn about the connection between stellar winds and coronal \mbox{X-ray} activity from changes in the solar wind over the solar cycle.

An alternative method that is used in several models in the literature is to scale either the wind terminal velocity, $v_\infty$, or the base wind temperature with the surface escape velocity, $v_{\text{esc}}$. 
For example, the model of \mbox{\citet{2011ApJ...741...54C}} makes the assumption that \mbox{$v_\infty = v_{\text{esc}}$}.
This assumption is based on the similarity between the terminal velocities of the solar wind and the surface escape velocity of 618~\kms. 
The base temperature of the wind can then be calculated as the value required to reproduce this velocity.  
Similarly, \mbox{\citet{2008ApJ...678.1109M}} and \mbox{\citet{2012ApJ...754L..26M}} set the base temperature in their 2D MHD wind models by assuming that the base sound speed is a fixed fraction of the surface escape velocity.
In most of their simulations, they assume that $c_s/v_{\text{esc}} = 0.222$, which is similar to the values of 0.329 and 0.478 for the slow and fast solar wind models that we presented in Section~\ref{sect:solarwindmodel}.
This assumption means that the base temperature of the wind is determined by the stellar mass and radius only by $T_0 \propto M_\star / R_\star$.
Given the example of the solar wind, these models have some plausibility, and the assumption is further supported by observational constraints on the winds of other types of stars, which tend to have terminal velocities that are within a factor of a few of the surface escape velocities (\citealt{1992ASPC...26..403J}).
Since the expansion of the Sun on the main-sequence has been very small, the surface escape velocity has decreased only slightly with age, and therefore both models would imply that the temperature, and therefore the wind speed, of the solar wind has remained approximately constant.
This would imply a very different history of the solar wind to the one implied by scaling wind temperature with coronal temperature.

Unfortunately, the observational and theoretical constraints on stellar wind properties are currently not able to confirm or rule out any of the above models. 
We therefore explore two stellar wind models that make radically different assumptions about the wind base temperature.

\begin{itemize}

\item
\emph{Model~A}: we follow the method of \citet{2007A&A...463...11H} and assume that the base wind temperature scales with coronal temperature, $\bar{T}_{\text{cor}}$, as inferred from coronal X-ray observations. 
We define $\bar{T}_{\text{cor}}$ as the emission measure weighted average coronal temperature (\citealt{2007A&A...468..353G}).
We therefore calculate the wind base temperature from

\begin{equation}
T_0 = T_{0,\odot} \left( \frac{\bar{T}_{\text{cor}}}{\bar{T}_{\text{cor},\odot}} \right),
\end{equation}

\noindent where $T_{0,\odot}$ is 1.8~MK for the slow wind and 3.8~MK for the fast wind.
This means that wind temperature is calculated as a function of stellar mass, radius, and rotation rate, with more rapidly rotating stars having higher wind temperatures. 

\vspace{2mm}

\item
\emph{Model~B}: we follow the method of \mbox{\citet{2012ApJ...754L..26M}} and assume that the base sound speed is a fixed fraction of the escape velocity.
We therefore calculate the wind temperature as 

\begin{equation} \label{eqn:T0ModelB}
T_0 = \frac{2 G  \mu m_\text{p} }{\gamma k_\text{B}} \left( \frac{c_s}{v_{\text{esc}}} \right)^2 \left( \frac{M_\star}{R_\star} \right),
\end{equation}

\noindent where the fraction $c_s/v_{\text{esc}}$ is 0.329 for the slow wind and 0.478 for the fast wind, and $\gamma=5/3$. 
Unlike in Model~A, the wind temperature in this model does not depend on the stellar rotation rate, but only on the mass and radius. 
Since $M_\star / R_\star$ is only a weak function of stellar mass and age, this model is similar to that of (\citealt{2014arXiv1409.1237S}) who assumed that the wind temperature was the same for all stars.

\end{itemize}

\noindent A note of caution is needed for Model~A. 
We are not only assuming that wind temperatures and coronal temperatures are linked, we are also assuming that wind temperature scales specifically with the \emph{average} coronal temperature. 
There is evidence that the mechanism that heats the fast component of the solar wind is the same as the mechanism that heats the quiet solar corona (\citealt{2011Natur.475..477M}).
This might imply that the temperatures of quiet regions of stellar coronae should be used in Model~A instead of the coronal average temperature.
Given the difficulty in measuring the temperatures of the quiet regions of stellar coronae, we assume that $\bar{T}_{\text{cor}}$ can be used as a proxy.

These two models lead to radically different wind properties. 
However, we stress that neither of these models allow us to predict the mass loss rates in the wind since we also need to specify the base density.
A faster wind does not imply a higher mass flux, as the example of the slow and fast components of the solar wind shows.

In the case of Model~A, since we scale wind temperature with coronal temperature, it is important that we understand how coronal temperature can be calculated for difference stars.
All low-mass main-sequence stars have hot magnetically confined coronae that emit strongly in X-rays.
Although the mechanisms that heat coronae are poorly understood, it is known empirically that coronal temperature scales well with X-ray activity, with the most active stars having hotter coronae than less active stars.
Johnstone \& G\"{u}del (2015) showed that one universal relation exists between $\bar{T}_\text{cor}$ and $F_\text{X}$ for all low-mass main-sequence stars.
They derived the following scaling law

\begin{equation} \label{eqn:telleschi}
\bar{T}_{\text{cor}} \approx 0.11 F_\text{X}^{0.26},
\end{equation}

\noindent where $\bar{T}_{\text{cor}}$ is in MK and $F_\text{X}$ is in erg~s$^{-1}$~cm$^{-2}$.

The value of $F_\text{X}$ for any low-mass main-sequence star can be estimated relatively easily from its basic parameters. 
At slow rotation, the brightness of a star in X-rays is determined primarily by its mass, radius, and rotation rate, and at fast rotation, the rotation dependence saturates (\citealt{1981ApJ...248..279P}; \citealt{1984A&A...133..117V}).
Above the saturation threshold, the X-ray emission no-longer depends on rotation.
The dependence of X-ray emission on rotation is well represented as a correlation between $R_\text{X}$ and Rossby number, $Ro=P_{\text{rot}} / \tau_{\text{conv}}$, where $R_\text{X} \equiv L_\text{X} / L_\text{bol}$ is the X-ray luminosity normalised by the bolometric luminosity and $\tau_{\text{conv}}$ is the convective turnover time.
The relation between $R_\text{X}$ and $Ro$ can be represented as

\begin{equation} \label{eqn:wrightlaw}
R_\text{X} = \left \{
\begin{array}{ll}
R_{\text{X},\text{sat}}, & \text{if } Ro \le Ro_{\text{sat}},\\
C Ro^\beta, & \text{if } Ro \ge Ro_{\text{sat}},\\
\end{array} \right.
\end{equation}

\noindent where $Ro_{\text{sat}}$ is the saturation Rossby number and $R_{\text{X},\text{sat}}$ is the level at which saturation occurs.
\citet{2011ApJ...743...48W} compiled a database of stars with known rotation periods and X-ray luminosities, and derived \mbox{$\log_{10} R_{\text{X},\text{sat}}=-3.13$}, \mbox{$Ro_{\text{sat}} = 0.13$}, and \mbox{$\beta=-2.18$}, which is consistent with other determinations in the literature (e.g. \citealt{2003A&A...397..147P}).
This implies that \mbox{$C = 8.68 \times 10^{-6}$}. 
There is significant uncertainty in the exact value of $\beta$. 
\citet{2011ApJ...743...48W} argued that the sample of stars is likely to suffer from selection biases, and so they constructed an unbiased subsample of stars and found $\beta = -2.7$, consistent with the determination of \citet{1997ApJ...483..947G} from solar analogues.
On the other hand, \citet{2014arXiv1408.6175R} reanalysed this sample of stars and found no evidence of a bias in the results. 
We therefore assume that $\beta=-2.18$. 
We warn that assuming that $R_{X,\text{sat}}$, $Ro_{\text{sat}}$, and $\beta$ are constants for all stars is an approximation.
As can be seen in the analysis of \citet{2003A&A...397..147P}, these parameters are likely to have at least weak dependences on stellar mass.
We also ignore the large scatter around the best fit relation between $R_\text{X}$ and $Ro$ that can be seen in the data, and assume instead that every star lies exactly on the best fit line.  
For consistency, we use the determination of the mass-dependence of the convective turnover time from \citet{2011ApJ...743...48W}.
This gives larger values of $\tau$ for lower mass stars: for a 1~M$_\odot$ star, $\tau \approx 15$~days and for a 0.5~M$_\odot$ star, $\tau \approx 35$~days. 
This means that the X-ray emission for low-mass stars saturates at slower rotation than for high-mass stars. 
A important consequence of this is that the saturation value of $F_\text{X}$ is lower for lower mass stars, meaning that they are never able to become as active as the rapidly rotating higher mass stars.

If a measurement for $F_\text{X}$ is available for a given star, then $T_\text{cor}$ can be estimated from Eqn.~\ref{eqn:telleschi}.
Otherwise, if the rotation period is known, Eqn.~\ref{eqn:telleschi} and Eqn.~\ref{eqn:wrightlaw} can be combined to estimate coronal temperatures for any low-mass main-sequence star, which is how we estimate $T_\text{cor}$ for Model~A for the rest of this paper.
For the Sun, we would predict a coronal average temperature of 2.4~MK, which is within the temperatures calculated for cycle minimum and cycle maximum by \citet{2000ApJ...528..537P}.
For a star of a given mass, the temperature depends on rotational angular velocity approximately as $\bar{T}_{\text{cor}} \propto \Omega_\star^{0.7}$ in the unsaturated regime.
At a given rotation rate, lower mass stars tend to have higher coronal temperatures than higher mass stars in the unsaturated regime, though the mass dependence is very weak for the slow rotators.
This is similar to the dependence of $\bar{T}_{\text{cor}} \propto \Omega_\star^{0.6}$ used in the models of \cite{2007A&A...463...11H}.

\subsection{Scaling base density} \label{sect:scalingn0}

For a given star, once the temperature of the wind is set, we can use our wind model to derive the wind speeds and $\dot{M}_\star / n_0$.
At this point, the wind densities far from the star and the mass loss rate in our wind model are directly proportional to $n_0$. 
Determining the base density for a given star is very difficult. 
As we show in Fig.~\ref{fig:fastdens}, the base density in our solar wind model is unrealistically small compared to measured values.
This discrepancy is likely due to our unrealistic assumptions about the geometry of the wind within $\sim$3~R$_\odot$.
Therefore, our base density does not represent a real physical density within the stellar corona, but is a fudge factor that can be scaled to reproduce the correct mass fluxes once everything else about the wind is known.

The simplest assumption that we could make is that the base density is a constant and \emph{then we could} determine its value from the solar wind. 
However, this assumption is unlikely to be realistic.
Another possible approach is to scale wind density with coronal density.
\citet{2007A&A...463...11H} assumed that wind densities have a power law dependence on stellar rotation only, such that $n_0 \propto \Omega_\star^{n_n}$, where $n_n \approx 0.6$, as suggested for the coronae of Sun like stars by \citet{2003ApJ...599..516I}. 
Similarly, \citet{2014arXiv1409.1237S} assumed that the wind base density is given by $n_0 = \bar{n}_c / 10$, where $\bar{n}_c$ is the average closed coronal density.
They argued that $L_\text{X} \propto \bar{n}_c^2 R_\star^3$, which combined with Eqn.~\ref{eqn:wrightlaw} implies that $\bar{n}_c \propto \Omega_\star^{1.1}$ for a given stellar mass and radius. 
Observationally, it is very difficult to constrain coronal densities. 
Measurements of coronal densities were reported for Sun-like stars by \citet{2004ApJ...617..508T} and \citet{2004A&A...427..667N}.
The general trend is that more active stars have higher coronal densities; however, the measurements suffer from significant biases towards measuring the densities of low temperature and high density plasma (\citealt{2004A&ARv..12...71G}).
Furthermore, existing measurements are mostly either upper limits or have large error bars. 
For this reason, and because of the disconnect between the densities in closed and open regions of the solar corona, we do not attempt to use coronal density as a proxy for wind density.

Alternatively, if the mass flux is known, then the base densities could be constrained once the wind temperature has been set.  
An interesting possibility for estimating the mass loss rates of low-mass stars is to use rotational evolution.
As low-mass main-sequence stars age, their rotation rates slow with time due to the removal of angular momentum in the wind.
The rate at which a star loses angular momentum depends strongly on the wind mass loss rate, and so observational constraints on main-sequence rotational evolution could be used to estimate wind mass loss rates.
This possibility has been explored previously on the main-sequence (\citealt{2000GeoRL..27..501G}) and on the pre-main-sequence (\citealt{2008ApJ...681..391M}).

In Paper~II of this series, we construct a physical model for the rotational evolution of low-mass stars between 100~Myrs and 5~Gyrs, which we then fit to the observed rotation periods of over 2000 stars.
One of the important parameters for determining the angular momentum loss from a magnetised stellar wind is the mass loss rate.
In the absence of any method for predicting the mass loss rates in our rotational evolution model, we make the assumption that the mass loss rate per unit surface area has power law dependences on rotation and stellar mass, such that

\begin{equation} \label{eqn:generalMdot}
\dot{M}_\star = \dot{M}_\odot \left( \frac{R_\star}{R_\odot} \right)^2  \left( \frac{\Omega_\star}{\Omega_\odot} \right)^a  \left( \frac{M_\star}{M_\odot} \right)^b,
\end{equation}

\noindent where $a$ and $b$ are free parameters in our model.
An advantage of this approach is that we do not need to consider the mechanisms of the wind driving or the details of how the evolution of the mass loss rate is determined by the evolution of the stellar magnetic field.
In the above formula, values of \mbox{$\dot{M}_\odot = 1.4 \times 10^{-14}$~M$_\odot$~yr$^{-1}$} and \mbox{$\Omega_\odot = 2.67 \times 10^{-6}$~rad~s$^{-1}$} should be used. 
We further assume that at high rotation rates, the mass loss rate saturates. 
Using this model, we can then calculate the wind base density once the wind temperature is set by simply scaling it to reproduce the mass loss rate predicted by the above formula.

In Paper~II, we include Eqn.~\ref{eqn:generalMdot} in our rotational evolution model and treat $a$ and $b$ as free parameters.
We therefore determine the values of $a$ and $b$ by fitting our rotational evolution model to the observational constraints and find best fits of $a=1.33$ and $b=-3.36$. 
Assuming that stellar radius scales with $M_\star^{0.8}$, this means that $\dot{M}_\star \propto \Omega_\star^{1.33} M_\star^{-1.76}$.
Therefore, at a given rotation rate, lower mass stars have higher mass loss rates than higher mass stars by factors of a few. 
For example, at a given rotation rate in the unsaturated regime, a 0.5~M$_\odot$ star will have a mass loss rate that is 3.4 times higher than a 1.0~M$_\odot$ star.
In Paper~II, we also treat the rotation rate at which the mass loss saturates as a free parameter in our rotational evolution model, and find that 

\begin{equation}
\Omega_{sat} = 15 \Omega_\odot \left( \frac{M_\star}{M_\odot} \right)^{2.3}.
\end{equation}

\noindent This implies that in the saturated regime, $\dot{M}_\star \propto M_\star^{1.3}$.

\subsection{The wind speeds of rapidly rotating stars} \label{sect:magnetorotation}

\begin{figure}
\centering
\includegraphics[width=0.49\textwidth]{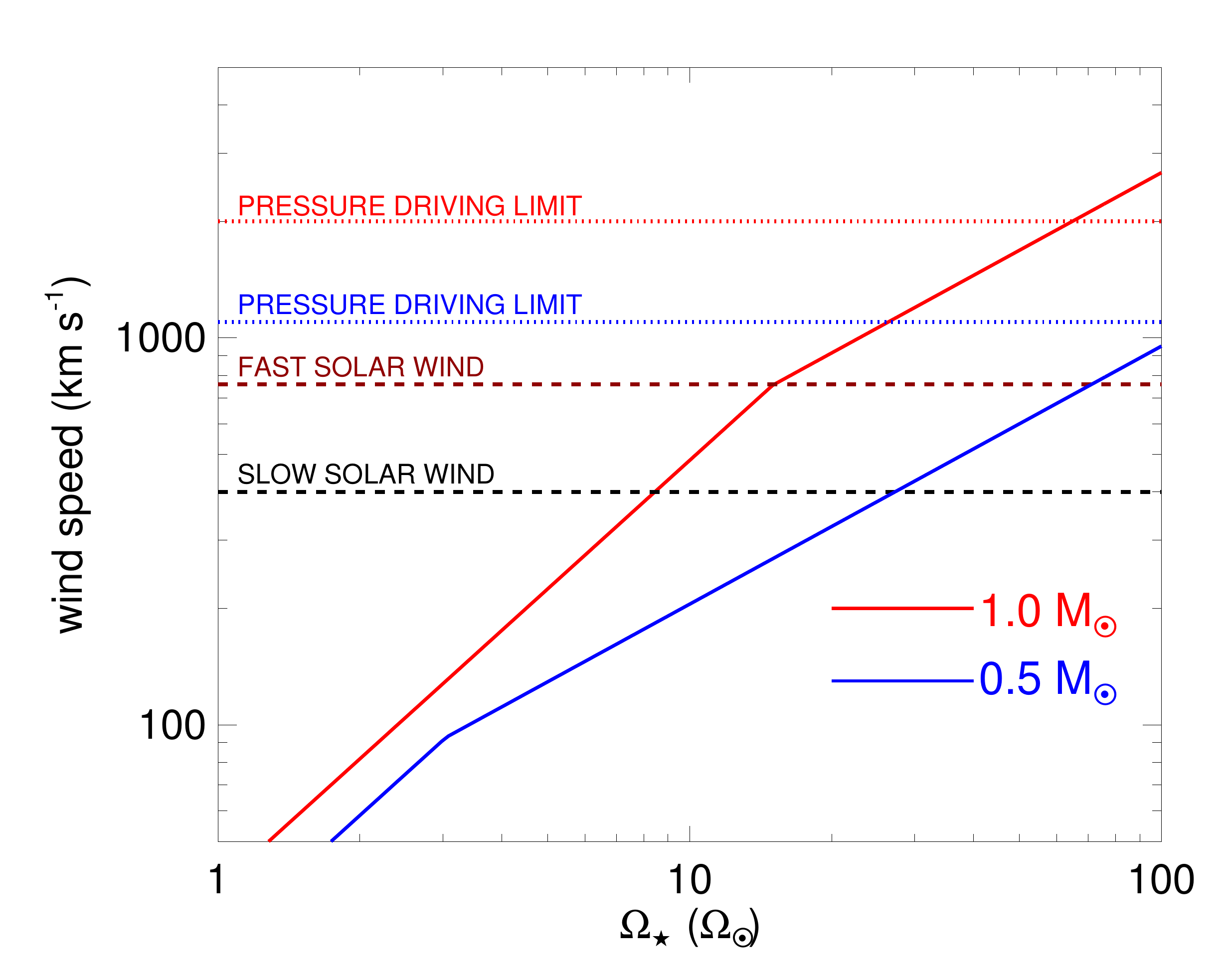}
\caption{
Figure showing the Michel velocity, given by Eqn.~\ref{eqn:MichelVel}, as a function of stellar rotation rate.
The solid blue and red lines are for stellar masses of 0.5~M$_\odot$ and 1.0~M$_\odot$ respectively, and the dashed black and red lines show the speeds of the slow and fast components of the solar wind respectively.
The dotted lines show our estimates for the highest wind speeds possible for purely thermal pressure driven winds, calculated assuming Model~A for the wind temperature.
}
 \label{fig:MichelVelocity}
\end{figure}

A mechanism that we do not consider in our solar wind model is the acceleration of the wind due to the coupling of the wind to the rotating star's magnetic field. 
As the wind expands away from the star, it is given angular momentum and kinetic energy by the stressed magnetic field, which can increase the wind speeds far from the star and, for the fastest rotators, even the mass loss rates can be increased. 
The basic magnetic rotator theory was developed by \citet{1967ApJ...148..217W} who applied their model to the Sun and showed that rotation had a negligible influence on the radial outflow speeds of the solar wind.
On the other hand, \citet{1976ApJ...210..498B} discussed the Weber-Davis model in the context of more rapidly rotating stars with stronger magnetic fields and showed that this mechanism can significantly alter the wind speeds, estimating a wind speed of $\sim$4000~\kms\hspace{0mm} for the young Sun in the equatorial plane.
\citet{1976ApJ...210..498B} defined two regimes: in the \emph{slow magnetic rotator} (SMR) regime, magneto-rotational effects are negligible, and in the \emph{fast magnetic rotator} (FMR) regime, such effects strongly influence the wind speeds and sometimes even the mass loss rates.

In the FMR regime, the wind structure necessarily becomes highly latitude dependent, and properly taking into account magneto-rotational effects requires 2D or 3D wind models (\citealt{1973ApJ...184...17S}). 
Increasing the complexity of our wind model to take into account the latitude dependence of the wind structure would be undesirable in this paper.
On the other hand, given the importance of magneto-rotational effects for the winds of rapidly rotating stars, ignoring them completely would be inappropriate. 
We therefore use a simplistic model for estimating the influence of the magneto-rotational acceleration of the winds \emph{in the equatorial plane} and leave more detailed analysis for later work.

In the FMR regime, the terminal wind speed in the equatorial plane is approximately given by the Michel velocity, $v_{\text{M}}^3 = r^4 B_r^2 \Omega_\star^2/\dot{M}_\star$, where $B_r$ is the strength of the radial component of the magnetic field at radius $r$ (\citealt{1976ApJ...210..498B}). 
Given that far from the star, a star's magnetic field is approximately radial, $B_r$ is given by $B_r \propto r^{-2}$, and therefore $v_\text{M}$ is independent of radius. 
The Michel velocity can therefore be calculated at any $r$, so taking \mbox{$r=R_\star$} we find 

\begin{equation} \label{eqn:MichelVel}
v_{\text{M}} = \left( \frac{R_\star^4 B_\star^2 \Omega_\star^2}{\dot{M}_\star} \right)^{\frac{1}{3}},
\end{equation}

\noindent where $B_\star$ is a measure of the star's magnetic field strength.
It is important to be clear exactly what $B_\star$ means; the acceleration of the wind due to the magnetic field is determined by the magnetic field strength close to the equatorial plane far from the stellar surface.  
Within the stellar corona, the magnetic field can be decomposed into the sums of different components, such as dipoles, quadrupoles, and octupoles (\citealt{2002MNRAS.333..339J}; \citealt{2010RPPh...73l6901G}). 
Close to the stellar surface, the more complex components dominate, but further out in the corona, the dipole field is dominant (see Fig.~5 of \citealt{2014MNRAS.437.3202J}).
Therefore, on large scales within the corona, the magnetic field strength can be approximated by the dipole field strength, $B_{\text{dip}}$, with a dependence on distance of $r^{-3}$. 
At a certain radius, the magnetic field takes on a radial structure due to the wind, and therefore has a $r^{-2}$ dependence.
The magnetic field strength far from the stellar surface in the equatorial plane can be approximated as 

\begin{equation}
B_r \approx \frac{1}{2} B_{\text{dip}} \left( \frac{R_{\text{ss}} }{ R_\star} \right)^{-3} \left(  \frac{r}{R_{\text{ss}} } \right)^{-2},
\end{equation}

\noindent where $B_\text{dip}$ is the dipole field strength at the stellar pole and $R_\text{ss}$ is the radius of the `source surface' where the magnetic field is assumed to become radial due to the stellar wind. 
The factor of $1/2$ in the above equation is due to the fact that the magnetic field strength at the pole of a dipole field is a factor of two stronger than at the equator. 
Based on the solar example, we make the approximation that \mbox{$R_\text{ss} = 2.5 R_\star$}.
Therefore, we approximate $B_\star$ in Eqn.~\ref{eqn:MichelVel} as \mbox{$B_\star \approx 1.25 B_\text{dip}$}. 
This is not a real field strength at the stellar surface, but a field strength that leads to approximately the correct field strengths outside of the stellar corona if it is assumed that the field strength decreases with $r^{-2}$ at all $r$ above the stellar surface.
In Paper~II, we derive an average \emph{polar} field strength for the Sun of 2.7~G and assume that this value scales to other stars in the unsaturated regime as $B_\text{dip} \propto Ro^{-1.32}$, where $Ro$ is the Rossby number.
This latter scaling law was derived by \citet{2014MNRAS.441.2361V} using a large sample of measured magnetic field strengths.
Since these field strengths were based on Zeeman-Doppler~Imaging reconstructions of the magnetic fields across the surfaces of stars, which is only sensitive to the large scale magnetic field strengths, the scaling laws of \citet{2014MNRAS.441.2361V} are likely to be the best approximations of dipole field strength available in the literature.  

Although the $\dot{M}_\star$ term in Eqn.~\ref{eqn:MichelVel} corresponds to the mass flux in the equatorial plane of the star (i.e. it is the mass loss rate that a star would experience if the mass flux in all directions was equal to the mass flux in the equatorial plane), we make the assumption that the magneto-rotational mechanism does not significantly influence the mass flux at the base of the wind, and therefore the real stellar mass loss rate.
This might be unrealistic for the most rapidly rotating stars, meaning that the mass loss rate does not truly saturate at rapid rotation as we assume in our model. 
As discussed in Paper~II, one indication that the mass loss rates do saturate is the rotational evolution of the most rapidly rotating stars. 
The exponential shape of the spin down curves for rapidly rotating stars implies approximately that $d\Omega_\star / dt \propto \Omega_\star$ and the slow spin down of such stars implies that the wind torque increases with rotation slowly above the saturation threshold for the magnetic field.
Both of these facts are difficult to explain without saturation of the mass loss rate. 
Similarly, \citet{1976ApJ...210..498B} analysed the Weber-Davis model and found that although the wind speed was strongly dependent on rotation, little change in the mass loss rate was found.
However, we still expect that at the fastest rotation rates, the mass loss rates could be influenced by magneto-rotational driving.

In Fig.~\ref{fig:MichelVelocity}, we show the dependence of $v_{\text{M}}$ on rotation for 0.5~M$_\odot$ and 1.0~M$_\odot$ stars. 
We estimate that the current solar wind has a Michel velocity of $\sim$40~\kms, and is therefore in the SMR regime.
This is similar to the value of 60-90~\kms\hspace{0mm} estimated by \citet{1976ApJ...210..498B}.
The transition from the SMR regime to the FMR regime happens approximately where $v_{\text{M}}$ exceeds the speed that the wind would have had in the absence of rotation (\citealt{1976ApJ...210..498B}). 
We assume that the wind speed at 1~AU is given by $v_{\text{M}}$ in the FMR regime and is given by our thermal pressure driven wind model in the SMR regime, with the mass flux being determined in both cases by Eqn.~\ref{eqn:generalMdot}. 
This means that a solar mass star with a rotation rate close to 100~$\Omega_\odot$ would have wind speeds in the equatorial plane of~$\sim3000$~\kms. 
One interesting advantage of this approach is that in the FMR regime, the uncertainties discussed in Section~\ref{sect:scalingT0} for how the wind temperatures should be calculated become irrelevant.

\section{Results} \label{sect:results}

\subsection{A grid of models} \label{sect:grid}

\begin{figure*}
\centering
\includegraphics[width=0.49\textwidth]{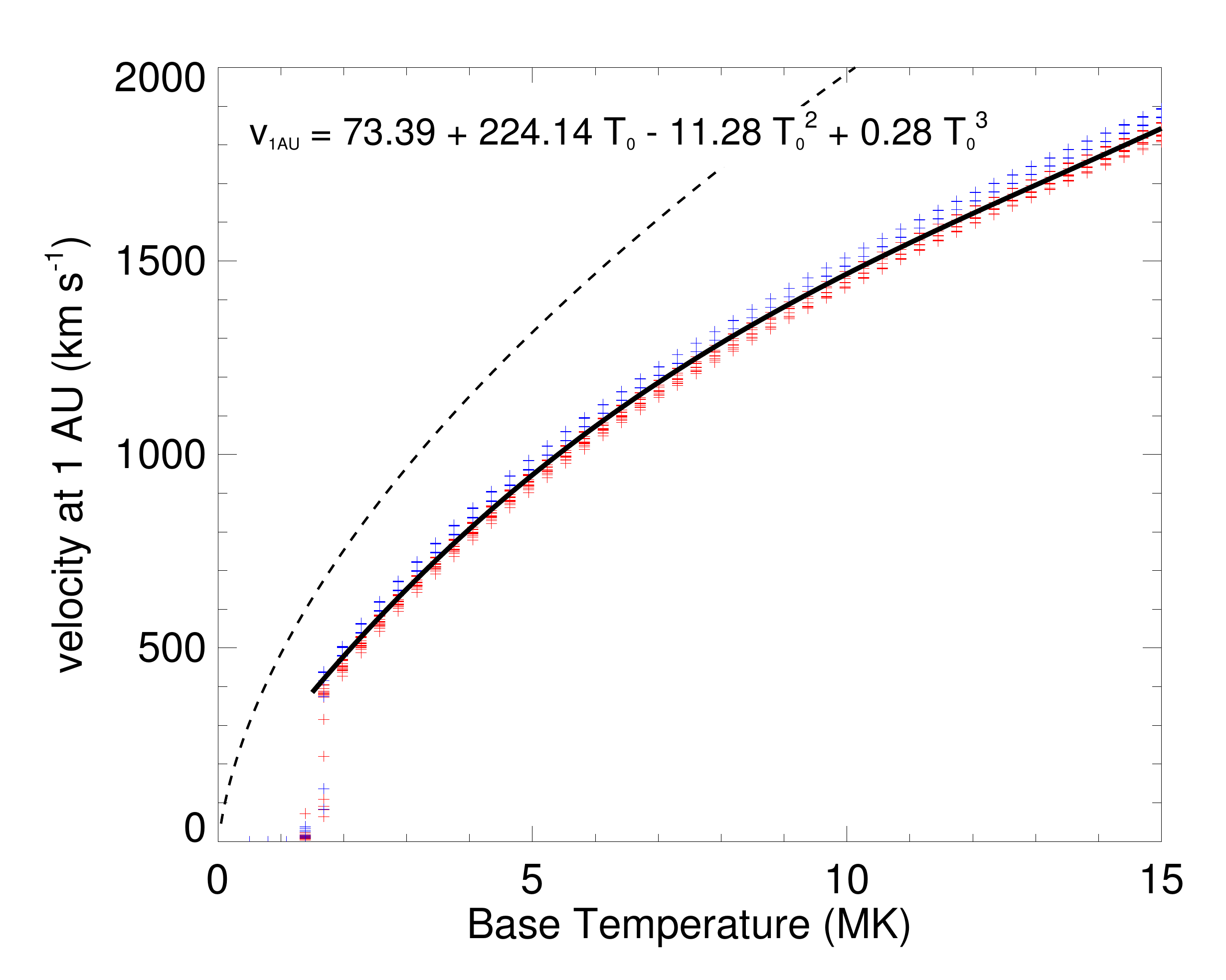}
\includegraphics[width=0.49\textwidth]{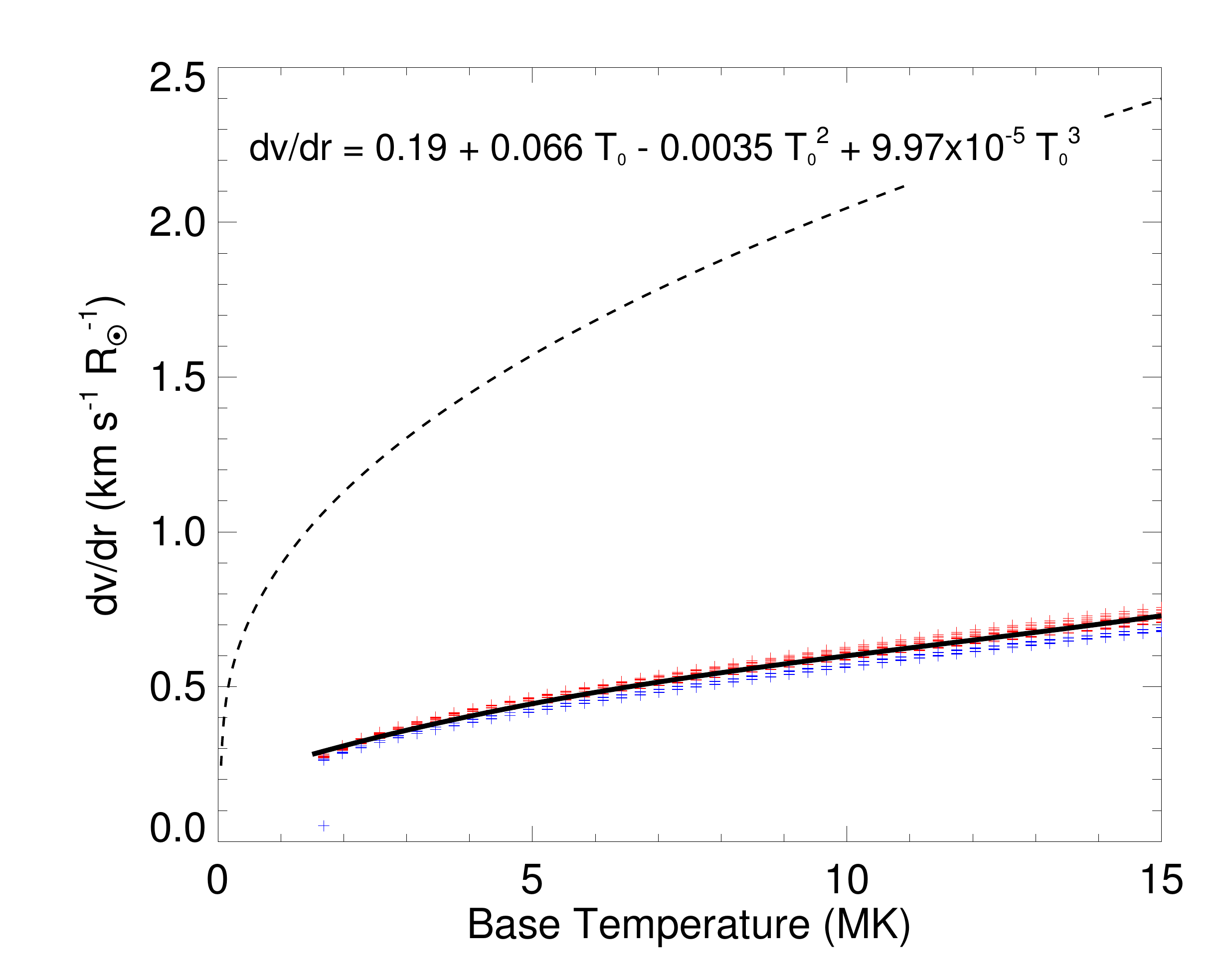}
\includegraphics[width=0.49\textwidth]{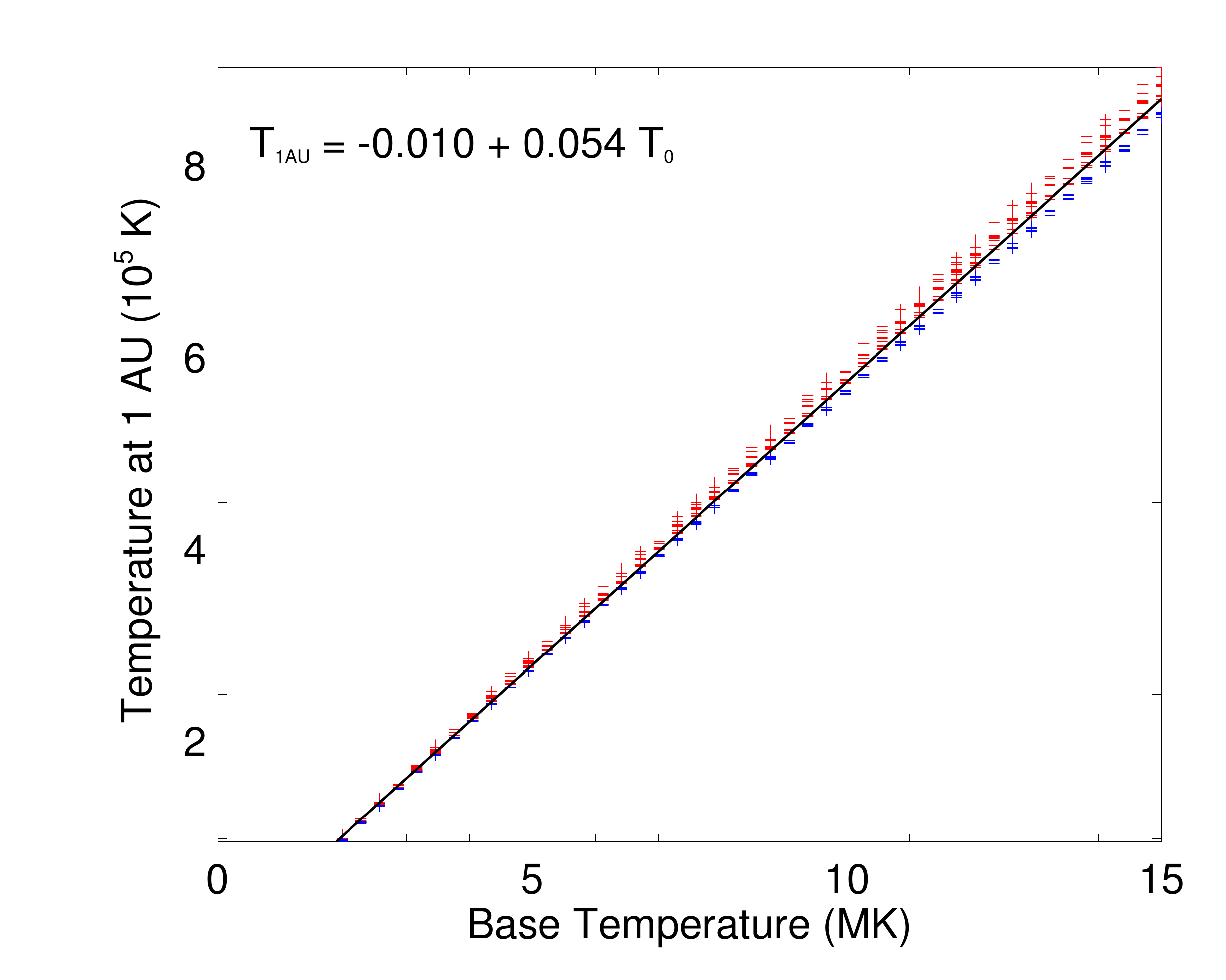}
\includegraphics[width=0.49\textwidth]{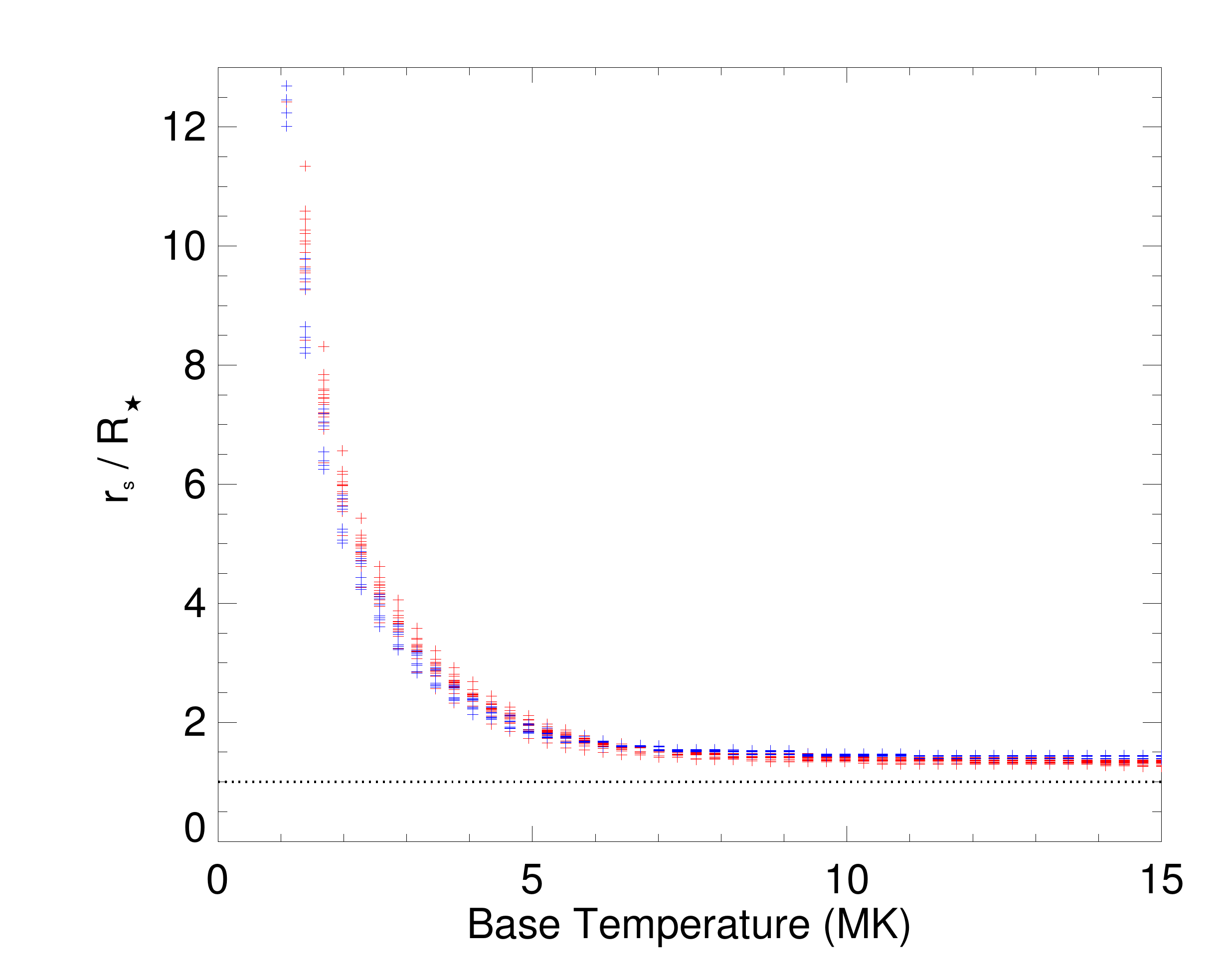}
\caption{
Figures showing the results of the grid of 1D hydrodynamic wind models with differing base temperatures, stellar masses, and stellar radii, as discussed in Section~\ref{sect:grid}.
The plots show wind velocity at 1~AU (\emph{upper left panel}), the average wind acceleration between 30~$R_\star$ and 1~AU (\emph{upper right panel}), the wind temperature at 1~AU (\emph{lower left panel}), and the radius at which the wind becomes supersonic (\emph{lower right panel}), against wind base temperature for all models. 
Clearly temperature is the most important parameter in determining the properties of the wind. 
The small spreads in each parameter at each temperature is a result of differing surface escape velocities for stars with different masses and radii.
At low temperatures, the wind does not have enough energy to expand and no wind is possible. 
The red and blue points represent stellar masses above and below 0.65~M$_\odot$ respectively. 
The dashed lines in the upper panels show the predictions for an isothermal Parker wind from a 1~M$_\odot$ and 1~R$_\odot$ star.
}
 \label{fig:GRIDresults}
\end{figure*}


Our physical wind model has only four input parameters: these are the stellar mass, the stellar radius, the wind base temperature, and the wind base density. 
The first two of these are basic properties of the star and the latter two are determined by our assumptions discussed in Section~\ref{sect:scalingT0} and Section~\ref{sect:scalingn0}.
In order to study the effects of these parameters on our winds, we produce a grid of 1200 wind models.
Each model is run using our hydrodynamic solar wind model developed in Section~\ref{sect:solarwindmodel} by varying the stellar mass, the stellar radius, and the wind temperature.
In our grid, we assume stellar masses, $M_\star / M_\odot$, of 0.5, 0.6, 0.7, 0.8, 0.9, and 1.0, and for each mass calculate models for four different radii equally spaced about the minimum and maximum radii that main-sequence stars of these masses will have between 100~Myrs and 5~Gyrs. 
Then for each combination of mass and radius, we calculate 50 models with base temperatures equally spaced between 0.5~MK and and 15~MK. 
For all models, we assume a base proton density of 10$^7$~cm$^{-3}$, which is chosen arbitrarily.
We do not explore the influence of the base density in this section since it does not influence the wind acceleration\footnotemark and the density at all points in the winds, and therefore the mass loss rates, are simply directly proportional to the base density once the other parameters are set. 

\footnotetext{
It is important to note that the base density does not influence the wind acceleration in 1D hydrodynamic wind models.
In 2D or 3D magnetohydrodynamic wind models, the base density will influence the \mbox{plasma-$\beta$} and the Alfv\'{e}n surface, and therefore have some influence on the dynamics of the wind.
}

In Fig.~\ref{fig:GRIDresults}, we show the dependence of wind speed, wind acceleration far from the star, temperature at 1~AU, and the location of the sonic point on the base temperature of the wind for all of our models. 
Although at a given base temperature, there is a small spread in these parameters, all of the models are on very tight tracks.
This is unsurprising given the weak relation between surface escape velocity and stellar mass on the main-sequence (approximately, $v_{\text{esc}} \propto M_\star^{0.1}$). 

We find that the wind speed at 1~AU can be well described by

\begin{equation} \label{eqn:windspeedgrid}
v_{1\text{AU}} \approx 73.39 + 224.14 T_0 - 11.28 T_0^2 + 0.28 T_0^3,
\end{equation}

\noindent where $T_0$ is in MK and $v_{1\text{AU}}$ is in km~s$^{-1}$.
As we showed for the solar wind model in Section~\ref{sect:solarwindmodel}, the wind never really comes to a terminal velocity, though beyond 30~R$_\odot$, the wind acceleration is small.
In the upper right panel of Fig.~\ref{fig:GRIDresults}, we show the average wind acceleration between 30~R$_\star$ and 1~AU for each of our models.
To be clear, when we discuss `acceleration', we are referring to the radial gradient of the wind speed at a given location, $dv/dr$, and not the temporal gradient of the wind speed for a given particle. 
We find that $dv/dr$ in this region is given by 

\begin{equation} \label{eqn:windaccelgrid}
\frac{dv}{dr} \approx 0.19 + 0.066 T_0 - 0.0035 T_0^2 + 9.97 \times 10^{-5} T_0^3,
\end{equation}

\noindent where $T_0$ is in MK and $dv/dr$ is in km~s$^{-1}$~R$_\odot^{-1}$.
For a comparison, we also show in Fig.~\ref{fig:GRIDresults} the predictions for an isothermal Parker wind from a 1~M$_\odot$ and 1~R$_\odot$ star. 
As we have already shown, at a given wind temperature, the isothermal model overestimates the wind speeds, and this can clearly be seen in Fig.~\ref{fig:GRIDresults}. 
However, more interesting is the shape of the relation between $v_{1\text{AU}}$ and $T_0$ since this determines how well the isothermal model can be used to scale the solar wind to other stars.
The isothermal model predicts a steeper increase in wind speed with increasing wind temperature than we predict from our more sophisticated model; however, the relations are similar and the isothermal model is an acceptable first approximation.
On the other hand, the isothermal model provides a poor description for the wind acceleration far from the stellar surface.

\begin{figure}
\centering
\includegraphics[width=0.49\textwidth]{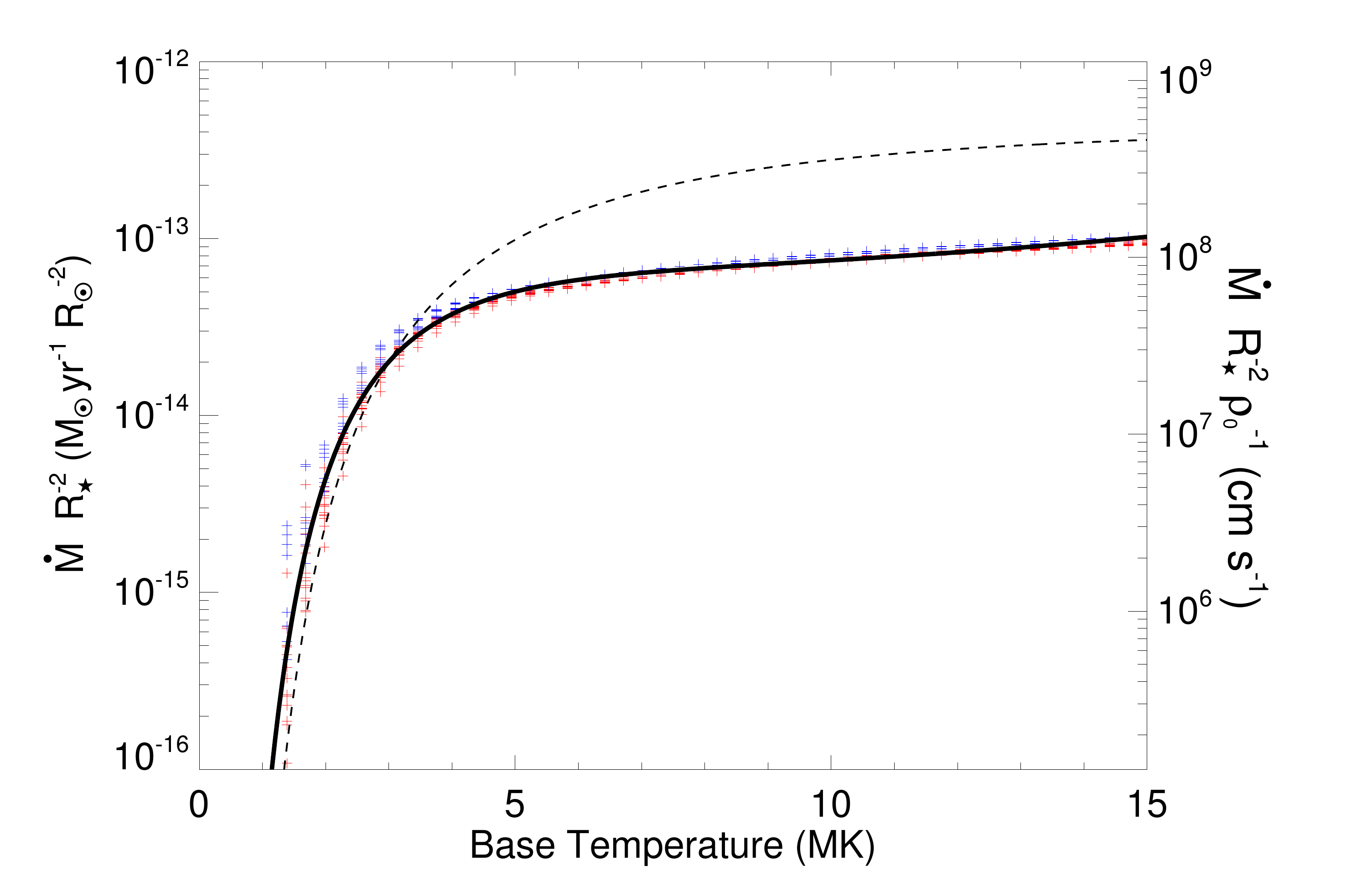}
\caption{
Mass loss rate as a function of wind base temperature for our grid of 1D hydrodynamic wind models discussed in Section~\ref{sect:grid}.
The scale on the left \mbox{$y$-axis} shows $4\pi$ times the mass loss rate per unit surface area, $\dot{M}_\star R_\star^{-2}$, assuming a base proton density of $10^{-7}$~cm$^{-3}$.
The scale on the right axis shows \mbox{$\dot{M}_\star R_\star^{-2} \rho_0^{-1}$}, where $\rho_0$ is the mass density at the base of the wind. 
The models differ by stellar mass, stellar radius, and base temperature, but clearly the most important parameter is the base temperature.
The red and blue points represent stellar masses above and below 0.65~M$_\odot$ respectively. 
The solid line shows our best fit given by Eqn.~\ref{eqn:MdotResult} and the dashed line shows what would be expected for an isothermal wind from a 1~M$_\odot$ and 1~R$_\odot$ star.
}
 \label{fig:GRIDresultsMdot}
\end{figure}

The temperature at 1~AU depends linearly on the base temperature as 

\begin{equation} \label{eqn:windtempgrid}
T_{1\text{AU}} \approx 0.054 T_0 - 0.010,
\end{equation}

\noindent where both temperatures are in MK.
The lower panel of Fig.~\ref{fig:GRIDresults} shows the radius of the point where the wind becomes supersonic. 
This is determined by two competing factors; higher temperatures lead to the wind being accelerated faster, but also lead to higher sound speeds.
The former effect dominates, and therefore hotter winds tend to have sonic points closer to the stellar surface.
Similarly, for an isothermal Parker wind, the radius of the sonic point at a given stellar mass is given by $r_s \propto T^{-1}$ (\citealt{1999isw..book.....L}). 

Probably the most interesting parameter that characterises the stellar wind is the mass loss rate. 
In our wind model, the mass loss rate is determined by four parameters: the base density of the wind, the wind temperature, the surface area of the star, and the surface escape velocity. 
At a given base temperature and surface escape velocity, the mass loss rate of the wind is directly proportional to both the stellar surface area and the base density.  
Fig.~\ref{fig:GRIDresultsMdot} shows the dependence of $\dot{M}_\star R_\star^{-2}$ on base temperature, assuming a base proton density of 10$^{7}$~cm$^{-3}$.
Alternatively, the relation shown in Fig.~\ref{fig:GRIDresultsMdot} can be considered a dependence of $\dot{M}_\star R_\star^{-2} \rho_0^{-1}$ on wind base temperature.
Once again, although there is some spread in the values at a given temperature, it is clear that the most important parameter is the base temperature. 
The best fit line shown in Fig.~\ref{fig:GRIDresultsMdot} is given by

\begin{equation} \label{eqn:MdotResult}
\begin{multlined}
\log \left( \frac{\dot{M}_\star}{\rho_0 R_\star^2} \right) \approx  3.42 + 10.20 \log T_0 \\- 1.94 \left(\log T_0\right)^2 + 3.79 \left( \log T_0 \right)^3,
\end{multlined}
\end{equation}

\noindent where $T_0$ is in MK, $R_\star$ is in R$_\odot$, $\rho_0$ is in g~cm$^{-3}$, and $\dot{M}_\star$ is in M$_\odot$~yr$^{-1}$.
To be clear, since we do not have a model that we can use to predict $\rho_0$, we do not use this relation later in the paper; instead we calculate $\dot{M}_\star$ from Eqn.~\ref{eqn:generalMdot}.
If necessary, it is possible to predict $\rho_0$ for a given star by inserting Eqn.~\ref{eqn:generalMdot} into Eqn.~\ref{eqn:MdotResult}.

For comparison, we also show in Fig.~\ref{fig:GRIDresultsMdot} the relation between $\dot{M}_\star$ and wind temperature for an isothermal wind from a 1~M$_\odot$ and 1~R$_\odot$ star. 
The dependence in our wind model is similar to that of the isothermal model, probably because in our wind model, the polytropic index of 1.05 is used in most of the subsonic wind region. 
However, there is some disagreement between our model and the isothermal model at high wind temperatures, indicating that using the isothermal model to scale the solar wind to other stars is likely to lead to an overestimation of the wind flux at high wind temperatures.

\begin{figure*}[!htbp]
\centering
\includegraphics[trim= 0cm 0cm 2cm 0cm, width=0.6\textwidth]{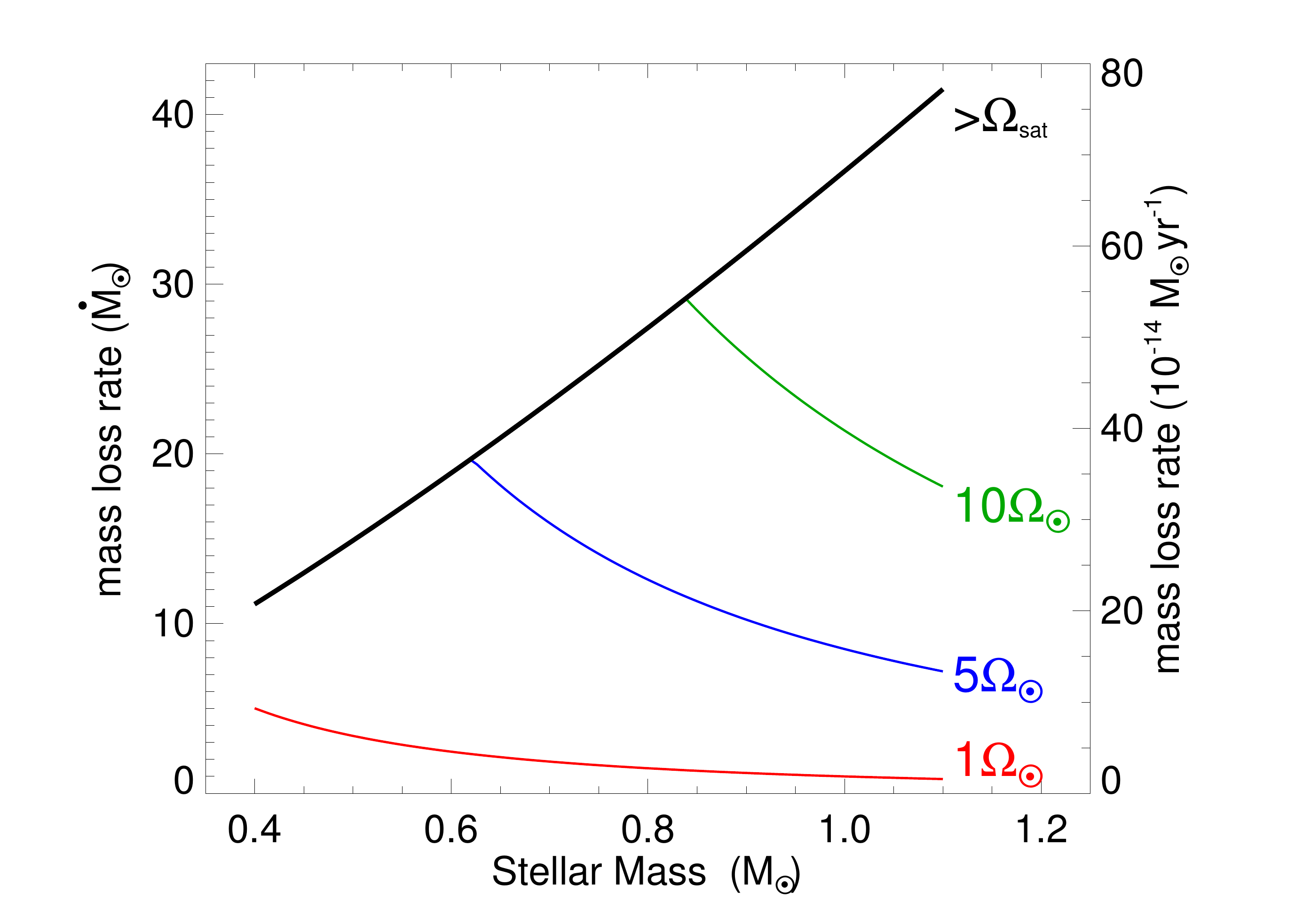}
\includegraphics[trim=-7cm 2cm 5cm 2cm, width=0.4\textwidth]{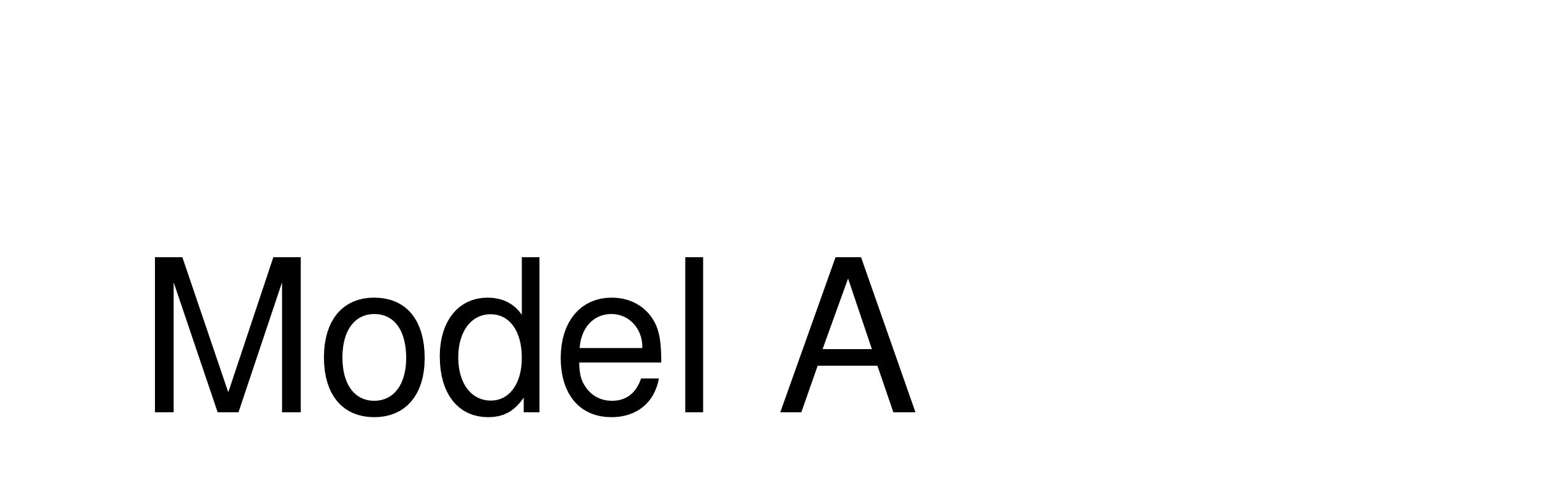}
\includegraphics[trim=-7cm 2cm 5cm 2cm, width=0.4\textwidth]{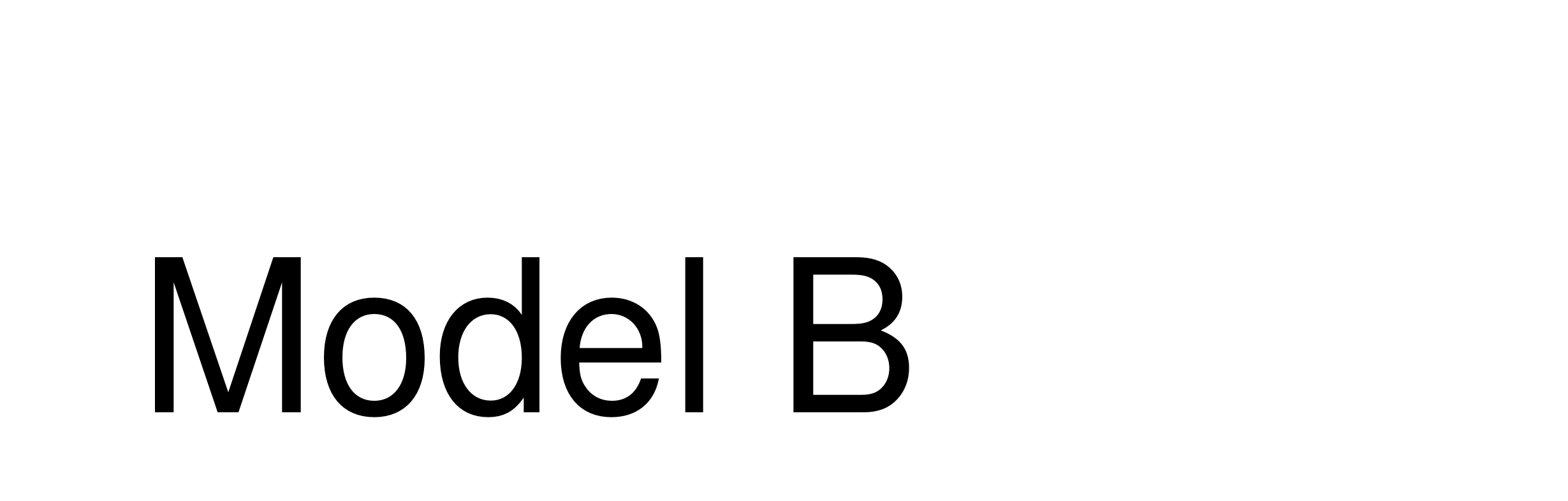}
\includegraphics[width=0.41\textwidth]{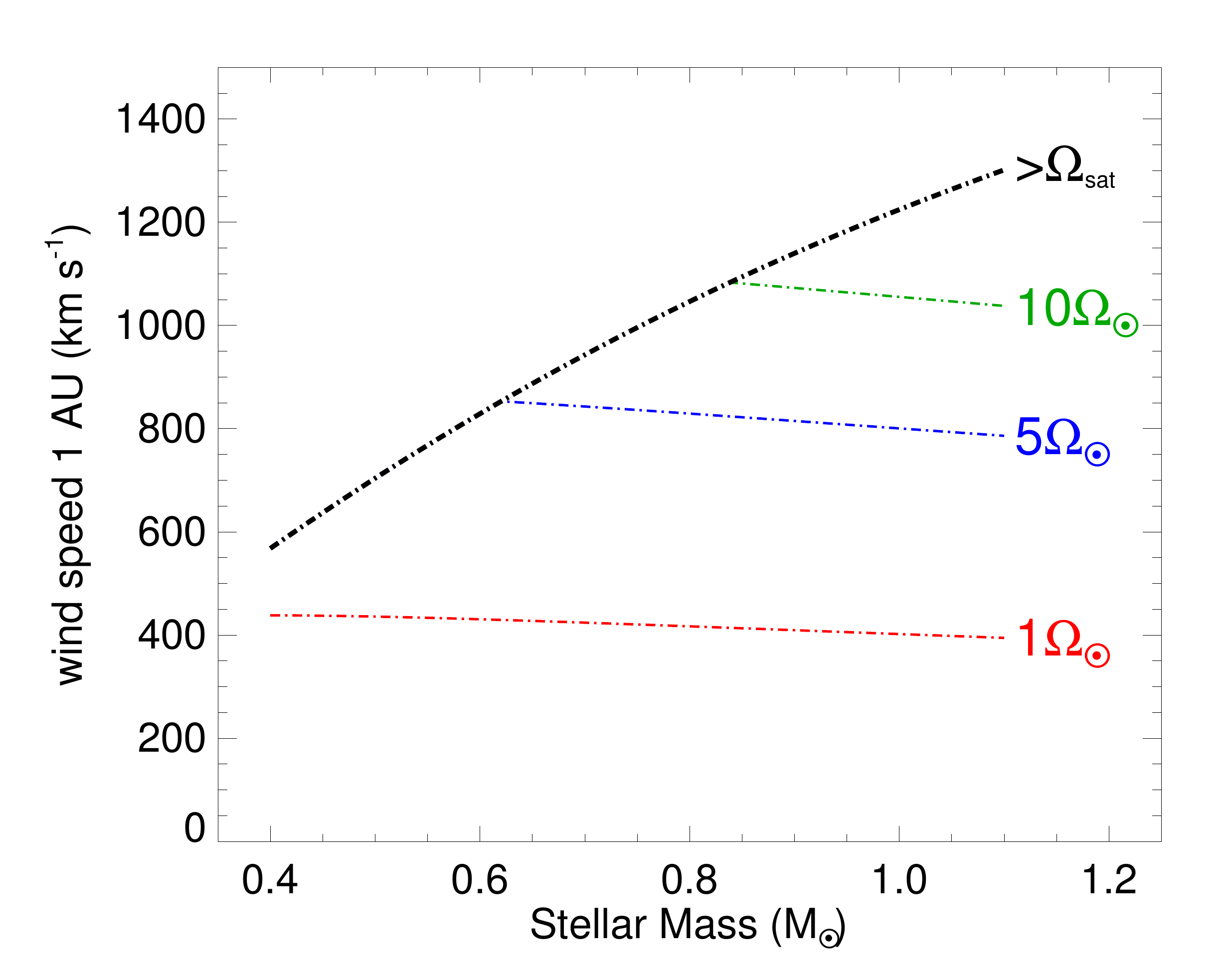}
\includegraphics[width=0.41\textwidth]{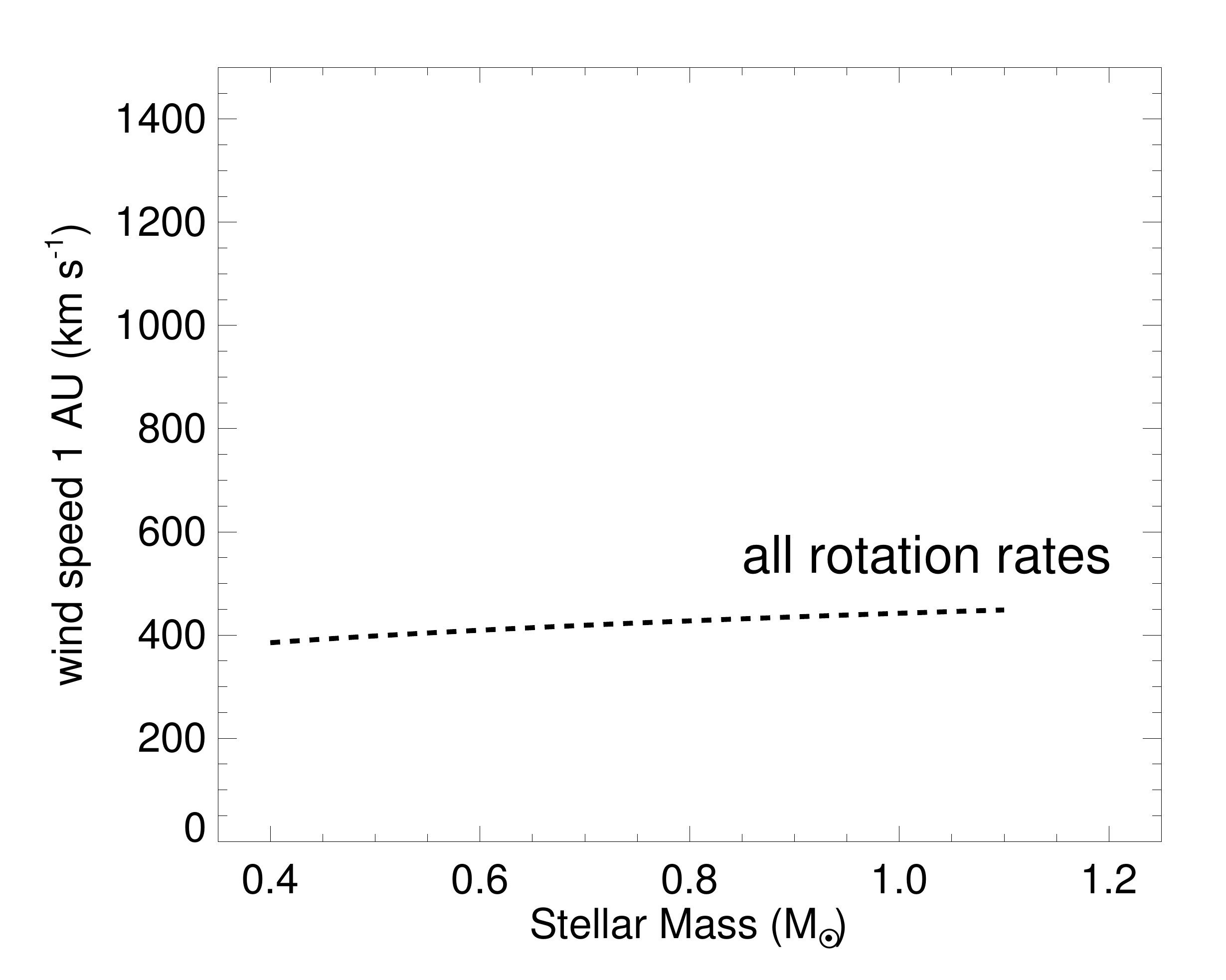}
\includegraphics[width=0.41\textwidth]{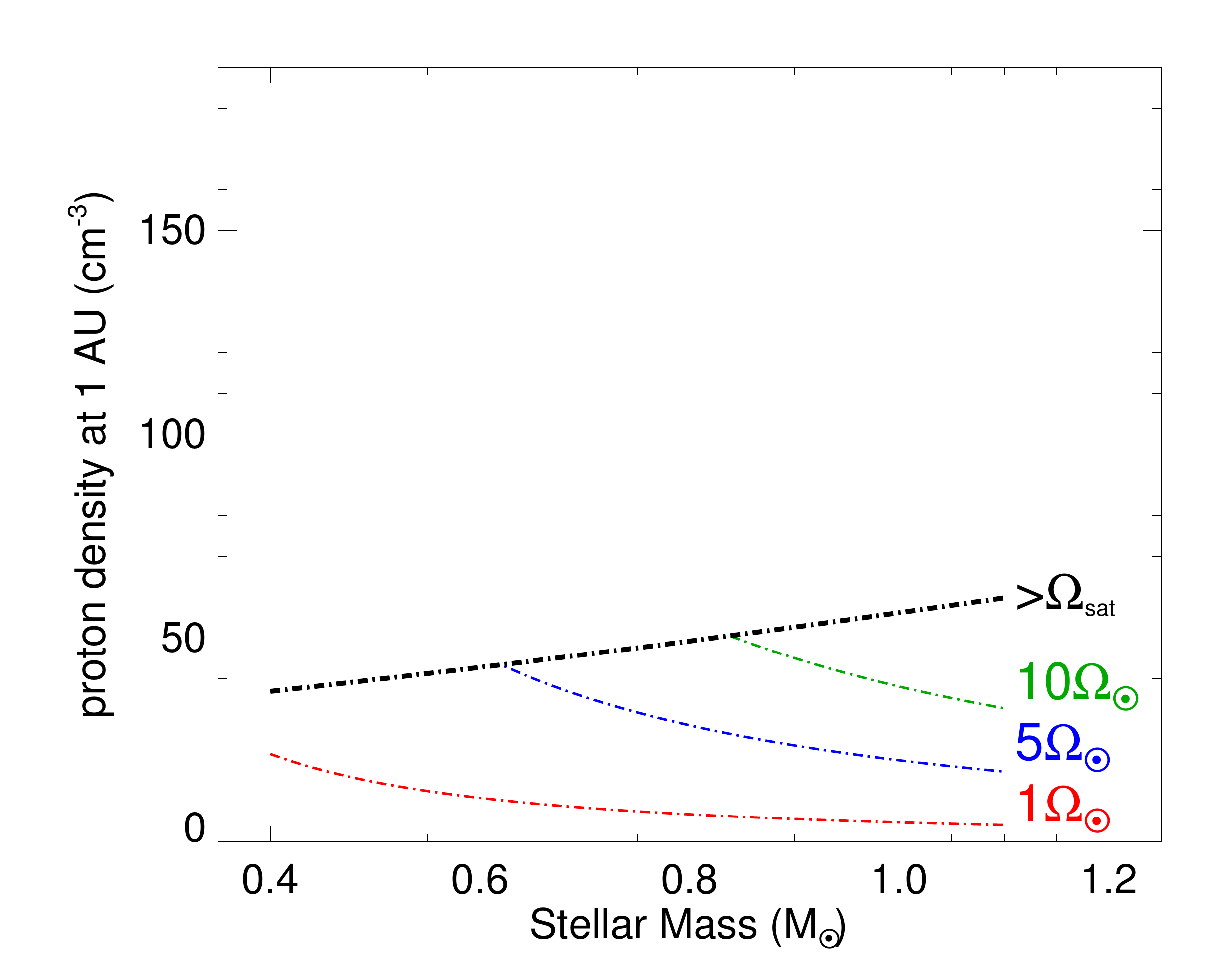}
\includegraphics[width=0.41\textwidth]{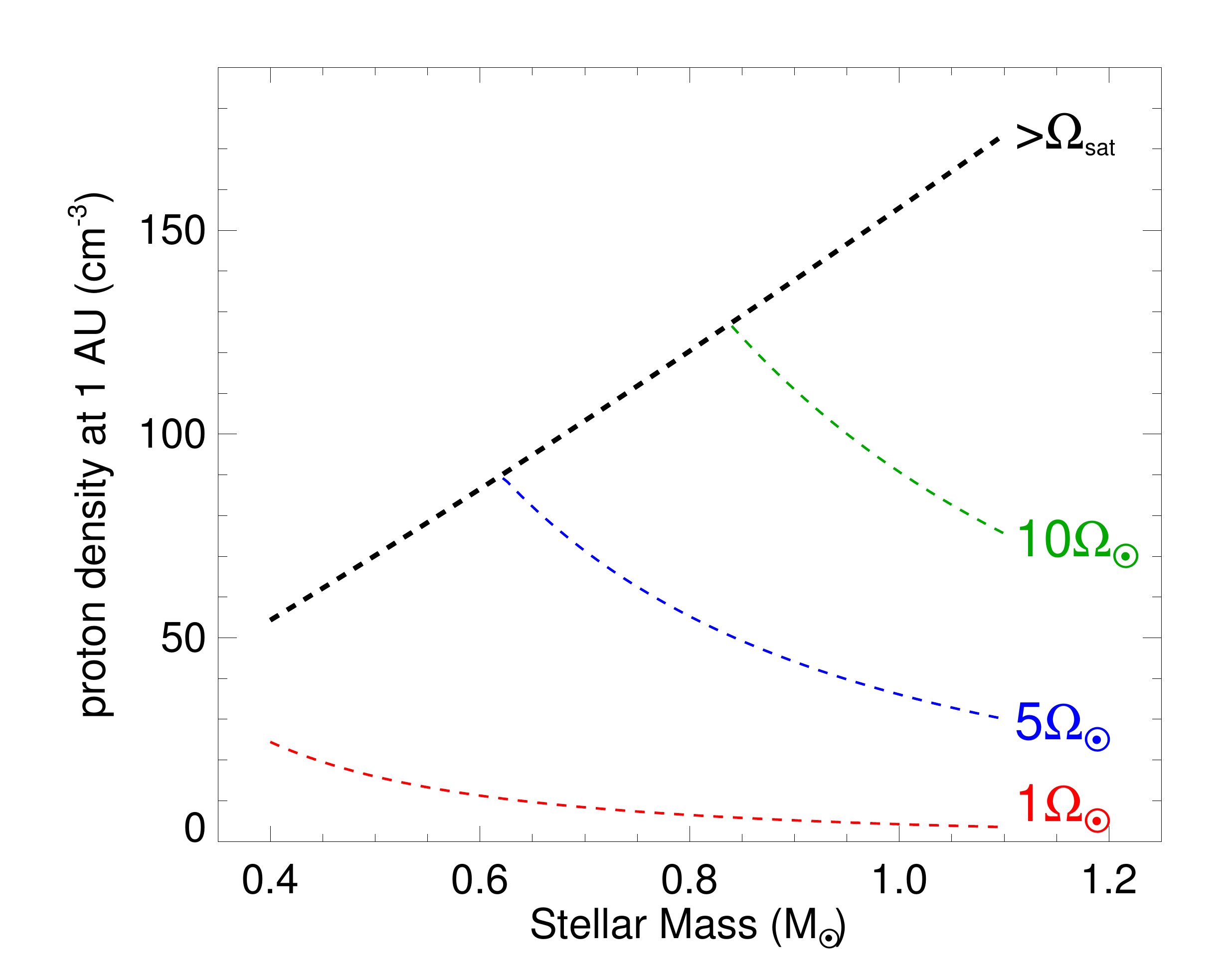}
\caption{
Figures comparing the wind properties at 1~AU as predicted from Model~A and Model~B.
The upper panel shows wind mass loss rate as a function of stellar mass at several different rotation rates.
We predict the wind mass loss rate using $\dot{M}_\star \propto R_\star^2 \Omega_\star^{1.33} M_\star^{-3.36}$, which we derive from rotational evolution models in Paper~II. 
The middle panels show wind speed at 1~AU as a function of mass for each rotation rate predicted by our two wind models.
The lower panels show proton density at 1~AU as a function of mass for each rotation rate, calculated by combining the mass loss rate and wind speed predictions. 
}
 \label{fig:ModelAModelBcomparison}
\end{figure*}

At base temperatures below 5~MK, the wind mass loss rate is strongly temperature dependent.
However, at higher temperatures, this dependence becomes much weaker and by 10~MK, there is almost no dependence on wind temperature.
This is likely because as the temperature increases, the sonic point moves closer to the stellar surface.
The mass flux in a hydrodynamic thermal pressure driven wind is determined by heating that takes place below the sonic point.
Heating that takes place above the sonic point can only contribute to the final kinetic energy of the wind, but not to the mass flux.
As the sonic point moves closer to the surface, the fraction of heating that contributes to the mass flux in the wind decreases, and therefore, we get the saturation of the wind mass loss rate at high temperatures.
This can also be seen in the upper left panel of Fig.~\ref{fig:GRIDresults} where the wind speeds at 1~AU continue to increase rapidly with increasing wind temperature, even beyond base temperatures of 10~MK.
However, we stress that the saturation of $\dot{M}_\star$ occurs when we assume a constant base density; varying the base density with stellar rotation rate for example could lead to changes in the saturation mass loss rate.

Another interesting feature of the results that can be seen in Fig.~\ref{fig:GRIDresults} and Fig.~\ref{fig:GRIDresultsMdot} is that at low temperatures, the wind terminal velocities and mass loss rates tend to zero. 
This is because the winds no longer have enough energy to expand and remain bound to the stars.
In our simulations, this happens at a temperature of approximately 1.5~MK. 
Although it is physically realistic that at low temperatures, the winds are not able to expand, it is likely that our models overestimate this threshold for real winds given that our winds are driven only by thermal pressure gradients. 
As we discuss in Section~\ref{sect:numericalmodel}, in real winds, it is likely that other forces, such as wave-pressure forces, are partly responsible for the driving of the wind.

\subsection{A comparison between Model A and Model B} \label{sect:resultsAandB}


In Section~\ref{sect:scalingT0}, we develop two radically different methods for scaling the base temperature in our stellar wind model.
Since the temperature is the main parameter that determines the acceleration of the wind, these models can directly predict the wind speed far from the stellar surface.
However, neither model can be used to predict the mass flux in the wind since the base density is still a parameter that needs to be determined.
As discussed in Section~\ref{sect:scalingn0}, we use the rotational evolution model developed in Paper~II of this series to constrain the mass loss rates as a function of stellar radius, mass, and rotation rate, which we then use to scale the base density in the wind.  
Therefore, the mass flux, $\rho v$, in our wind is not dependent on which assumption we make for the base temperature of the wind. 
The difference between Model~A and Model~B is in how this mass flux is achieved.

In the upper panel of Fig.~\ref{fig:ModelAModelBcomparison}, we show how mass loss rate depends on stellar mass and rotation, assuming for simplicity that $R_\star \propto M_\star^{0.8}$. 
At slow rotation, the mass loss rate has a moderate dependence on stellar mass, with lower mass stars having slightly higher mass loss rates than higher mass stars, but the main parameter is clearly stellar rotation rate.
On the other hand, in the saturated regime, there is a strong dependence of mass loss rate on stellar mass, with low-mass stars having lower mass loss rates than high-mass stars. 
This is due entirely to the mass dependence of the saturation threshold.
The lower saturation threshold for low-mass stars means that they are never able to achieve the high mass loss rates of the most rapidly rotating solar-mass stars.

In the middle and lower panels of Fig.~\ref{fig:ModelAModelBcomparison}, we show how the changes in the mass flux are distributed between changes in wind speed and wind density at 1~AU.
For simplicity, we only show the results that we get for scaling our slow solar wind model, and we emphasise that these results represent the wind speeds considering acceleration from thermal pressure only, and not considering acceleration due to magneto-rotational effects, as discussed in Section~\ref{sect:magnetorotation}. 
In the case of Model~B, the wind velocity is not a function of rotation rate and only a weak function of stellar mass.
Therefore, the changes in the mass flux of the wind are entirely due to changes in the wind density. 
On the other hand, in Model~A, the wind speed is a strong function of rotation.
In the saturated regime, the wind speed is also a strong function of stellar mass due to the strong mass dependence of the saturation threshold.
We therefore predict in Model~A a much weaker dependence of the wind density on rotation than predicted in Model~B.

\begin{figure}
\centering
\includegraphics[width=0.49\textwidth]{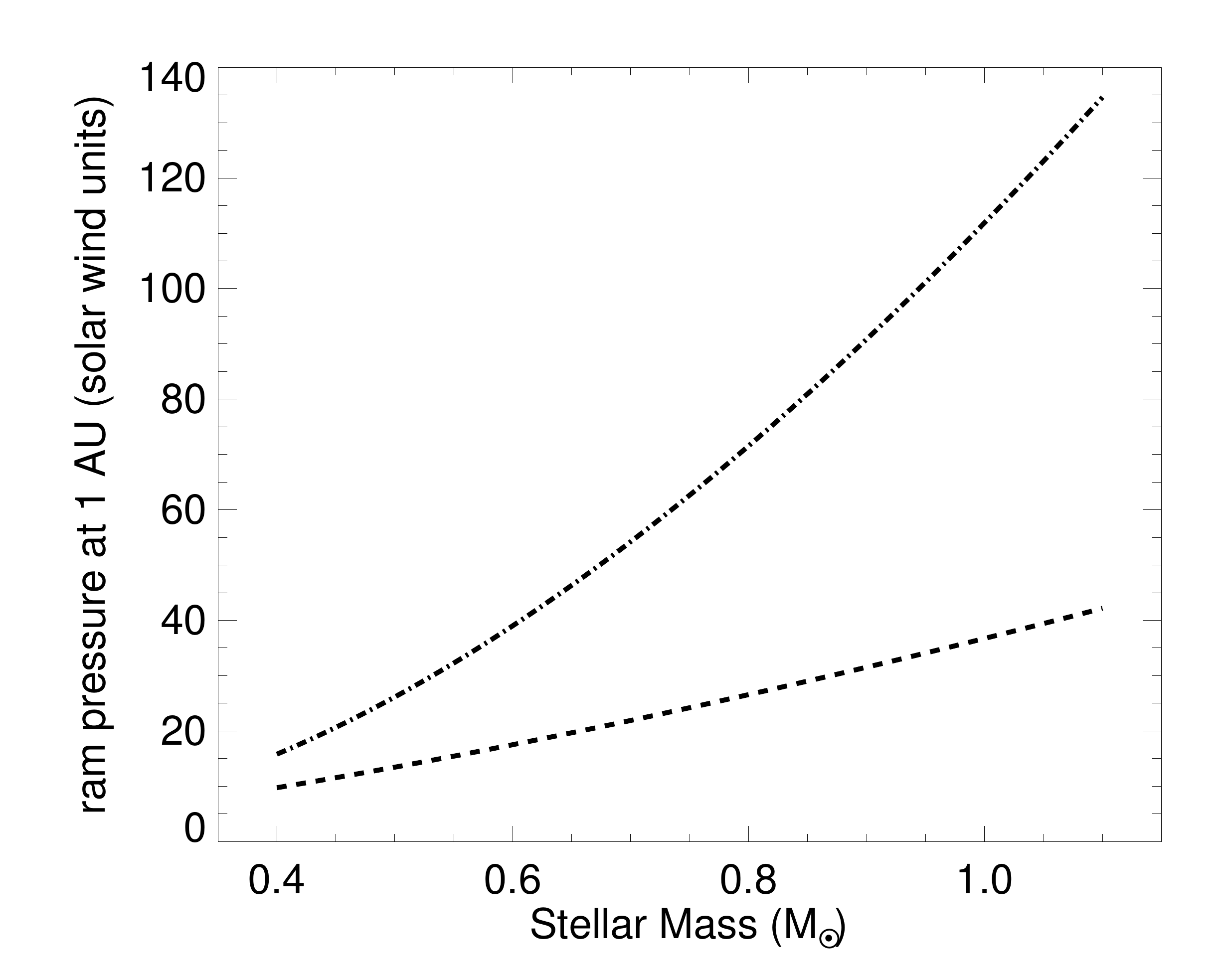}
\includegraphics[width=0.49\textwidth]{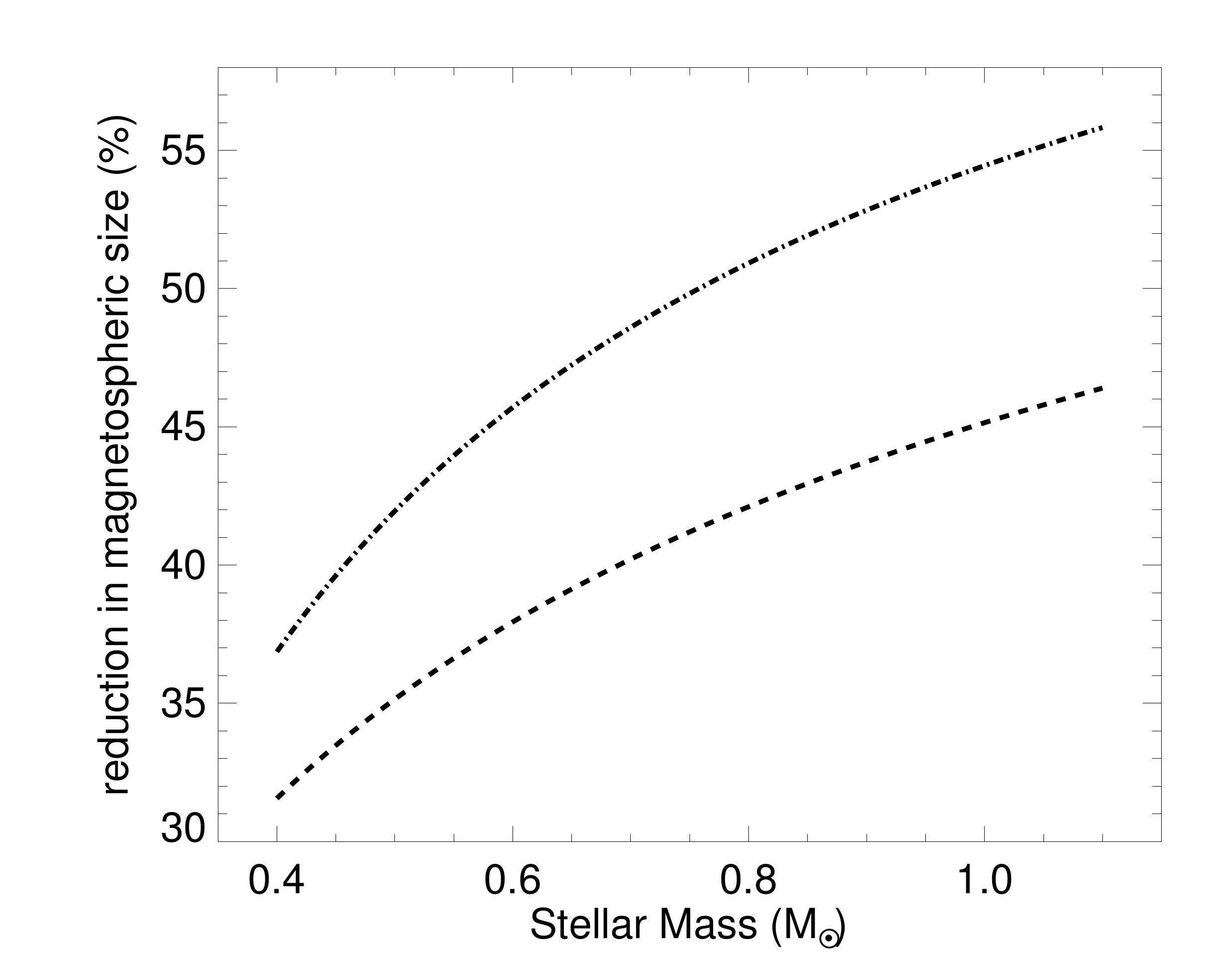}
\caption{
Figures showing the saturation levels for the ram pressure at 1AU (\emph{upper panel}) and the reduction in a planet's magnetospheric standoff distance that this would lead to relative to the standoff distance that the magnetosphere would have in the current slow solar wind (\emph{lower panel}). 
The ram pressure values are given in units of the ram pressure of the current slow solar wind at 1~AU.
As in Fig.~\ref{fig:ModelAModelBcomparison}, the dot-dashed lines correspond to the predictions of Model~A and the dashed lines correspond to the predictions of Model~B. 
}
 \label{fig:ModelAModelBcomparison2}
\end{figure}

\begin{figure}
\centering
\includegraphics[width=0.49\textwidth]{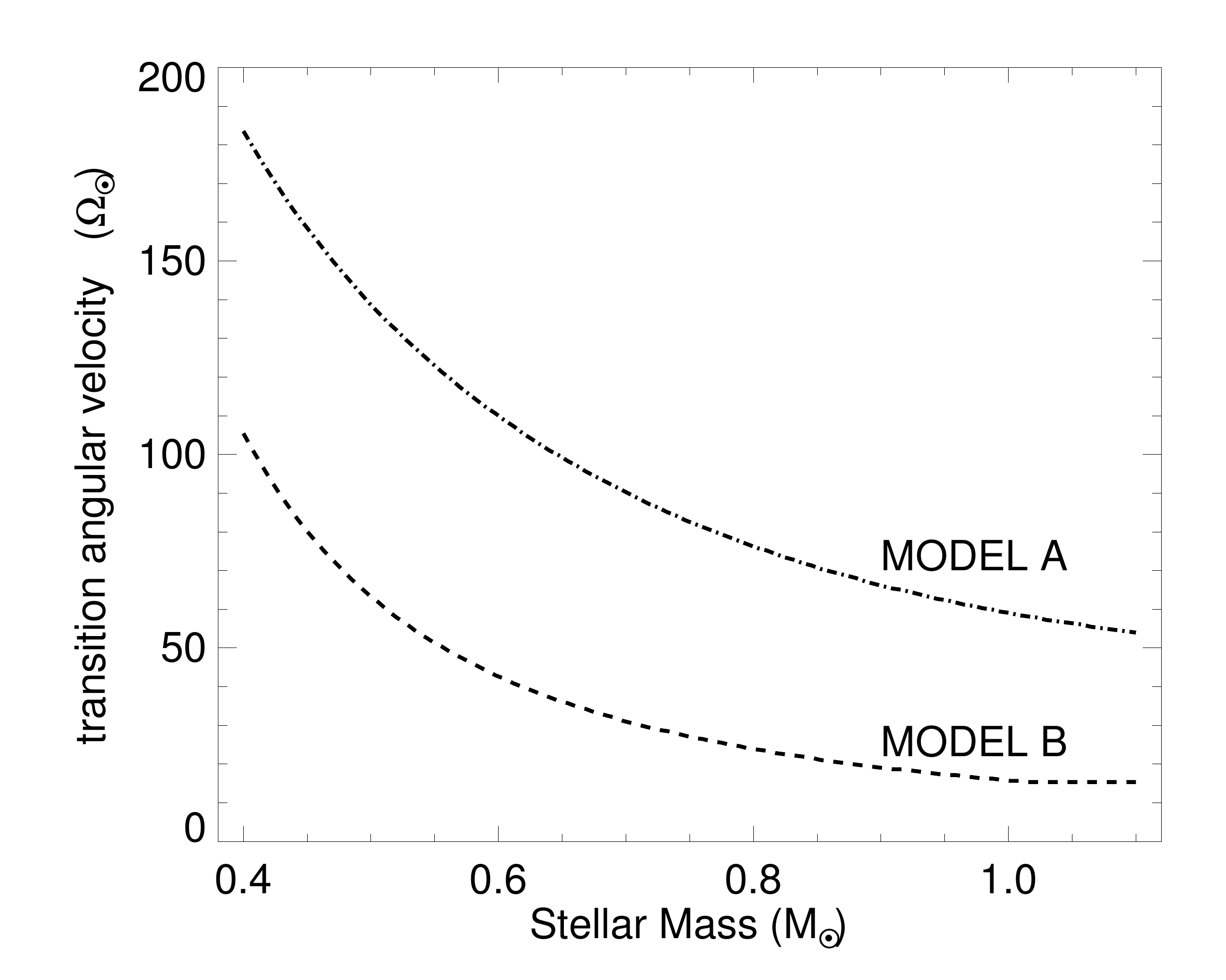}
\caption{
Figure showing the stellar rotation rates at which stars transition from the SMR regime to the FMR regime as a function of stellar mass for both Model~A and Model~B. 
This transition is assumed to happen when the Michel velocity becomes larger than the speed that the wind would have due only to acceleration by thermal pressure gradients.
}
 \label{fig:FMRtransition}
\end{figure}

Although the mass fluxes are the same in our two wind models, given the differences in the wind speeds and densities, our models can predict radically different wind ram pressures.
In the upper panel of Fig.~\ref{fig:ModelAModelBcomparison2}, we show the slow wind ram pressure at 1~AU as a function of mass for the two models.
Since it represents the most extreme case of the difference between the two models, we only show the values for stars at the saturation threshold. 
In the case of Model~B, the wind ram pressure for a rapidly rotating solar mass star is only a factor of $\sim$40 higher than it is in the current solar wind. 
For Model~A, since the increased mass flux at rapid rotation is a result of both increased density and increased wind speed, there is a much larger change in the ram pressure, with a rapidly rotating solar mass star having a ram pressure that is a factor of $\sim$120 higher than it is in the current solar wind.
This factor of three larger ram pressure in Model~A in comparison to Model~B is a result of the wind speed being a factor of three larger for rapidly rotating solar mass stars than the current solar wind speed in Model~A, as can be seen in the middle left panel of Fig.~\ref{fig:ModelAModelBcomparison}. 

To demonstrate why this uncertainty in the wind ram pressure can be important, we calculate the reduction in the standoff distance for a planetary magnetosphere that would be caused by these winds relative to the standoff distance that the magnetosphere would have in the current slow solar wind.
This is shown in the lower panel of Fig.~\ref{fig:ModelAModelBcomparison2}.
Although wind-planet interactions are highly complicated and cannot simply be represented by the magnetospheric standoff distance, this is a quantity that is easily calculated using simple pressure balance arguments. 
The standoff distance is approximately where the magnetic pressure of the planetary magnetic field is equal to the ram pressure in the wind, though in some cases, the magnetic pressure from the star's magnetic field might also have some influence (\citealt{2013A&A...557A..67V}).
Since for a dipolar magnetic field, the planetary magnetic pressure, $P_\text{B}$, depends on distance from the planet as $P_{\text{B}} \propto r^{-6}$, this means that for a given planetary magnetic field, the magnetospheric standoff distance, $r_{\text{M}}$, depends on the wind ram pressure, $P_{\text{ram}}$, as $r_{\text{M}} \propto P_{\text{ram}}^{-1/6}$.
A two orders of magnitude increase in the wind ram pressure leads to a 50\% decrease in the magnetospheric size.  
As can be seen in Fig.~\ref{fig:ModelAModelBcomparison2}, the higher ram pressures predicted by Model~A lead to much smaller planetary magnetospheres around rapidly rotating stars than in Model~B.

Although we do not consider magneto-rotational acceleration of the wind in detail, it is interesting to estimate at which rotation rates it becomes important. 
As we describe in Section~\ref{sect:magnetorotation}, we assume stars enter the FMR regime when the Michel velocity becomes larger than the speed that the wind would have simply due to acceleration by thermal pressure gradients.
In Model~A, due to the higher wind speeds for the winds of rapidly rotating stars, this transition will be at higher rotation rates than in Model~B.
In Fig.~\ref{fig:FMRtransition}, we show at which angular velocity the transition from the SMR regime to the FMR regime happens as a function of mass for both stars.
For a solar mass star, this happens at $\sim 60\Omega_\odot$ and $\sim 15\Omega_\odot$ for Model~A and Model~B respectively. 
For a 0.5~M$_\odot$ star, this happens at $\sim 135\Omega_\odot$ and $\sim 60\Omega_\odot$ for Model~A and Model~B respectively. 
It is therefore unlikely that magneto-rotational acceleration will be significant in the unsaturated regime at any mass.
In the saturated regime, however, it is clearly very important for the most rapid rotators.

\section{Summary and Discussion} \label{sect:summaryfinal}

Despite their importance, the properties of the winds of low-mass stars are poorly constrained. 
Significant progress has been made in recent years in modelling these winds, with particular attention being given to the dynamical interplay between the coronal magnetic field and the wind (e.g. \citealt{2009ApJ...699..441V}; \citealt{2012ApJ...754L..26M}; \citealt{2014ApJ...783...55C}).
In this paper, we concentrate on the scaling of the free parameters in the models to the stellar winds of stars based on their fundamental parameters and how these scaling laws influence the propagation of the stellar wind in interplanetary space.

Our wind model is based on a 1D thermal pressure driven model for the solar wind run using the \emph{Versatile Advection Code}. 
In order to heat the wind as it expands, we assume a polytropic equation of state with a spatially varying polytropic index, $\alpha$. 
We use observations of the solar wind close to the solar surface and \mbox{\emph{in situ}} measurements of the solar wind from four spacecraft to constrain the values of $\alpha$ and the base densities and temperatures in our wind for both the slow and fast components of the solar wind. 
Our model provides an excellent description of the solar wind far from the solar surface.
However, likely due to our assumption that the wind is driven by thermal pressure gradients and the fact that we neglect the influence of the solar magnetic field geometry on the wind expansion, our model is unrealistic at small distances from the solar surface.
For example, in order to reproduce the fast wind speed at 1~AU, we need to assume an unrealistically high wind temperature close to the solar surface.

Scaling the solar wind model to other stars is a difficult problem that has not been solved.
We have treated the base density and base temperature in our solar wind model as free parameters and constrained them using \mbox{\emph{in situ}} measurements of the solar wind. 
However, no such measurements of the winds of other stars are available.
We must therefore constrain the free parameters in our wind model using more indirect methods.
A significant input into our wind model is developed in Paper~II of this series, where we construct a rotational evolution model and derive a scaling law for the mass loss rate based on fitting our rotational evolution model to the observational constraints.
We find that the mass loss rate can be described as $\dot{M}_\star \propto R_\star^{2} \Omega_\star^{1.33} M_\star^{-3.36}$.  
The advantage of this approach is that it is not dependent on arbitrary assumptions about how the free parameters scale to other stars.
However, there are also disadvantages to this approach, as we discuss in detail in Section~5.3 of Paper~II.

Knowing the mass loss rate alone tells us the mass flux in the wind, i.e. $\rho v$, but does not allow us to predict the values of $\rho$ and $v$ separately. 
In order to disentangle these two quantities, we produce two models for scaling the solar wind model to other stars corresponding to two radically different assumptions about how the wind temperature depends on the stellar properties.
In Model~A, we assume that the wind temperature scales linearly with the coronal temperature, and in Model~B, we assume that the sound speed at the base of the wind is a fixed fraction of the surface escape velocity.
Both of these models are motivated by the example of the solar wind but we do not know which of them is more realistic. 
For Model~A, we use the scaling relation between coronal temperature and X-ray surface flux derived by Johnstone \& G\"{u}del (2015) to calculate coronal temperatures.

The wind temperature is the most important parameter in determining the properties of thermal pressure driven winds.
In order to test this dependence, we produce a grid of 1200 wind models with different stellar masses, radii, and wind temperatures.
Although the surface escape velocity, proportional to \mbox{$\sqrt{M_\star / R_\star}$}, is significant for determining the wind properties, it can be considered approximately constant for all low-mass main-sequence stars, and we therefore find that our wind properties can be predicted from the base temperature, the base density, and the stellar surface area only. 
In Section~\ref{sect:grid}, we derive scaling laws for the wind speed and temperature as a function of base temperature, and a scaling law for the mass loss rate as a function of base temperature, base density, and stellar surface area.

%
%

Our model can easily be applied by anyone wanting to estimate the properties of stellar winds far from the stellar surface using the scaling laws given in the previous sections without having to run hydrodynamic simulations. 
This requires that the stellar mass and rotation rate are known and is only possible in the region far from the stellar surface after the wind has come to approximately its terminal velocity. 
The solar wind comes close to its terminal velocity by approximately 30~$R_\odot$, though hotter winds will accelerate faster. 
This can be done for the slow and fast wind models separately.
In order to do this, a decision must be made for which of the two models for scaling the wind base temperature is more reasonable. 
Once this is done, the following steps can be taken to predict the wind density, velocity, and temperature:

\begin{enumerate}

\item
Firstly, based on the assumption of either Model~A or Model~B, the base temperature of the wind must be estimated.
For Model~A, the base temperature is calculated from the coronal temperature, which is calculated from the surface X-ray flux as $\bar{T}_{\text{cor}} \approx 0.11 F_\text{X}^{0.26}$, where $\bar{T}_{\text{cor}}$ is in MK and $F_\text{X}$ is in \mbox{erg s$^{-1}$ cm$^{-2}$}.
If no measurements of $F_\text{X}$ for the star are available, it can be estimated using the scaling relations for X-ray emission derived by \citet{2011ApJ...743...48W} and given in Eqn.~\ref{eqn:wrightlaw}.
The base temperatures for the slow and fast winds are then given by $T_0 = 0.75 \bar{T}_{\text{cor}}$ and $T_0 = 1.58 \bar{T}_{\text{cor}}$ respectively. 
For Model~B, the base temperature of the wind can be calculated from Eqn.~\ref{eqn:T0ModelB}, assuming $c_s/v_{\text{esc}}$ has values of 0.329 and 0.478 for the slow and fast winds respectively.

\vspace{3mm}
\item
When $T_0$ is known, the wind speed at the desired distance from the centre of the star, $r$, should be estimated. 
Firstly, the value at 1~AU should be estimated using Eqn.~\ref{eqn:windspeedgrid}.
The wind speed can crudely be assumed to be constant far from the stellar surface, and so the 1~AU value can be assumed.
More accurate would be to assume a constant acceleration of the wind, and therefore the wind speed at 1~AU can be scaled to $r$ by 

\begin{equation}
v(r) = v_{1\text{AU}} + \left( r - r_{1\text{AU}} \right) \frac{dv}{dr},
\end{equation}

\noindent where $dv/dr$ is the wind acceleration far from the star and $r_{1\text{AU}} = 1$~AU.
The wind acceleration can be estimated from $T_0$ using Eqn.~\ref{eqn:windaccelgrid}.
In the case of rapidly rotating stars, the Michel velocity, given in Eqn.~\ref{eqn:MichelVel}, should be calculated.
If this is much larger than the wind speeds at 1~AU, then our thermal pressure driven model is not likely to be appropriate for determining the wind speed.
If only the wind conditions in the equatorial plane are desired, either the Michel velocity can be assumed for the wind speed far from the star, or the full equations for the Weber-Davis model can be solved.
For detailed descriptions of these equations, see \citet{1967ApJ...148..217W}, \citet{1976ApJ...210..498B}, \citet{1999isw..book.....L}, and \citet{2005A&A...434.1191P}.

\vspace{3mm}
\item
Next, the wind mass loss rate should be estimated.
The mass loss rate can be calculated from 

\begin{equation}
\dot{M}_\star = \dot{M}_\odot R_\star^{2} \Omega_\star^{1.33} M_\star^{-3.36},
\end{equation}

\noindent where \mbox{$\dot{M}_\odot = 1.4 \times 10^{-14}$~\mdot} is the solar wind mass loss rate, and the other quantities are in solar units (where the Carrington rotation rate of \mbox{$\Omega_\odot = 2.67 \times 10^{-6}$~rad~s$^{-1}$} should be used for the solar rotation rate).
In fact, it is not necessary to use our method in this step.
If an alternative method is preferred, such as the model of \citet{2011ApJ...741...54C} or the constraints of \citet{2005ApJ...628L.143W}, then these mass loss rates can be used instead.

\vspace{3mm}
\item
With the mass loss rate and the wind speed at $r$, the wind density is easily calculated from \mbox{$\rho(r) = \dot{M}_\star / 4 \pi r^2 v(r)$}.
This can be done for both the slow and fast wind models using the same value of $\dot{M}_\star$ given that they have approximately the same mass fluxes.

\vspace{3mm}
\item
Finally, the wind temperature at $r$ can be estimated.
Firstly, the wind temperature at 1~AU should be calculated from the base temperature using Eqn.~\ref{eqn:windtempgrid}, such that $T_{1\text{AU}} = 0.054 T_0 - 0.010$, where both temperatures are in MK. 
The temperature at $r$ is then determined by the assumption of a polytropic equation of state, such that $T(r) = T_{1\text{AU}} \left[ \rho(r) / \rho_{1\text{AU}}  \right]^{\alpha-1}$, where $\alpha = 1.51$ and $\rho_{1\text{AU}}$ can be estimated using the previous step for $r_{1\text{AU}}$.

\end{enumerate}

\noindent The above method provides a good approximation for the wind properties that we calculate from our hydrodynamic wind model beyond $\sim$30~R$_\star$. 
We emphasise that this simple method should not be applied for calculating the wind properties close to the star.

In Paper~II, we discuss in detail the uncertainties in our determinations of the wind mass loss rate as a function of stellar mass and rotation. 
Assuming these mass loss rates are approximately correct, the largest uncertainty in our wind model is how the base temperature should be scaled to the winds of other stars.
As we show in Section~\ref{sect:resultsAandB}, for slow rotators, the scaling of the wind temperature is not likely to be a significant problem.
However, for rapidly rotating stars, different assumptions about the wind temperature lead to radically different predictions for the wind speeds and densities for a given mass flux. 
The assumption that wind temperature scales approximately with coronal temperature, which is certainly plausible, leads to fast winds with low densities. 
The assumption that the wind temperature is approximately constant leads to slow dense winds for the same mass flux. 
These assumptions lead to very different predictions for the ram pressures.
It is therefore critical that a solution to this problem is found. 
It is likely that the solution will eventually come from a detailed physical understanding of the solar wind applied to other stars, and some progress has already been made in this direction (\citealt{2011ApJ...741...54C}; \citealt{2013PASJ...65...98S}; \citealt{2014ApJ...790...57C}).
These solutions, however, will still require observational confirmation, which means that it is necessary to measure and disentangle the properties of the winds from several stars. 

One interesting possibility for doing this is the comparison of astrospheric Ly$\alpha$ absorption with radio interferometric measurements of thermal emission from the wind. 
In the case of astrospheric Ly$\alpha$ absorption, the measurements are mostly sensitive to the wind ram pressure, $\dot{M}_\star v_{\text{w}}$, whereas the observations at radio wavelengths are sensitive to the wind density, $\dot{M}_\star / v_{\text{w}}$ (\citealt{1975MNRAS.170...41W}; \citealt{1975A&A....39....1P}).
Measurements of both of these quantities for a given stellar wind would allow us to disentangle the wind density and velocity. 
It is not clear, however, if this method will be helpful.
In order to determine how the wind temperature scales to other stars, we need speed and density measurements ideally for a wind from an active rapidly rotating star.
However, it is unclear whether wind measurements from astrospheric Ly$\alpha$ absorption can be achieved for highly active stars, with current measurements for the most active stars providing unclear results (\citealt{2005ApJ...628L.143W}; \citealt{2014ApJ...781L..33W}).
Additionally, detections of thermal radio emission from winds is difficult, with previous attempts leading to non-detections (\citealt{2000GeoRL..27..501G}).
Where detections are achieved, it is likely to be difficult to distinguish between emission from the stellar corona and emission from the wind.
This is especially problematic for the most active stars given that their coronae emit strongly at radio wavelengths (\citealt{1993ApJ...405L..63G}; \citealt{1994A&A...285..621B}).

Another possibility for measuring stellar wind properties that has received attention in the literature recently is from stellar observations during planetary transits. 
Interactions between a stellar wind and a transiting planet with a small orbital distance can lead to measurable effects on the transits themselves.
One such possibility is the measurements of the Ly$\alpha$ emission line in and out of transit.
Charge exchange between neutral hydrogen atoms from the planetary atmosphere with stellar wind protons leads to the formation of neutral hydrogen tails behind the planet (\citealt{2014A&A...562A.116K}).
These neutral hydrogen atoms can then absorb large amounts of the star's Ly$\alpha$ emission line, leading to strong absorption in Ly$\alpha$ during and slightly after the planetary transit. 
\citet{2014ApJ...786..132K} measured such absorption during the transit of the M-dwarf GJ~436 by the hot Neptune GJ~436b. 
It is possible that detailed modelling of the wind interactions with the planetary magnetosphere and atmosphere can be used to derive the properties of the stellar wind at the planetary orbit (\citealt{2011Ap&SS.335....9L}; Kislyakova et al. 2014). 
Another possibility for inferring wind properties from planetary transits comes from the recent discovery that the transit of the hot Jupiter \mbox{WASP-12b} was deeper and began earlier in near-UV than in optical wavelengths (\citealt{2010ApJ...714L.222F}; \citealt{2012ApJ...760...79H}). 
This could be due to the presence of dense material located several planetary radii in front of the planet as it orbits its host star.  
Currently, several explanations for the existence of such material have been proposed (\citealt{2010ApJ...721..923L}; \citealt{2010ApJ...722L.168V}; \citealt{2011MNRAS.416L..41L}; \citealt{2013ApJ...764...19B}).
In cases like that of \mbox{WASP-12b}, the form of the transit lightcurve could be dependent on the wind conditions near the planet (\citealt{2013MNRAS.436.2179L}), and so such transits might provide important constraints on the wind properties.

Winds from low-mass main-sequence stars have rightfully received a lot of attention in recent years.
The subject has become of particular importance due to the discoveries of hundreds of planets orbiting other stars (at the time of writing, more than 1800 planets have confirmed detections), and the recent attention in the literature to the astrophysical conditions required for the formation of habitable planetary environments (\citealt{2014arXiv1407.8174G}).
Given the sensitivity of the development of planetary atmospheres to interactions with their host stars, understanding the formation and propagation of stellar winds is of primary importance.
We attempt in this paper to construct a stellar wind model that can be used to predict the wind properties of other stars at all distances within a few AU of the stellar surface, although many open questions still remain.
In Paper~II of this series, we couple our wind model to observationally driven models for the evolution of rotation to explore how wind properties from stars with a range of masses evolve with time on the main-sequence.


\section{Acknowledgments} 

The authors thank the referee, Brian Wood, for providing useful comments that helped us to improve the manuscript.
CPJ thanks Kristina Kislyakova and Aline Vidotto for their useful comments on the manuscript.
CPJ, MG, and TL acknowledge the support of the FWF NFN project S116601-N16 ``Pathways to Habitability: From Disks to Active Stars, Planets and Life'', and the related FWF NFN subproject S116604-N16 ``Radiation \& Wind Evolution from T~Tauri Phase to ZAMS and Beyond''. 
This publication is supported by the Austrian Science Fund (FWF).


\appendix

\section{Method for determining the variations in the polytropic constant with radius} \label{appendix:polyK}

\begin{figure}
\centering
\includegraphics[trim = 0mm 4mm 7mm 0mm, clip=true,width=0.49\textwidth]{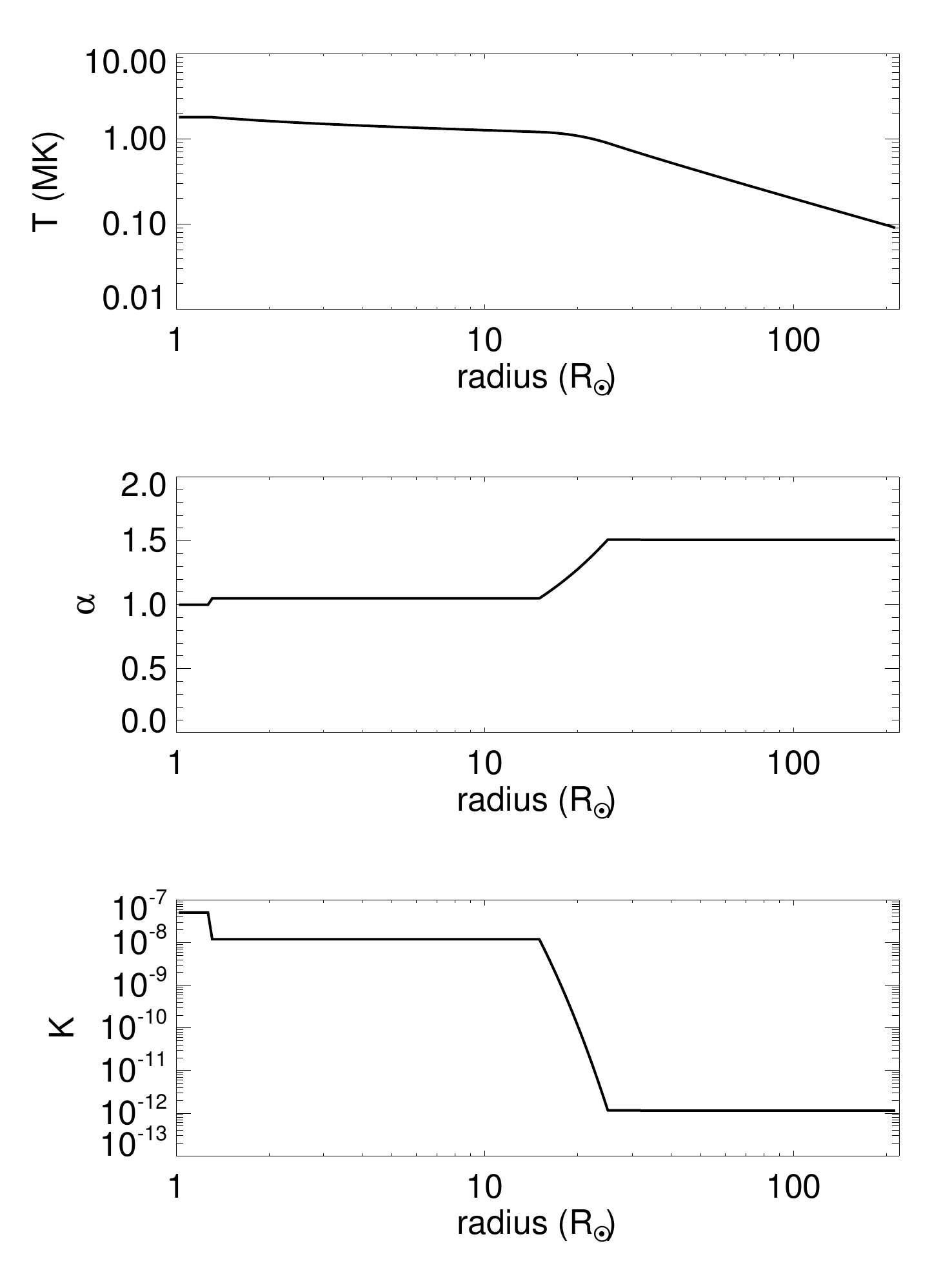}
\caption{
Plots demonstrating the radial structure of the wind temperature (\emph{upper panel}), the polytropic index (\emph{middle panel}), and the polytropic constant (\emph{lower panel}) for the slow solar wind simulation presented in Section~\ref{sect:solarwindmodel}.
The units of the polytropic constant are \mbox{R$_\odot^{3\alpha-1}$ kg$^{1-\alpha}$}.
}
 \label{fig:alphaKdemonstration}
\end{figure}

In our solar wind model, we heat the wind by assuming a polytropic equation of state, where the pressure, $p$, is related to the mass density, $\rho$, by $p = K \rho^\alpha$.  
In order to accurately reproduce the solar wind, we assume that $\alpha$ is a function of distance from the Sun as described in Section~\ref{sect:solarwindmodel}.
We assume $\alpha (r)$ is given by

\begin{equation} \label{eqn:polytropicindex}
\alpha (r) = \left \{
\begin{array}{ll}
1.0, & \text{if } r \le 1.3 R_\star,\\
\alpha_{\text{in}}, & \text{if } 1.3 R_\star < r \le R_{\text{in}},\\
\alpha_{\text{in}} + \left( \frac{\alpha_{\text{out}} - \alpha_{\text{in}}}{R_{\text{out}} - R_{\text{in}}} \right) \left( r - R_{\text{in}} \right), & \text{if } R_{\text{in}} < r \le R_{\text{out}},\\
\alpha_{\text{out}}, & \text{if } r > R_{\text{out}},
\end{array} \right.
\end{equation}

\noindent where $R_{\text{in}}$ and $R_{\text{out}}$ are the radii between which $\alpha$ changes linearly. 
The structure of $\alpha (r)$ is shown in Fig.~\ref{fig:alphaKdemonstration}.

When assuming a polytropic equation of state, the temperature is given by $T = \frac{\mu m_\text{p}}{k_\text{B}} K \rho^{\alpha-1}$.
It is normal for the polytropic constant, $K$, to be uniform over all space.
However, when $\alpha$ has significant spatial variations, this leads to an unphysical temperature structure.
Therefore, at 1.3~R$_\star$ and between $R_{\text{in}}$ and $R_{\text{out}}$, we need to vary $K$ in response to the changes in $\alpha$. 
Since it is therefore a function of time in our hydrodynamic simulations before the simulations relax to a steady state, we must recalculate $K$ at the beginning of each time step. 
Our method for calculating the radial structure of $K$ does not introduce any new free parameters into the model, but it does require making arbitrary assumptions about how the temperature changes between $R_{\text{in}}$ and $R_{\text{out}}$. 
\emph{We stress however, that our final result is only weakly sensitive to exactly which assumptions we make, as long as our assumptions lead to reasonably smooth temperature variations.}

At 1.3~R$_\star$, the necessary change in $K$ is easily calculated. 
If we assume that a quantity with the subscript $i$ represents the value of that quantity at the $i$th grid cell, our algorithm for calculating $K(r)$ at the beginning of each time step is as follows. 
We start by calculating $K_{\text{surface}}$, which is the polytropic constant for each grid point where $r_i < 1.3 R_\star$.
This is given by  

\begin{equation}
K_{\text{surface}} = \frac{k_\text{B}}{\mu m_\text{p}} T_0,
\end{equation}

\noindent where $T_0$ is the base temperature. 
We then calculate $K_{\text{in}}$, which is the polytropic constant in the region between 1.3~R$_\star$ and $R_{\text{in}}$, by assuming that the temperature at the first grid cell where $r_i > 1.3 R_\star$ is equal to the temperature of the previous cell. 
This is given by

\begin{equation}
K_{\text{in}} = \frac{k_\text{B}}{\mu m_\text{p}} T_0 \rho_i^{1-\alpha_i}.
\end{equation}

The region where $\alpha$ changes linearly with radial distance is more complicated.
In regions where $\alpha$ is uniform, the wind temperature structure as a function of distance can be approximated as a power law, where $T \propto r^{\delta}$, and the corresponding radial gradient in temperature is $dT/dr = T r^{-1} \delta$.
In regions where $\delta$ varies with $r$, $\delta$ must also vary and the gradient of temperature is not as simple.
For simplicity, we assume that between $R_{\text{in}}$ and $R_{\text{out}}$, $dT/dr = T r^{-1} \delta$ and $\delta$ varies linearly between these two radii, giving

\begin{equation} \label{eqn:deltaassumption}
\delta (r) = \delta_{\text{in}} + \left( \frac{\delta_{\text{out}} - \delta_{\text{in}}}{R_{\text{out}} - R_{\text{in}}} \right) \left( r - R_{\text{in}} \right).
\end{equation}

\noindent This ensures that the wind temperature structure changes smoothly between $R_{\text{in}}$ and $R_{\text{out}}$. 
We calculate values of $\delta_{\text{in}}$ and $\delta_{\text{out}}$ at the beginning of each time step based on the wind structure calculated in the previous time step. 
The value of $\delta_{\text{in}}$ is based on all radii between 5~R$_\star$ and 15~R$_\star$ and the value of $\delta_{\text{out}}$ is based on all radii between 25~R$_\star$ and 35~R$_\star$
Then starting at the first grid cell where \mbox{$r > R_{\text{in}}$}, we calculate $K_i$ by the following three steps:

\begin{enumerate}

\item
We first calculate the temperature gradient at the previous grid cell using

\begin{equation}
\frac{dT_{i-1}}{dr} = T_{i-1} r_{i-1}^{-1} \delta_{i-1},
\end{equation}

\noindent where $T_{i-1}$ is given by 

\begin{equation}
T_{i-1} = \frac{\mu m_\text{p}}{k_\text{B}} K_{i-1} \rho_{i-1}^{\alpha_{i-1} - 1},
\end{equation}

\noindent and $\alpha_{i-1}$and $\delta_{i-1}$ are calculated from Eqn.~\ref{eqn:polytropicindex}  and Eqn.~\ref{eqn:deltaassumption} respectively.

\vspace{2mm}

\item
We then calculate the temperature at the $i$th grid point from

\begin{equation}
T_i = T_{i-1} + \frac{dT_{i-1}}{dr} \left( r_i - r_{i-1} \right).
\end{equation}

\vspace{2mm}

\item
Finally, we calculate $K_i$ from

\begin{equation}
K_i = \frac{k_\text{B}}{\mu m_\text{p}} T_i \rho_i^{1-\alpha_i}.
\end{equation}

\end{enumerate}

\noindent These three steps are equivalent to 

\begin{equation}
K_i = K_{i-1} \left[ 1 + \delta_{i-1} \left( \frac{r_i}{r_{i-1}} - 1 \right)  \right] \frac{\rho_i^{1-\alpha_i}}{\rho_{i-1}^{1-\alpha_{i-1}}}.
\end{equation}

\noindent We then repeat this for each grid cell moving outwards until the first grid cell where $r_i > R_{\text{out}}$. 
For each grid cell beyond $R_{\text{out}}$, we assume $K_i = K_{\text{out}}$, where $K_{\text{out}}$ is simply taken as the value of $K$ at the final cell where $r_i < R_{\text{out}}$. 

Our wind simulations start with a uniform density and temperature gas filling the entire domain.
A shock forms at the stellar surface and moves out through the domain.
During this time, the temperature structure between $R_{\text{in}}$ and $R_{\text{out}}$ calculated with the above method is physically unrealistic. 
However, once the simulation has settled to a steady state, we are left with a reasonable temperature structure, as can be seen in Fig.~\ref{fig:alphaKdemonstration}.

\bibliographystyle{aa}
\bibliography{mybib}

\end{document}